\documentclass[notitlepage,a4paper,aps,prd,onecolumn,superscriptaddress,nofootinbib,groupedaddress]{revtex4}

\usepackage{amsmath,comment}
\usepackage{amsfonts,color}
\usepackage{amsmath}
\usepackage{amsthm}
\usepackage{graphicx}
\graphicspath{{Figures/}}
\usepackage{pdfpages}
\usepackage{amsmath}
\usepackage{empheq}
\usepackage{amsfonts,color}
\usepackage{amssymb,float}
\usepackage{physics}
\usepackage{dsfont}
\usepackage{booktabs,multirow}
\usepackage{titlesec}
\usepackage[none]{hyphenat}
\titleformat{\paragraph}
{\normalfont\normalsize\bfseries}{\theparagraph}{1em}{}
\titlespacing*{\paragraph}
{0pt}{3.25ex plus 1ex minus .2ex}{1.5ex plus .2ex}

\usepackage[none]{hyphenat}

\usepackage{amsmath}
\usepackage[utf8]{inputenc}
\allowdisplaybreaks
\usepackage{color}
\usepackage{hyperref}
\hypersetup{
    colorlinks=true,
    linkcolor=blue,
    filecolor=magenta,      
   citecolor=blue
}

\usepackage{enumitem}

\usepackage{tikz}
\usetikzlibrary{shapes.geometric}
\usetikzlibrary{arrows.meta,arrows}

\usepackage{booktabs}
\newcommand{\branchcell}[2]{%
\begin{minipage}[t][#1][c]{\linewidth}
\vspace{0pt}
\centering
\bfseries #2
\end{minipage}%
}

\begin{document}
\title{Black holes with torsion hair in cubic Holst-type Poincaré gauge gravity:\newline from singular to regular geometries}

\author{Sebastian Bahamonde}
\email{sbahamondebeltran@gmail.com}
\affiliation{Cosmology, Gravity, and Astroparticle Physics Group, Center for Theoretical Physics of the Universe,
Institute for Basic Science (IBS), Daejeon, 34126, Korea.}

\author{Jorge Gigante Valcarcel}
\email{jorgevalcarcel@ibs.re.kr}
\affiliation{Center for Geometry and Physics, Institute for Basic Science (IBS), Pohang 37673, Korea.}

\author{Matteo Magi}
\email{mmagi@ibs.re.kr}
\affiliation{Cosmology, Gravity, and Astroparticle Physics Group, Center for Theoretical Physics of the Universe,
Institute for Basic Science (IBS), Daejeon, 34126, Korea.}

\begin{abstract}

Motivated by the singularity theorems of Poincaré Gauge (PG) theory, we investigate extensions of the Holst quadratic model by introducing cubic order invariants constructed from the curvature and torsion tensors into the gravitational action. Such models are characterised by a kinetic structure that is governed by a pseudoscalar mode, whereas the remaining irreducible modes of torsion contribute through nonlinear interactions that can have important implications for the space-time geometry. In particular, in line with other well-known models of PG theory, the Birkhoff theorem does not hold in general, allowing for new exact static and spherically symmetric black hole solutions with dynamical torsion. Across the different torsion sectors, corresponding to the irreducible modes and parity components of the torsion field involved in the analysis, we find both singular and regular configurations. Among the singular solutions, we obtain Kiselev-like and Boulware-Deser-like geometries, as well as new geometries with distinct algebraic and Lambert $W$ metric corrections. In addition, we find regular black holes with both primary and secondary torsion hair, which evade the singularity theorems through violations of the causal convergence conditions induced by the nonlinear torsion interactions. Therefore, we show that Holst-type PG models can support a rich variety of black hole geometries, while providing explicit mechanisms for evading the singularity theorems of PG theory.

\end{abstract}

\maketitle
\section{Introduction}

Since its inception, General Relativity (GR) has provided the standard description of gravitation, successfully explaining a wide range of phenomena, including the formation of black holes, the large-scale evolution of the universe and the emission of gravitational waves from compact binary mergers~\cite{Will:2014kxa,LIGOScientific:2016aoc,LIGOScientific:2017vwq}.

In GR, the gravitational interaction has a geometrical character: it is encoded in the space-time curvature, which is mathematically represented by a Riemannian manifold equipped with the metric tensor and the Levi-Civita connection. Indeed, according to the Fundamental Theorem of Riemannian geometry, the latter is uniquely determined, being symmetric and metric-compatible.

Despite its remarkable success, several considerations motivate extensions of GR. In particular, the coupling of gravity to matter fields with intrinsic spin points out that the space-time geometry may possess not only curvature but also torsion~\cite{Hehl:1971qi,Yasskin:1980bu,Audretsch:1981xn,Obukhov:2014fta,Cembranos:2018ipn,Fadeev:2020gjk,Trukhanova:2022utt,Bahamonde:2025fei,Bahamonde:2026vqm}. In this case, the gravitational field is described by a gauge field associated with the external rotations and translations of the Poincaré group~\cite{Hehl:1976kj,Obukhov:1987tz,Blagojevic:2013xpa,ponomarev2017gauge,Obukhov:2022khx}. Thereby, a large number of scalar invariants constructed from the curvature and torsion tensors can be included into an action functional to endow these modes with dynamics. In this sense, different restrictions imposed on the Lagrangian coefficients lead to a broad class of gravitational models, for which a wide range of fundamental properties and phenomenological implications may arise. Thus, one of the most relevant aspects of this construction concerns stability, as distinct types of ghostly, gradient and tachyonic instabilities arise for generic values of these coefficients~\cite{Neville:1978bk,Sezgin:1979zf,Sezgin:1981xs,Miyamoto:1983bf,Fukui:1984gn,Fukuma:1984cz,Battiti:1985mu,Kuhfuss:1986rb,Blagojevic:1986dm,Baikov:1992uh,Yo:1999ex,Yo:2001sy,Lin:2018awc,Jimenez:2019qjc,Bahamonde:2024sqo}.

Among the different choices of quadratic couplings, a particularly interesting subclass is given by the Holst quadratic model. Specifically, this model can be considered as a quadratic order correction to the standard Holst linear action, which has been shown to play an important role in nonperturbative quantisation schemes to reproduce the Bekenstein-Hawking entropy law~\cite{Holst:1995pc,Ashtekar:2004eh,Perez:2005pm}. Due to its relatively simple structure, the pathological interactions that generally plague the quadratic models of Poincaré Gauge (PG) theory are directly excluded from the action~\cite{Yo:1999ex,Yo:2001sy,Jimenez:2019qjc}, while at the same time it admits an interesting phenomenology~\cite{Chen:2009at,delaCruzDombriz:2021nrg,Casado-Turrion:2023omz}.

Following these lines, in the present work we investigate extensions of the Holst quadratic model of PG theory by including into the gravitational action cubic order invariants defined from mixing terms of the curvature and torsion tensors. The inclusion of cubic invariants considerably enriches the structure of the theory, which makes the model particularly suitable for exploring departures from the standard Riemannian picture in empty space-times, where torsion remains as a propagating field. In this regard, a central consequence is the absence of a general Birkhoff theorem, which allows us to obtain different exact static and spherically symmetric black hole solutions, exhibiting both singular and regular geometries. Indeed, the search for such regular space-times is deeply motivated by the singularity theorems of PG theory~\cite{Cembranos:2016xqx}, which in the present extension of the Holst quadratic model can be evaded through the nonlinear interactions provided by the mixing terms of the curvature and torsion tensors. Thus, the detailed study of these novel space-times provides a useful framework for understanding how higher order invariants in PG theory can resolve gravitational singularities. By exploring these exact solutions, we aim to shed light on the physical implications of the torsion field and the viability of these regular geometries as theoretical alternatives to the standard black hole paradigm.

This paper is organised as follows. In Sec.~\ref{sec:RC} we establish the definitions and conventions adopted throughout this work for the main quantities arising in Riemann-Cartan (RC) geometry, which constitutes a suitable framework for describing a space-time endowed with curvature and torsion. Then, in Sec.~\ref{sec:model} we construct the theoretical framework by extending the Holst quadratic model of PG theory with cubic order invariants, presenting the mixing terms between the curvature and torsion tensors, as well as the resulting field equations. These equations are restricted to static and spherically symmetric RC space-times in Sec.~\ref{sec:BHs}, where we impose a first set of regularity conditions to prevent torsion singularities at the event horizons, which allows us to simplify the problem and determine the possible contributions of torsion to the metric structure in the reciprocal case. Building on this result, we systematically solve the field equations for different sectors of the theory, corresponding to the specific irreducible modes and parity components of the torsion field involved in the analysis. Thus, in Sec.~\ref{sec:Kiselev} we first consider the sector given by the odd-parity components of the tensor mode, which leads to Kiselev-like black hole solutions. We then turn to the even-parity vector-tensor sector in Sec.~\ref{subsec:PQmixed}, which admits two distinct branches of solutions, respectively featuring algebraic and Lambert $W$ metric corrections. Focusing on a different even and odd-parity axial-tensor sector allows us to obtain in Sec.~\ref{sec:EGB} two more black hole configurations: the former displaying further Lambert $W$ metric corrections and the latter exhibiting a Boulware-Deser-like geometry. In all these cases the central singularities persist, but the present class of Holst-type PG models also admits regular black hole solutions. In particular, in Sec.~\ref{general-regular-axial-tensor} we find in the odd-parity axial-tensor sector exact black hole solutions that are free of curvature and torsion singularities. Remarkably, these solutions include an independent torsion hair, on top of the ADM mass; therefore representing regular black holes with primary hair. In addition, in Sec.~\ref{general-regularAll} we demonstrate the existence of regular Hayward-like black holes in the general vector-axial-tensor sector. In this case, the torsion parameter is constrained by the mass and the coupling constants of the theory, in order for the field equations to be satisfied; thereby corresponding to a secondary hair. For completeness, we examine in Appendix~\ref{app:horizon-structure} the horizon structure of the new geometries obtained in our work, identifying the conditions under which event horizons form and characterising these black hole configurations. Finally, we present our conclusions in Sec.~\ref{sec:conclusions}.

We work in natural units $c=G=\hbar=1$ and consider the metric signature $(+,-,-,-)$. In addition, we use a tilde accent to denote quantities defined from the affine connection with torsion. On the other hand, Latin and Greek indices run from $0$ to $3$, referring to anholonomic and coordinate bases, respectively.

\section{Definitions and conventions}\label{sec:RC}

The description of gravity in a RC space-time identifies the antisymmetric part of the affine connection as an additional property of the gravitational field:
\begin{equation}
    T^{\lambda}{}_{\mu \nu}=2\tilde{\Gamma}^{\lambda}{}_{[\mu \nu]}\,,
\end{equation}
which generalises the covariant derivative acting on an arbitrary vector $v^{\lambda}$ as
\begin{equation}
\tilde{\nabla}_{\mu}v^{\lambda}=\nabla_{\mu}v^{\lambda}+K^{\lambda}{}_{\rho\mu}v^{\rho}\,,
\end{equation}
with
\begin{equation}
    K^{\lambda}{}_{\rho\mu}=\frac{1}{2}\left(T^{\lambda}{}_{\rho \mu}-T_{\rho}{}^{\lambda}{}_{\mu}-T_{\mu}{}^{\lambda}{}_{\rho}\right).
\end{equation}

The corresponding RC curvature tensor can then be expressed as the sum of the Riemann tensor and further post-Riemannian corrections
\begin{equation}
\tilde{R}^{\lambda}{}_{\rho\mu\nu}=R^{\lambda}{}_{\rho\mu\nu}+\nabla_{\mu}K^{\lambda}{}_{\rho \nu}-\nabla_{\nu}K^{\lambda}{}_{\rho \mu}+K^{\lambda}{}_{\sigma \mu}K^{\sigma}{}_{\rho \nu}-K^{\lambda}{}_{\sigma \nu}K^{\sigma}{}_{\rho \mu}\,,
\end{equation}
whose algebraic symmetries allow defining the Ricci tensor by the contraction
\begin{equation}
\tilde{R}_{\mu\nu}=\tilde{R}^{\lambda}{}_{\mu\lambda\nu}\,.
\end{equation}

In addition, the trace of the Ricci tensor provides the scalar curvature
\begin{equation}
    \tilde{R}=g^{\mu\nu}\tilde{R}_{\mu\nu}\,,
\end{equation}
whereas the pseudotrace of the curvature tensor gives rise to the so-called Holst pseudoscalar
\begin{equation}
    \ast\tilde{R}=\varepsilon^{\lambda\rho\mu\nu}\tilde{R}_{\lambda\rho\mu\nu}\,,
\end{equation}
with $\varepsilon_{\lambda\rho\mu\nu}
=\sqrt{-g}\,\epsilon_{\lambda\rho\mu\nu}$ denoting the Levi-Civita tensor,
being $\epsilon_{\lambda\rho\mu\nu}$ the totally antisymmetric symbol
normalised as $\epsilon_{0123}=+1$.
For stability and phenomenological considerations, it is essential to carry out an irreducible decomposition of the torsion tensor under the four-dimensional pseudo-orthogonal group, which includes vector, axial and tensor modes as~\cite{McCrea:1992wa}:
\begin{equation}
    T^{\lambda}{}_{\mu \nu}=\frac{1}{3}\left(\delta^{\lambda}{}_{\nu}T_{\mu}-\delta^{\lambda}{}_{\mu}T_{\nu}\right)+\frac{1}{6}\,\varepsilon^{\lambda}{}_{\rho\mu\nu}S^{\rho}+t^{\lambda}{}_{\mu \nu}\,,
\end{equation}
where
\begin{align}\label{Tdec1}
    T_{\mu}&=T^{\nu}{}_{\mu\nu}\,,\\
    S_{\mu}&=\varepsilon_{\mu\lambda\rho\nu}T^{\lambda\rho\nu}\,,\\
    t_{\lambda\mu\nu}&=T_{\lambda\mu\nu}-\frac{2}{3}g_{\lambda[\nu}T_{\mu]}-\frac{1}{6}\,\varepsilon_{\lambda\rho\mu\nu}S^{\rho}\,.\label{Tdec3}
\end{align}

Indeed, these modes can play an important role in a variety of physically relevant processes and configurations. For instance, it is worthwhile to emphasise the interaction of the axial vector to Dirac spinors under minimal coupling, as well as the cosmological implications of both vector and axial parts in spatially homogeneous and isotropic space-times, where the additional tensor mode of torsion is excluded by the symmetry~\cite{Hehl:1971qi,Cembranos:2018ipn,Bahamonde:2026vqm,tsamparlis1979cosmological}.

The dynamics of these geometrical degrees of freedom (dof) can then be encoded in an action functional, whose invariant structure leads to a broad class of gravitational models. Interestingly, the propagating character of torsion demands the presence of higher order curvature terms in the gravitational action, at the cost of considerably increasing the complexity of the theory. Thus, in the following sections we shall focus on the widely studied Holst quadratic model of PG theory and consider its extension by introducing cubic order invariants in the gravitational action.

\section{Holst-type PG models with cubic order invariants}\label{sec:model}

From the class of quadratic PG models that reduce to the Einstein-Hilbert action in the absence of dynamical torsion
\begin{align}
S=&\,\frac{1}{16\pi}\int
\Bigl[
-R-\frac{1}{2}\left(2c_{1}+c_{2}\right)\tilde{R}_{\lambda\rho\mu\nu}\tilde{R}^{\mu\nu\lambda\rho}+c_{1}\tilde{R}_{\lambda\rho\mu\nu}\tilde{R}^{\lambda\rho\mu\nu}+c_{2}\tilde{R}_{\lambda\rho\mu\nu}\tilde{R}^{\lambda\mu\rho\nu}+d_{1}\tilde{R}_{\mu\nu}\bigl(\tilde{R}^{\mu\nu}-\tilde{R}^{\nu\mu}\bigr)
\Bigr.\,
\nonumber\\
\Bigl.
&
+\frac{1}{2}m^2_TT_\mu T^\mu+\frac{1}{2}m^2_S S_\mu S^\mu +\frac{1}{2}m^2_tt_{\lambda\mu\nu}t^{\lambda\mu\nu}\Bigr]\sqrt{-g}\,d^4x\,.
\end{align}
the Holst quadratic model results from the choice of Lagrangian coefficients
\begin{equation}
    c_{2}=-\,4c_{1}\,, \quad d_{1}=0\,,\label{Lagcoeff}
\end{equation}
which simplifies the action to
\begin{align}
S=&\,\frac{1}{16\pi}\int
\Bigl(
-R-\frac{1}{4}c_{1}\ast\!\tilde{R}^{2}+\frac{1}{2}m^{2}_{T}T_\mu T^\mu+\frac{1}{2}m^2_{S} S_\mu S^\mu +\frac{1}{2}m^2_{t}t_{\lambda\mu\nu}t^{\lambda\mu\nu}\Bigr)\sqrt{-g}\,d^4x\,.\label{quadratic_action}
\end{align}

Under this stringent constraint on the parameter space of quadratic PG theory, the model dynamically isolates a massive pseudoscalar mode, ensuring the absence of fundamental instabilities~\cite{Yo:1999ex,Yo:2001sy,Jimenez:2019qjc}. While this quadratic baseline successfully provides a theoretically well-behaved framework, extending the action to include cubic order invariants introduces nonlinear couplings that offer a richer phenomenological landscape in the strong-field regime.

Thereby, moving from quadratic to cubic PG theory, we consider the action
\begin{align}
S=&\,\frac{1}{16\pi}\int
\Bigl(
-R-\frac{c_{1}}{4}\ast\!\tilde{R}^{2}+\mathcal{L}_{\rm curv-tors}^{(3)}+\frac{1}{2}m^2_{T}T_\mu T^\mu+\frac{1}{2}m^2_{S} S_\mu S^\mu +\frac{1}{2}m^2_{t}t_{\lambda\mu\nu}t^{\lambda\mu\nu}\Bigr)\sqrt{-g}\,d^4x\,,\label{cubic_action}
\end{align}
where the cubic order Lagrangian contains six different types of mixing terms of the curvature and torsion tensors
\begin{equation}
    \mathcal{L}_{\rm curv-tors}^{(3)}=\mathcal{L}^{(3)}_{\tilde{R}TT}+\mathcal{L}^{(3)}_{\tilde{R}SS}+\mathcal{L}^{(3)}_{\tilde{R}tt}+\mathcal{L}^{(3)}_{\tilde{R}TS}+\mathcal{L}^{(3)}_{\tilde{R}Tt}+\mathcal{L}^{(3)}_{\tilde{R}St}\,,
\end{equation}
with
\begin{align}
    \mathcal{L}^{(3)}_{\tilde{R}TT}=&\;h_{1}\tilde{R}_{\mu\nu}T^{\mu}T^{\nu}+h_{2}\tilde{R}T_{\mu}T^{\mu}\,,\\
    \mathcal{L}^{(3)}_{\tilde{R}SS}=&\;h_{3}\tilde{R}_{\mu\nu}S^{\mu}S^{\nu}+h_{4}\tilde{R}S_{\mu}S^{\mu}\,,\\
    \mathcal{L}^{(3)}_{\tilde{R}tt}=&\;h_5 \tilde{R}_{\lambda\rho\mu\nu}t_{\sigma}{}^{\lambda\rho}t^{\sigma\mu\nu}+h_6\tilde{R}_{\lambda\rho\mu\nu}t_{\sigma}{}^{\lambda\mu}t^{\sigma\rho\nu}+h_7\tilde{R}_{\lambda\rho\mu\nu}t^{\lambda\rho}{}_{\sigma}t^{\sigma\mu\nu}+h_{8}{} \tilde{R}_{\lambda\rho\mu\nu}t^{\lambda\mu}{}_{\sigma}t^{\sigma\rho\nu}+h_{9}{}\tilde{R}_{\lambda\rho\mu\nu}t^{\lambda\mu}{}_{\sigma}t^{\rho\nu\sigma}\nonumber\\
&+\,h_{10}{}\tilde{R}_{\lambda\rho}t_{\mu\nu}{}^{\lambda}t^{\rho\mu\nu}+h_{11}{}\tilde{R}_{\lambda\rho}t_{\mu\nu}{}^{\lambda}t^{\mu\nu\rho}+h_{12}{}\tilde{R}t_{\lambda\rho\mu}t^{\lambda\rho\mu}\,,\\
    \mathcal{L}^{(3)}_{\tilde{R}TS}=&\;h_{13}{}\varepsilon^{\lambda\rho\mu\nu}\tilde{R}_{\lambda\rho\mu\nu}T_{\sigma}S^{\sigma}+h_{14}{} \varepsilon_{\nu}{}^{\lambda\rho\sigma}\tilde{R}_{\lambda\rho\mu\sigma}T^{\mu}S^{\nu}+h_{15}{}\varepsilon^{\lambda\rho\mu\nu}\tilde{R}_{\lambda\rho}T_{\mu}S_{\nu}\,,\\
    \mathcal{L}^{(3)}_{\tilde{R}Tt}=&\;h_{16}{}\tilde{R}_{\lambda\rho\mu\nu}T^{\nu}t^{\lambda\rho\mu}+h_{17}{} \tilde{R}_{\lambda\rho\mu\nu}T^{\rho}t^{\lambda\mu\nu}+h_{18}{}\tilde{R}_{\lambda\rho}T_{\mu}t^{\mu\lambda\rho}+h_{19}{} \tilde{R}_{\lambda\rho}T_{\mu}t^{\lambda\rho\mu}\,,\\
    \mathcal{L}^{(3)}_{\tilde{R}St}=&\;h_{20}{}\varepsilon_{\alpha\rho\mu\nu}\tilde{R}_{\tau}{}^{\rho\mu\nu}S^{\gamma}t^{\alpha\tau}{}_{\gamma}+h_{21}{}\varepsilon_{\alpha\rho\mu\nu}\tilde{R}_{\tau}{}^{\rho\mu\nu}S^{\gamma}t_{\gamma}{}^{\alpha\tau}+h_{22}{}\varepsilon_{\alpha\rho}{}^{\mu\nu}\tilde{R}^{\rho}{}_{\mu\tau\nu}S^{\gamma}t_{\gamma}{}^{\alpha\tau}+h_{23}{}\varepsilon_{\alpha\rho}{}^{\mu\nu}\tilde{R}_{\gamma\mu\tau\nu}S^{\alpha}t^{\gamma\rho\tau}\nonumber\\
&+\,h_{24}{}\varepsilon_{\alpha\rho}{}^{\mu\nu}\tilde{R}_{\gamma\mu\tau\nu} S^{\alpha}t^{\rho\tau\gamma}+h_{25}{}\varepsilon_{\alpha\rho\tau\mu}\tilde{R}^{\mu}{}_{\gamma}S^{\alpha}t^{\rho\tau\gamma}+h_{26}{}\varepsilon_{\lambda\rho\mu\nu} \tilde{R}^{\lambda\rho}S_{\sigma}t^{\sigma\mu\nu}\,.
\end{align}

The field equations of this class of Holst-type PG models are then derived by performing variations with respect to the tetrad field and the spin connection, which according to the principle of least action take the form:
\begin{align}
    E^{\mu\nu}&=0\,,\label{tetrad_eq}\\
    E^{\lambda\mu\nu}&=0\,,\label{connect_eq}
\end{align}
where $E^{\mu\nu}$ and $E^{\lambda\mu\nu}$ are tensor quantities determined by the curvature and torsion tensors (see Appendix A in~\cite{Bahamonde:2025fei}).

\section{Static and spherically symmetric space-times in PG theory}\label{sec:BHs}

The search for black hole configurations requires solving the system of field equations~\eqref{tetrad_eq}-\eqref{connect_eq}, which can be simplified by the imposition of a set of space-time symmetries. In the simplest case, the metric and torsion tensors satisfy the same symmetry conditions provided by a Killing vector ${\xi}$, which in turn are consequently reflected in the curvature tensor:
\begin{equation}
    \mathcal{L}_{\xi}g_{\mu\nu}=\mathcal{L}_{\xi}T^{\lambda}{}_{\mu\nu}=0\,.
\end{equation}
In particular, for static and spherically symmetric space-times, these tensors are preserved under the action of the Killing vectors
\begin{align}
    \xi_{0}&=\partial_{t}\,,\\
    \xi_{1}&=\sin\varphi\,\partial_{\vartheta}+\cot\vartheta\cos\varphi\,\partial_{\varphi}\,,\\
    \xi_{2}&=-\,\cos\varphi\,\partial_{\vartheta}+\cot\vartheta\sin\varphi\,\partial_{\varphi}\,,\\
    \xi_{3}&=-\,\partial_{\varphi}\,,
\end{align}
which constrains their components as follows~\cite{Hohmann:2019fvf}:
\begin{align}\label{sph_metric}
    g_{tt}=\Psi_{1}(r)\,, \quad g_{rr}=-\,\frac{1}{\Psi_{2}(r)}\,, \quad g_{\vartheta\vartheta}=g_{\varphi\varphi}\csc^2\vartheta=-\,r^2\,,
\end{align}
\begin{align}\label{sph_torsion}
    T^t\,_{t r} &= t_{1}(r)\,, \quad T^r\,_{t r} = t_{2}(r)\,, \quad T^\vartheta\,_{t \vartheta} = T^\varphi\,_{t \varphi} = t_{3}(r)\,, \quad T^\vartheta\,_{r \vartheta} = T^\varphi\,_{r \varphi} =  t_{4}(r)\,, \\
    T^\vartheta\,_{t \varphi} &= T^\varphi\,_{\vartheta t} \sin^{2}\vartheta = t_{5}(r) \sin{\vartheta}\,, \quad T^\vartheta\,_{r \varphi} = T^\varphi\,_{\vartheta r} \sin^{2}\vartheta = t_{6}(r) \sin{\vartheta}\,, \\
    T^t\,_{\vartheta \varphi} &= t_{7}(r) \sin\vartheta\,, \quad T^r\,_{\vartheta \varphi} = t_{8}(r) \sin \vartheta\,,
\end{align}
where $(t,r,\vartheta,\varphi)$ denote spherical coordinates.

As for the field equations, in the static and spherically symmetric case there are ten independent equations, e.g.
\begin{equation}
    \bigl\{E^{trt}, E^{trr}, E^{t\vartheta\vartheta}, E^{r\vartheta\vartheta}, E^{\vartheta\varphi t}, E^{\vartheta\varphi r}, E^{t\vartheta\varphi}, E^{r\vartheta\varphi}, E^{tt}, E^{rr}\bigr\},
\end{equation}
corresponding to the ten independent dof of the metric and torsion tensors. Nevertheless, the highly nonlinear character of this system of equations requires additional constraints to find exact black hole solutions. Some remarkable examples include the use of the double duality ansatz~\cite{Mielke:1981xe}, which for a particular class of quadratic PG models has been shown to substantially simplify the field equations and facilitate the construction of solutions~\cite{Baekler:1981lkh,Bakler:1988nq}. For other models, this simplification may become overly restrictive, excluding nontrivial solutions to the field equations. Relevant examples arise for the class of models that reduce to GR in the absence of dynamical torsion, which were obtained by the implementation of a weak-field limit and the imposition of a set of regularity conditions to avoid divergences at the event horizons~\cite{Cembranos:2016gdt,Cembranos:2017pcs}.

For the present class of Holst-type PG models, it is worthwhile to stress that although the field equations derived from the gravitational action~\eqref{cubic_action} reduce to the vacuum Einstein equations in the absence of dynamical torsion, the choice of Lagrangian coefficients~\eqref{Lagcoeff} identically satisfies the weak-field limit. Hence, a significant reduction in the torsion dof can only be achieved by imposing suitable regularity conditions at the event horizons, which can be directly obtained by demanding that the torsion field $\mathcal{F}^{a}\,_{b c} = \vartheta^{a}\,_{\lambda}\vartheta_{b}\,^{\mu}\vartheta_{c}\,^{\nu}T^{\lambda}\,_{\nu\mu}\,$ expressed in the boosted basis
\begin{eqnarray}
\vartheta^{\hat{0}}&=&\frac{1}{2}{\biggl(\frac{\Psi_1(r)}{ \Psi_2(r)}\biggr)^{1/4}}\left\{ \left[\sqrt{\Psi_{1}(r)\Psi_{2}(r)}+1\right]\,dt+\left[1-\frac{1}{\sqrt{\Psi_{1}(r)\Psi_{2}(r)}}\right]\,dr \right\},\\
\vartheta^{\hat{1}}&=&\frac{1}{2}{\biggl(\frac{\Psi_1(r)}{ \Psi_2(r)}\biggr)^{1/4}}\left\{ \left[\sqrt{\Psi_{1}(r)\Psi_{2}(r)}-1\right]\,dt+\left[1+\frac{1}{\sqrt{\Psi_{1}(r)\Psi_{2}(r)}}\right]\,dr \right\},\\
\vartheta^{\hat{2}}&=&r\,d \vartheta \,,\\%
\vartheta^{\hat{3}}&=&r\sin\vartheta \, d \varphi\,,
\end{eqnarray}
does not present any singular term in the roots of the metric functions. Thereby, one readily finds that the divergent contributions to the components of the torsion field
\begin{align}
\mathcal{F}^{\hat{0}}\,_{\hat{0}\hat{1}} &=-\color{black}\,{\frac{1}{2}\biggl(\frac{\Psi_2(r)}{\Psi_1(r)}\biggr)^{1/4}}\left\{\left[1+\sqrt{\Psi_{1}(r)\Psi_{2}(r)}\right]t_{1}(r)+\left[1-\frac{1}{\sqrt{\Psi_{1}(r)\Psi_{2}(r)}}\right]t_{2}(r)\right\},\\
\mathcal{F}^{\hat{1}}\,_{\hat{0}\hat{1}} &=-\color{black}\,{\frac{1}{2}\biggl(\frac{\Psi_2(r)}{\Psi_1(r)}\biggr)^{1/4}}\left\{\left[1+\frac{1}{\sqrt{\Psi_{1}(r)\Psi_{2}(r)}}\right]t_{2}(r)-\left[1-\sqrt{\Psi_{1}(r)\Psi_{2}(r)}\right]t_{1}(r)\right\} ,\\
\mathcal{F}^{\hat{2}}\,_{\hat{0}\hat{2}} &=\mathcal{F}^{\hat{3}}\,_{\hat{0}\hat{3}} =-\color{black}\,{\frac{1}{2}\biggl(\frac{\Psi_2(r)}{\Psi_1(r)}\biggr)^{1/4}}\left\{\left[1+\frac{1}{\sqrt{\Psi_{1}(r)\Psi_{2}(r)}}\right]t_{3}(r)+\left[1-\sqrt{\Psi_{1}(r)\Psi_{2}(r)}\right]t_{4}(r)\right\},\\
\mathcal{F}^{\hat{2}}\,_{\hat{1}\hat{2}} &= \mathcal{F}^{\hat{3}}\,_{\hat{1}\hat{3}} =-\color{black}\,{\frac{1}{2}\biggl(\frac{\Psi_2(r)}{\Psi_1(r)}\biggr)^{1/4}}\left\{\left[1+\sqrt{\Psi_{1}(r)\Psi_{2}(r)}\right]t_{4}(r)-\left[1-\frac{1}{\sqrt{\Psi_{1}(r)\Psi_{2}(r)}}\right]t_{3}(r)\right\} ,\\
\mathcal{F}^{\hat{3}}\,_{\hat{0} \hat{2}} &= -\,\mathcal{F}^{\hat{2}}\,_{\hat{0} \hat{3}}=\color{black} \frac{1}{2}\,\biggl(\frac{\Psi_2(r)}{\Psi_1(r)}\biggr)^{1/4}\left\{\left[1+\frac{1}{\sqrt{\Psi_{1}(r)\Psi_{2}(r)}}\right]t_{5}(r)+\left[1-\sqrt{\Psi_{1}(r)\Psi_{2}(r)}\right]t_{6}(r)\right\} ,\\
 \mathcal{F}^{\hat{2}}\,_{\hat{1} \hat{3}} &= -\,\mathcal{F}^{\hat{3}}\,_{\hat{1} \hat{2}}= \color{black} \frac{1}{2}\,\biggl(\frac{\Psi_2(r)}{\Psi_1(r)}\biggr)^{1/4}\left\{\left[1-\frac{1}{\sqrt{\Psi_{1}(r)\Psi_{2}(r)}}\right]t_{5}(r)-\left[1+\sqrt{\Psi_{1}(r)\Psi_{2}(r)}\right]t_{6}(r)\right\} ,\\
\mathcal{F}^{\hat{0}}\,_{\hat{2} \hat{3}} &=-\color{black} \,{\frac{1}{2r^2}\biggl(\frac{\Psi_1(r)}{\Psi_2(r)}\biggr)^{1/4}}\left\{\left[1+\sqrt{\Psi_{1}(r)\Psi_{2}(r)}\right]t_{7}(r)+\left[1-\frac{1}{\sqrt{\Psi_{1}(r)\Psi_{2}(r)}}\right]t_{8}(r)\right\} ,\\
\mathcal{F}^{\hat{1}}\,_{\hat{2} \hat{3}} &= -\,{\frac{1}{2r^2}\biggl(\frac{\Psi_1(r)}{\Psi_2(r)}\biggr)^{1/4}}\left\{\left[1+\frac{1}{\sqrt{\Psi_{1}(r)\Psi_{2}(r)}}\right]t_{8}(r)-\left[1-\sqrt{\Psi_{1}(r)\Psi_{2}(r)}\right]t_{7}(r)\right\},
\end{align}
cancel provided that
\begin{align}
     t_{2}(r)&=t_{1}(r)\sqrt{\Psi_{1}(r)\Psi_{2}(r)}\,,\quad  t_{3}(r)=-\,t_{4}(r)\sqrt{\Psi_{1}(r)\Psi_{2}(r)}\,,\label{reg1}\\
     t_{5}(r)&=-\,t_{6}(r)\sqrt{\Psi_{1}(r)\Psi_{2}(r)}\,, \quad t_{8}(r)=t_{7}(r)\sqrt{\Psi_{1}(r)\Psi_{2}(r)}\,,\label{reg2}
\end{align}
which in fact implies that the corresponding torsion scalars are regular and vanish identically.

In order to simplify the analysis of the field equations, which take a cumbersome form expressed in terms of the remaining dof, let us introduce the variables
\begin{eqnarray}
    t_1(r)=\frac{P(r)+2Q(r)}{\sqrt{\Psi_1(r)\Psi_2(r)}}\,,\quad  t_4(r)=\frac{Q(r)}{\sqrt{\Psi_1(r)\Psi_2(r)}}\,,\quad t_6(r)=\frac{U(r)}{\sqrt{\Psi_1(r)\Psi_2(r)}}\,,\quad t_7(r)=\frac{Y(r)+2r^2U(r)}{\Psi_1(r)}\,,
\end{eqnarray}
which re-express the torsion modes as
\begin{align}
    T_{\mu} &=P(r)
    \left(
        1,
        -\,\frac{1}{\sqrt{\Psi_1(r)\Psi_2(r)}},
        0,0
    \right),\label{eq:Tvector_PQ}\\
    S_{\mu} &=\frac{2Y(r)}{r^2}
    \left(
        1,
        -\,\frac{1}{\sqrt{\Psi_1(r)\Psi_2(r)}},
        0,0
    \right),\label{eq:Svector_PQ}\\
    t_{\lambda\mu\nu} &=\left(
\begin{array}{cccccc}
t_{ttr} & 0 & 0 & 0 & 0 & t_{t\vartheta\varphi} \\
t_{rtr} & 0 & 0 & 0 & 0 & t_{r\vartheta\varphi} \\
0 & t_{\vartheta t\vartheta} & t_{\vartheta t\varphi} & t_{\vartheta r\vartheta} & t_{\vartheta r\varphi} & 0 \\ 0 & t_{\varphi t\vartheta} & t_{\varphi t\varphi} & t_{\varphi r\vartheta} & t_{\varphi r\varphi} & 0
\end{array} \right),\label{eq:ttensor_PQ}
\end{align}
where
\begin{align}
    t_{ttr}&=\frac{2\left(P(r)+3Q(r)\right)\sqrt{\Psi_{1}(r)}}{3\sqrt{\Psi_{2}(r)}}\,, \quad t_{t\vartheta\varphi}=\frac{2}{3}\left(Y(r)+3r^{2}U(r)\right)\sin\vartheta\\
    t_{rtr}&=-\,\frac{2\left(P(r)+3Q(r)\right)}{3\Psi_{2}(r)}\,, \quad t_{r\vartheta\varphi}=-\,\frac{2\left(Y(r)+3r^{2}U(r)\right)\sin\vartheta}{3\sqrt{\Psi_{1}(r)\Psi_{2}(r)}}\\
    t_{\vartheta t\vartheta}&= \frac{1}{3}r^2\left(P(r)+3Q(r)\right)\,, \quad t_{\vartheta t\varphi}=\frac{1}{3}\left(Y(r)+3r^{2}U(r)\right)\sin\vartheta\,,\\
    t_{\vartheta r\vartheta}&= -\,\frac{r^2(P(r)+3Q(r))}{3\sqrt{\Psi_{1}(r)\Psi_{2}(r)}}\,, \quad t_{\vartheta r\varphi}=-\,\frac{\bigl(Y(r)+3r^{2}U(r)\bigr)\sin\vartheta}{3\sqrt{\Psi_{1}(r)\Psi_{2}(r)}}\,,\\
    t_{\varphi t\vartheta}&=-\,\frac{1}{3}\left(Y(r)+3r^{2}U(r)\right)\sin\vartheta \,, \quad t_{\varphi t\varphi}=\frac{1}{3}r^2\left(P(r)+3Q(r)\right)\sin^{2}\vartheta\,,\\
    t_{\varphi r\vartheta}&= \frac{\left(Y(r)+3r^{2}U(r)\right)\sin\vartheta}{3\sqrt{\Psi_{1}(r)\Psi_{2}(r)}}\,, \quad t_{\varphi r\varphi}=-\,\frac{r^2\left(P(r)+3Q(r)\right)\sin^{2}\vartheta}{3\sqrt{\Psi_{1}(r)\Psi_{2}(r)}}\,.
\end{align}
Then, the combinations $E^{trr}(r)-E^{trt}(r)\sqrt{\Psi_1(r)\Psi_2(r)}$, $E^{r\vartheta\vartheta}(r)-E^{t\vartheta\vartheta}(r)\sqrt{\Psi_1(r)\Psi_2(r)}$, $E^{r\vartheta\varphi}(r,\vartheta)-E^{\vartheta\varphi r}(r,\vartheta)+\bigl(E^{\vartheta\varphi t}(r,\vartheta)-E^{t\vartheta\varphi}(r,\vartheta)\bigr)\sqrt{\Psi_1(r)\Psi_2(r)}$ and $E^{\vartheta\varphi r}(r,\vartheta)
-E^{\vartheta\varphi t}(r,\vartheta)\sqrt{\Psi_1(r)\Psi_2(r)}$ of the connection field equations respectively lead to the following equations:
\begin{align}
    &\left(\alpha_1 P(r)+\alpha_2 Q(r)\right)\left(\ln\frac{\Psi_1(r)}{\Psi_2(r)}\right)'=0\,,\label{non_rec1}\\
    &\left[
\left(2\alpha_1-\alpha_2\right)P(r)
+\left(9\alpha_1-4\alpha_2-162h_1\right)Q(r)
\right]
\left(\ln\frac{\Psi_1(r)}{\Psi_2(r)}\right)'=0\,,\label{non_rec2}\\
    &\left(\alpha_3r^2U(r)+\alpha_4 Y(r)\right)\left(\ln\frac{\Psi_1(r)}{\Psi_2(r)}\right)'=0\,,\label{non_rec3}\\
    &\left[
9\left(\alpha_3-2\alpha_4\right)r^{2}U(r)
+\left(648h_3+2\alpha_3-3\alpha_4\right)Y(r)-81c_1 rY'(r)
\right]\left(\ln\frac{\Psi_1(r)}{\Psi_2(r)}\right)'\nonumber\\
&+162c_{1}\left(rY''(r)-2Y'(r)\right)=0\,,\label{non_rec4}
\end{align}
with
\begin{align}
\alpha_1&=
2\left(
9h_1+4h_{10}+2h_{11}-3h_{16}+6h_{17}+6h_{19}
+4h_5+2h_6-2h_7-h_8+4h_9
\right),
\\
\alpha_2&=
3\left(
8h_{10}+4h_{11}-3h_{16}+6h_{17}+6h_{19}
+8h_5+4h_6-4h_7-2h_8+8h_9
\right),
\\
\alpha_3&=
9\left(
4h_{10}-2h_{11}-4h_5-2h_6+2h_7+h_8+4h_9
\right),
\\
\alpha_4&=
3\left(
4h_{10}-2h_{11}-3h_{23}+6h_{24}-6h_{25}
-4h_5-2h_6+2h_7+h_8+4h_9
\right).\label{alpha4}
\end{align}
Thus, these equations impose different restrictions on the torsion dof for the general and reciprocal cases, the latter being characterised by
\begin{equation}
    \Psi_1(r) = \Psi_2(r) \equiv \Psi(r)\,.\label{reciprocal_metric}
\end{equation}
As a guiding principle for solving the field equations of the model, we then focus exclusively on the search for exact solutions compatible with conventional Coulomb electric and magnetic fields, as in the standard Einstein-Maxwell theory. Accordingly, we restrict the analysis to the reciprocal case, in order to satisfy the Maxwell's equations in the RC manifold. In that case, Eqs.~\eqref{non_rec1}-\eqref{non_rec3} are trivially satisfied, whereas Eq.~\eqref{non_rec4} can be straightforwardly integrated, yielding for $c_1\neq0$ 
\begin{eqnarray}
    Y(r)=\kappa_{1}+\kappa_{2} r^3\,,\label{eqYsol}
\end{eqnarray}
where $\kappa_{1}$ and $\kappa_{2}$ are integration constants. Throughout the following sections, these two constants retain this universal meaning, whereas any additional integration constants are labelled starting from $\kappa_{3}$ independently within each solution sector.

On the other hand, the combination $r^{2}E^{tt}(r)-r^{2}E^{rr}(r)/\Psi^2(r)-2\left(r^2\Psi(r)E^{trt}(r)\right)'/\Psi(r)-4r^{3}E^{t\vartheta\vartheta}(r)$ of the tetrad and connection equations provides the following first-order equation for the metric function:
\begin{align}
&\Bigl[
\frac{r}{2}\left(1-\Psi(r)\right)
-\frac{\alpha_1}{36}\,rP^2(r)
-\frac{\alpha_2}{18}\,rP(r)Q(r)
-\frac{\beta_1}{2}\,rQ^2(r)
\nonumber\\
&+\frac{\alpha_3}{18}\,rU^2(r)
+\frac{\alpha_4}{9r}\,U(r)Y(r)
-\frac{\beta_2}{18r^3}\,Y^2(r)
+\frac{c_1}{12r}\,Y'{}^2(r)
\Bigr]'=0\,,\label{eq:metricfirstintegral}
\end{align}
with
\begin{equation}
    \beta_1=
    9h_1-\frac{\alpha_1}{2}+\frac{\alpha_2}{3}\,,
    \quad
    \beta_2=
    36h_3+\frac{\alpha_3}{9}-\frac{2\alpha_4}{3}\,.
\label{eq:beta12}
\end{equation}
Hence, the general form of the metric function expressed in terms of the torsion dof can be directly obtained from this equation, yielding
\begin{equation}
    \Psi(r)=1-\frac{2\mu}{r}
-\frac{\alpha_1}{18}P^2(r)
-\frac{\alpha_2}{9}P(r)Q(r)
-\beta_1 Q^2(r)
+\frac{\alpha_3}{9}U^2(r)+\frac{2\alpha_4}{9r^2}U(r)Y(r)-\frac{\beta_2}{9r^4}Y^2(r)
+\frac{c_1}{6r^2}Y'{}^2(r)\,,\label{eq:general_metric_integrated}
\end{equation}
where $\mu$ is an integration constant. Thus, it is clear that the space-time geometry is reduced to the Schwarzschild solution for Holst-type PG models satisfying
\begin{equation}
    c_1=\alpha_1=\alpha_2=\alpha_3=\alpha_4=\beta_1=\beta_2=0\,.
\end{equation}

With the torsion and metric functions $Y(r)$ and $\Psi(r)$ determined by Expressions~\eqref{eqYsol}-\eqref{eq:general_metric_integrated}, the remaining connection and tetrad equations to be solved can be re-expressed in a simpler form. Specifically, the tetrad equation is significantly simplified by considering the combination
\begin{align}
\mathcal E(r) \equiv{}&
E^{tt}(r)
-2\left(E^{trt}(r)\right)'
-\left(
\frac{4}{r}
+\frac{\Psi'(r)}{\Psi(r)}
+4t_1(r)
\right)E^{trt}(r)+
4r^2t_4(r)\,E^{t\vartheta\vartheta}(r)\nonumber\\
&-t_7(r)E^{\vartheta\varphi t}(r,\vartheta)\sin\vartheta
-\frac{t_7(r)}{\Psi(r)}
\left(
E^{\vartheta\varphi r}(r,\vartheta)
-2E^{r\vartheta\varphi}(r,\vartheta)
\right)\sin\vartheta\,,
\label{eq:reduced_tetrad_equation}
\end{align}
whereas the remaining connection equations can be rewritten and
denoted by
\begin{equation}
    \bigl(E^{trt}(r), E^{t\vartheta\vartheta}(r), E^{\vartheta\varphi t}(r,\vartheta)\sin\vartheta, \bigl(E^{\vartheta\varphi r}(r,\vartheta)-2E^{r\vartheta\varphi}(r,\vartheta)\bigr)\sin\vartheta\bigr)
    \equiv
    \bigl(\mathcal {C}^{(1)}(r), \mathcal{C}^{(2)}(r), \mathcal{C}^{(3)}(r),
    \mathcal{C}^{(4)}(r)\bigr)\,.
    \label{eq:reduced_equation_names}
\end{equation}
For generic values of the Lagrangian coefficients, these equations are independent. However, for certain choices of these coefficients, the system may exhibit degeneracies, with some of the equations becoming dependent or reducing to algebraic constraints. These special cases lead to a variety of singular and regular black hole solutions, which we shall derive in the following sections. In particular, these solutions emerge in different torsion sectors, depending on the irreducible modes and their parity components involved in the analysis. Accordingly, for the singular solutions obtained below, we distinguish the odd-parity tensor sector, the even-parity vector-tensor sector, as well as the even and odd-parity axial-tensor sector, whereas for the regular ones we consider the odd-parity axial-tensor sector and the general vector-axial-tensor sector.

\section{Odd-parity tensor sector: Kiselev-like black holes}\label{sec:Kiselev}

We first consider the backreaction effects provided by the odd-parity components of the tensor mode alone, described by the torsion function $U(r)$, i.e.
\begin{equation}
    P(r)=Q(r)=Y(r)=0\,,
    \quad
    U(r)\neq0\,.
    \label{eq:kiselev_support}
\end{equation}

In this case, the connection equations $\mathcal C^{(1)}=0$ and $\mathcal C^{(2)}=0$ can be written as
\begin{equation}
    \mathbb M_U
    \begin{pmatrix}
        U(r)\\ rU'(r)
    \end{pmatrix}=0\,,
    \quad
    \mathbb M_U=
    \begin{pmatrix}
        M_{11} & M_{12}\\
        M_{21} & M_{22}
    \end{pmatrix},
    \label{eq:kiselev_principal_system}
\end{equation}
where
\begin{align}
 M_{11}={}&
 12h_{10}-6h_{11}+6h_{16}-6h_{17}-18h_{18}
 +9h_{19}+2h_8+16h_9\,,
 \label{eq:kiselev_A1}\\
 M_{12}={}&
 -6h_{16}+6h_{17}-6h_{18}+3h_{19}
 -12h_5-6h_6+6h_7+5h_8-8h_9\,,
 \label{eq:kiselev_B1}\\
 M_{21}={}&
 -2\left(
 6h_{10}-3h_{11}+6h_{16}-6h_{17}-18h_{18}
 +9h_{19}
 -12h_5-6h_6+6h_7+2h_8+4h_9\right),
 \label{eq:kiselev_A2}\\
 M_{22}={}&
 -\left(
 24h_{10}-12h_{11}-12h_{16}+12h_{17}-12h_{18}
 +6h_{19}
 -12h_5-6h_6+6h_7+h_8+32h_9\right).
 \label{eq:kiselev_B2}
\end{align}

Hence, a nontrivial solution to these equations requires
\begin{equation}
    \det\mathbb M_U=M_{11}M_{22}-M_{12}M_{21}=0\,,
    \quad
    \mathbb M_U\neq0\,.
    \label{eq:kiselev_rank_one_condition}
\end{equation}
On this parameter subspace, the equations take then the simple form
\begin{equation}
 \frac{rU'(r)}{U(r)}
 =
 \gamma\,,
 \quad
 \gamma
 =
 -\,\frac{M_{11}}{M_{12}}
 =
 -\,\frac{M_{21}}{M_{22}}\,,\label{eq:kiselev_log_derivative}
\end{equation}
which leads to the solution
\begin{equation}
U(r)=\kappa_{3} r^\gamma\,,
\label{eq:kiselev_U_solution}
\end{equation}
where $\kappa_{3}$ is the integration constant associated with the tensor hair.

Substituting Eqs.~\eqref{eq:kiselev_rank_one_condition} and~\eqref{eq:kiselev_U_solution} into the remaining equations $\mathcal C^{(3)}=0$, $\mathcal C^{(4)}=0$ and $\mathcal E=0$ reduces the system to algebraic conditions on the Lagrangian coefficients. Specifically, the particular values $\gamma=-\,1/2$ and $\gamma=1$ respectively provide solutions on the conventional Schwarzschild and Schwarzschild-de Sitter geometries, whereas for general values of this parameter it is possible to find nontrivial configurations with the tensor hair scaling as $r^{2\gamma}$ in the metric function. 
A convenient $\gamma$-independent solution of the remaining algebraic condition is given by
\begin{align}
h_8&=2\left(2h_5+h_6+h_9-h_7\right), \quad
h_{11}=\frac14\left(3h_9-h_{10}\right),\label{eq:kiselev_h8}\\
h_{12}&=\frac1{16}\left(3h_{10}-h_9\right), \quad
h_{24}=\frac{1}{2}h_{23}\,, \quad h_{25}=m_t^2=0\,,
\label{eq:kiselev_algebraic_conditions}
\end{align}
which significantly simplifies the rank-one condition~\eqref{eq:kiselev_rank_one_condition} and the general expression of the parameter $\gamma$:
\begin{align}
h_5={}&
\frac{16\left(h_7-h_6\right)+h_9+9h_{10}}{32}
+\frac{
27\left(h_9+h_{10}\right)
\left(4h_{18}-2h_{19}-3h_9-3h_{10}\right)
}{
32\left(2h_{17}+2h_{18}-2h_{16}-h_{19}\right)
}\,,
\label{eq:kiselev_h5}\\
\gamma={}&
\frac{
2h_{17}+6h_{18}-3h_9-3h_{10}
-2h_{16}-3h_{19}
}{
3h_9+3h_{10}+2h_{17}+h_{19}
-2h_{16}-2h_{18}
}\,.
\label{eq:kiselev_gamma}
\end{align}

Thus, in the generic case $2h_{17}+2h_{18}-2h_{16}-h_{19}\neq0$ and $\gamma$ not singular, Expression~\eqref{eq:general_metric_integrated}~for the metric function becomes
\begin{equation}
\Psi(r)=
1-\frac{2m}{r}
+\frac92\left(h_9+h_{10}\right)
\kappa_{3}^2r^{2\gamma}\,,
\label{eq:kiselev_metric}
\end{equation}
in such a way that the integration constant $\mu$ can be interpreted as a mass parameter $m$.

On the other hand, the exceptional case
\begin{equation}
    h_{19}=2h_{18}-\frac{3}{2}\left(h_{9}+h_{10}\right), \quad h_{17}
=
h_{16}-\frac34\left(h_9+h_{10}\right),
\label{eq:kiselev_sigma_conditions}
\end{equation}
leads to the metric function
\begin{equation}
    \Psi(r)=1-\frac{2m}{r}+\frac{9}{2}\left(h_{9}+h_{10}\right)\kappa_{3}^{2}\,r^{2\sigma}\,,
\end{equation}
with
\begin{equation}
    \sigma=\frac{16h_5+8h_6+13h_9+9h_{10}-8h_7}{2\left(4h_{7}+7h_{9}+9h_{10}-8h_{5}-4h_{6}\right)}\,.
\end{equation}
As can be seen, this case stems from a structural bifurcation given by the simultaneous vanishing of the linear functionals of the parameter $\gamma$, which still provides nontrivial configurations beyond the Schwarzschild and Schwarzschild-de Sitter geometries. This property ceases to hold if $4h_{7}+7h_{9}+9h_{10}-8h_{5}-4h_{6}=0$, which demands $h_{10} = - \,h_{9}$ and cancels the backreaction effects of the tensor hair on the space-time geometry.

Overall, the solutions correspond to Kiselev-like black holes~\cite{Kiselev:2002dx,Visser:2019brz}, endowed with a primary tensor hair. In this sense, different black hole solutions with a primary torsion hair can also be found by switching on the vector and/or axial modes, as we shall show in the following sections.

\section{Even-parity vector-tensor sector: black holes with algebraic and Lambert $W$ metric corrections}
\label{subsec:PQmixed}

In the search for black hole solutions with multiple torsion modes, we begin by considering the even-parity vector-tensor sector, characterised by torsion functions
\begin{equation}
 U(r)=Y(r)=0\,,
 \quad
 P(r)\neq0\,,
 \quad
 P(r)+3Q(r)\neq0\,. \label{eq:PQmixed_support}
\end{equation}
In this case, the field equations directly reduce to the system $\mathcal C^{(1)}=0$, $\mathcal C^{(2)}=0$ and $\mathcal E=0$. For generic values of the Lagrangian coefficients, these three equations overdetermine the two functions $P(r)$ and $Q(r)$. We therefore restrict the parameter space to a subspace on which one of the three equations becomes dependent on the other two:
\begin{align}
 h_1&=h_2=h_{19}=m_T^2=m_t^2=0\,,\label{hcase2a}\\
 h_5&=2h_{12}-\frac{h_6}{2}+\frac{h_7}{2}+\frac{h_8}{4}-\frac{\eta}{4}\,,\quad h_9=\frac{\eta}{2}-4h_{12}\,,\quad h_{10}=\frac{3\eta}{2}+4h_{12}\,, h_{11}=\eta-4h_{12}\,,\\
 h_{16}&=\zeta+2h_{18}\,,\quad h_{17}=\frac{\zeta}{2}+h_{18}\,,\label{hcase2b}
\end{align}
with
\begin{equation}
 \eta=\frac{1}{2}\left(h_9+h_{10}\right),
 \quad
 \zeta=2\left(h_{17}-h_{18}\right),
 \quad
 \eta\zeta\neq 0\,.
 \label{eq:PQmixed_etazeta}
\end{equation}
Thereby, the metric function~\eqref{eq:general_metric_integrated} includes a correction given by the tensor mode as
\begin{equation}
 \Psi(r)=1-\frac{2\mu}{r}-\eta\,[P(r)+3Q(r)]^2\,,
 \label{eq:PQmixed_metric}
\end{equation}
whereas the remaining field equations reduce to
\begin{align}
&\zeta r Q(r) Q'(r)
+\eta P^2(r)
+\left(\zeta+6\eta\right)P(r)Q(r)
+3\left(\zeta+3\eta\right)Q^2(r)=0\,,
\label{eq:PQmixed_PQ1}\\
&2rQ(r)\left(P'(r)+3Q'(r)\right)
-\left(P(r)+Q(r)\right)\left(P(r)+3Q(r)\right)=0\,.
\label{eq:PQmixed_PQ2}
\end{align}

By taking
the ratio $P(r)/Q(r)$, Eqs.~\eqref{eq:PQmixed_PQ1}-\eqref{eq:PQmixed_PQ2} imply
\begin{align}
r\left(\frac{P(r)}{Q(r)}\right)'
={}&
\frac{1}{2\zeta}
\left(\frac{P(r)}{Q(r)}+3\right)
\left[
2\eta\left(\frac{P(r)}{Q(r)}\right)^2
+3\left(4\eta+\zeta\right)\frac{P(r)}{Q(r)}
+18\eta+7\zeta
\right],
\label{eq:PQmixed_ratio}
\end{align}
Hence, the radial equation can be integrated directly by separation of variables. In particular, the different solutions are determined by the roots of the quadratic polynomial in Eq.~\eqref{eq:PQmixed_ratio}. Thus, for their classification, we define the dimensionless quantity
\begin{equation}
\Delta \equiv 81+144\frac{\eta}{\zeta}\,,
\label{eq:PQmixed_discriminant}
\end{equation}
which is related to the discriminant of the quadratic polynomial by a rescaling factor $9/\zeta^2$.

Thereby, for $\Delta<0$, Eq.~\eqref{eq:PQmixed_ratio} can be directly integrated, leading to a real arctangent-exponential parametric solution. Note, however, that this case does not yield explicit expressions for the metric and torsion functions.

On the other hand, for $\Delta=0$, the ratio of the two combinations of the Lagrangian coefficients is given by
\begin{equation}
 \frac{\eta}{\zeta}=-\,\frac{9}{16}\,,
\end{equation}
namely
\begin{equation}
    h_9+h_{10}=-\,\frac{9}{4}\left(h_{17}-h_{18}\right).\label{eq:PQmixed_critical_relation}
\end{equation}
Then, Eq.~\eqref{eq:PQmixed_ratio} can
be directly integrated, yielding:
\begin{equation}
    \frac{P(r)}{Q(r)}=-\,\frac{1}{3}\left(5+\frac{4}{1+W_{k}\left(\kappa_{3}/r\right)}\right),
\end{equation}
where $\kappa_{3}$ is an integration constant and $W_{k}\left(\kappa_{3}/r\right)$, denotes the $k$-th branch of the
Lambert $W$ function~\cite{Corless:1996zz}:
\begin{equation}
 W_{k}\left(\kappa_{3}/r\right)e^{W_{k}\left(\kappa_{3}/r\right)}=\kappa_{3}/r\,.
\end{equation}
Inserting this result into Eqs.~\eqref{eq:PQmixed_PQ1}-\eqref{eq:PQmixed_PQ2} and integrating once more introduces a second constant $\kappa_4$, which determines the torsion functions as:
\begin{align}
 P(r)
 &=
 -\,\frac{\kappa_{4}}{4}
 \bigl(
 5W_{k}\left(\kappa_{3}/r\right)+9
 \bigr)e^{\frac{1}{3}W_{k}\left(\kappa_{3}/r\right)},
 \\
 Q(r)
 &=
 \frac{3\kappa_{4}}{4}
 \bigl(
 W_{k}\left(\kappa_{3}/r\right)+1
 \bigr)
 e^{\frac{1}{3}W_{k}\left(\kappa_{3}/r\right)},
 \label{eq:PQmixed_Lambert_profiles}
\end{align}
providing then the following metric function:
\begin{equation}
 \Psi(r)
 =
 1-\frac{2m}{r}
 +\frac{9\left(h_{17}-h_{18}\right)}{8} \kappa_{4}^2
 W_{k}^{2}\left(\kappa_{3}/r\right)
 e^{\frac{2}{3}W_{k}\left(\kappa_{3}/r\right)},
 \label{eq:PQmixed_Lambert_metric}
\end{equation}
where the mass parameter $m$ corresponds to the constant $\mu$ in Expression~\eqref{eq:PQmixed_metric}.

On the principal branch with $\kappa_{3}>0$, one has $W_0(\kappa_3/r)\approx\kappa_3/r$ at large distances, so that the torsion functions do not vanish at spatial infinity
\begin{equation}
P(r)\approx-\,\frac94\kappa_4 - \frac{2\kappa_3\kappa_4}{r}\,,
\quad
Q(r)\approx\frac34\kappa_4+ \frac{\kappa_3\kappa_4}{r}\,,
\end{equation}
which nevertheless yields an asymptotically flat metric structure, characterised by the function
\begin{equation}
    \Psi(r) \approx 1 - \frac{2m}{r} + \frac{9\left(h_{17}-h_{18}\right)\kappa_3^2 \kappa_4^2}{8r^2}\,.
\end{equation}

The horizon structure of this solution is analysed in Appendix~\ref{app:horizon-structure}. In particular, it is worth noting that on the globally real principal branch $W_0$ with $\kappa_{3}>0$, the solution possesses an outer event horizon and two inner horizons for a finite range of values of the black hole parameters. This parameter space is bounded by two extremal curves that meet at a critical configuration, where the three horizons coalesce.

Finally, for $\Delta>0$, the solutions can instead be written in algebraic form for particular values of the ratio $\eta/\zeta$. To parametrise this family, we introduce
\begin{equation}
 n\equiv
 \frac12\left(1+\frac{9}{\sqrt{\Delta}}\right).
 \label{eq:PQmixed_ndef}
\end{equation}
Using
\(
\Delta=81+144\eta/\zeta
\),
this is equivalent to
\begin{equation}
 \frac{\zeta}{\eta}
 =-\,\frac{4\left(2n-1\right)^2}{9n\left(n-1\right)}\,,
 \quad n\neq 1\,.
 \label{eq:PQmixed_nratio}
\end{equation}
For $n>1$, Eq.~\eqref{eq:PQmixed_ratio} can be integrated by parametrising the ratio $P(r)/Q(r)$ in terms of a function $q(r)$ as
\begin{equation}
 \frac{P(r)}{Q(r)}
 =
 \frac{5n+2+\left(7-5n\right)q(r)}
 {3\left[\left(n-1\right)q(r)-n\right]}\,.
 \label{eq:PQmixed_ratio_q}
\end{equation}
The resulting implicit equation for $q(r)$ is then
\begin{equation}
 q^{n}(r)
 -\frac{n^n}{\left(n-1\right)^{n-1}}
 \frac{r}{\kappa_{3}}\,q(r)
 +\frac{n^n}{\left(n-1\right)^{n-1}}
 \frac{r}{\kappa_{3}}
 =0\,,
 \label{eq:PQmixed_qequation}
\end{equation}
where $\kappa_{3}$ is an integration constant. Once $q(r)$ is determined, the general solution to Eqs.~\eqref{eq:PQmixed_PQ1}--\eqref{eq:PQmixed_PQ2} is
\begin{align}
 P(r)&=\frac{1}{3}\kappa_{4}\,
 q^{-\left(n+1\right)/3}(r)
 \left[
 5n+2+\left(7-5n\right)q(r)
 \right],
 \label{eq:PQmixed_Pn}\\
 Q(r)
 &=
 \kappa_{4}\,
 q^{-\left(n+1\right)/3}(r)
 \left[\left(n-1\right)q(r)-n
 \right],
 \label{eq:PQmixed_Qn}
\end{align}
where $\kappa_{4}$ is a second integration constant. Accordingly, Eq.~\eqref{eq:PQmixed_metric} provides the metric function
\begin{equation}
 \Psi(r)
 =
 1-\frac{2\mu}{r}
 -\frac{4\eta \kappa_{4}^2\left(2n-1\right)^2}{9}
 \left[q(r)-1\right]^2
 q(r)^{-2\left(n+1\right)/3}\,.
 \label{eq:PQmixed_algebraic_metric}
\end{equation}

For generic values of $n$, the function $q(r)$ is defined implicitly by Eq.~\eqref{eq:PQmixed_qequation}. Explicit expressions can, however, be obtained for particular choices of $n$. As a simple example, for $n=2$ and restricting to the asymptotically flat branch, one finds for $r>\kappa_{3}$
\begin{equation}
 q(r)=
 \frac{2}{
 1+\sqrt{1-\kappa_{3}/r}
 }\,.
 \label{eq:PQmixed_n2_q}
\end{equation}
Substitution into Eqs.~\eqref{eq:PQmixed_Pn}-\eqref{eq:PQmixed_Qn} then gives
\begin{equation}
 P(r)=\kappa_{4}\left(
 1+2\sqrt{1-\frac{\kappa_{3}}{r}}\,
 \right),
 \quad
 Q(r)=-\,\kappa_{4}\sqrt{1-\frac{\kappa_{3}}{r}}\,,
 \label{eq:PQmixed_n2_profiles}
\end{equation}
while, upon identifying $\mu=m$, the metric function reduces to
\begin{equation}
 \Psi(r)=
 1-\frac{2m}{r}
 -2\kappa_{4}^2\left(h_{17}-h_{18}\right)
 \left(
 \frac{\kappa_{3}}{2r}
 +\sqrt{1-\frac{\kappa_{3}}{r}}-1
 \right).
 \label{eq:PQmixed_sqrt_metric}
\end{equation}

As in the previous case, the present solution behaves at large distances as the Reissner-Nordstr\"om-like black hole
\begin{equation}
 \Psi(r)
 \approx
 1-\frac{2m}{r}
 +\frac{\left(h_{17}-h_{18}\right)\kappa_{3}^2\kappa_{4}^2}{4r^2}\,,
 \label{eq:PQmixed_sqrt_asymptotic}
\end{equation}
whereas the torsion functions take the nondecaying asymptotic form
\begin{equation}
    P(r) \approx 3\kappa_{4}-\frac{\kappa_{3}\kappa_{4}}{r}-\frac{\kappa_{3}^{2}\kappa_{4}}{4r^2}\,, \quad Q(r) \approx -\,\kappa_{4}+\frac{\kappa_{3}\kappa_{4}}{2r}+\frac{\kappa_{3}^{2}\kappa_{4}}{8r^2}\,.
\end{equation}
In fact, for arbitrary values of the parameter $n$ within the asymptotically flat branch of this family, Eq.~\eqref{eq:PQmixed_qequation} leads to the asymptotic solution
\begin{equation}
 q(r)
 \approx
 1+
 \frac{\left(n-1\right)^{n-1}}{n^n}
 \frac{\kappa_{3}}{r}\,,
\end{equation}
which yields
\begin{equation}
 \Psi(r)
 \approx
 1-\frac{2m}{r}
 -\eta \kappa_{4}^2
 \left[
 \frac{2\left(2n-1\right)\left(n-1\right)^{n-1}}{3n^n}
 \right]^2
 \frac{\kappa_{3}^2}{r^2}\,.
 \label{eq:PQmixed_algebraic_asymptotic}
\end{equation}
Thus, at large distances all of the asymptotically flat members of this algebraic family approach a Reissner--Nordstr\"om-like geometry.

\section{Even and odd-parity axial-tensor sector: black holes with Lambert $W$ metric corrections and
Boulware-Deser-like geometries}\label{sec:EGB}

In the even and odd-parity axial-tensor sector, the vector mode vanishes and the torsion functions read
\begin{equation}
    P(r)=0\,, \quad Y(r) \neq 0\,, \quad Q(r) \neq 0\,, \quad U(r) \neq 0\,.
\end{equation}
In order to solve the field equations within this sector, we first consider the backreaction effects provided by the even-parity components of the tensor mode alone, which are given by the torsion function $Q(r)$. Specifically, after substituting the metric function~\eqref{eq:general_metric_integrated}, the connection equations~\eqref{eq:Q_connection1} and~\eqref{eq:Q_connection2} form a coupled system of second-order differential equations for $Q(r)$. However, the second-order dependence in both equations enters only through the quantity $(Q(r)Q'(r))'$, which for a certain combination of the Lagrangian coefficients can be then eliminated to reduce these equations to a first-order system for $Q(r)$, without assuming any particular radial profile.

This procedure leads to two qualitatively different branches: a differential branch leading to a black hole solution with Lambert $W$ metric corrections and an algebraic branch giving rise to a Boulware-Deser-like geometry.

From these configurations, we then complete the torsion sector by switching on the axial mode and the odd-parity components of the tensor mode. Since the corresponding field equations become considerably more complicated in the presence of multiple modes, we simplify the analysis by restricting the torsion functions as $Y(r)+2r^2U(r)=0$. Furthermore, we additionally demand a decaying axial mode at large distances, which sets $\kappa_{2}=0$ in Expression~\eqref{eqYsol}. These assumptions make the system of equations tractable and allow us to obtain the respective exact solutions in a systematic way.

\subsection{First-order reduction of the system of equations for the even-parity components of the tensor mode}
\label{subsec:Q-first-order-reduction}

Let us solve first the system of equations for the even-parity components of the tensor mode in the absence of the axial components of this sector:
\begin{equation}
    U(r)=Y(r)=0\,,
    \quad
    Q(r)\neq0\,.
    \label{eq:Q_support}
\end{equation}
In this case, the connection equations $\mathcal C^{(3)}=0$ and $\mathcal C^{(4)}=0$ vanish identically, whereas the only independent equations are the connection equations $\mathcal C^{(1)}=0$ and $\mathcal C^{(2)}=0$, which can be respectively written as
\begin{align}
0=&\;3\mu
\left(
2\beta_1-8h_{10}-4h_{11}+3h_{16}-6h_{17}-4h_9\right)-\beta_1
\left(
2h_{10}-2h_{11}-24h_{12}-3h_{19}-2h_9
\right)rQ^2(r)
\nonumber\\
&+r^2Q(r)
\bigl[
6h_{11}-6h_{10}-6h_{16}-6h_{17}
+18h_{18}-15h_{19}-12h_9-2\beta_1
\nonumber\\
&\hspace{1.6cm}
-\beta_1
\left(
20h_{10}-20h_{11}-96h_{12}
+3h_{16}-6h_{17}+16h_9-2\beta_1
\right)Q'(r)
\bigr]
\nonumber\\
&+r^3
\bigl[\left(
6h_{16}-30h_{10}-6h_{11}-18h_{17}
+6h_{18}-9h_{19}-24h_9+7\beta_1
\right)Q'(r)-6m_t^2
\nonumber\\
&\hspace{1.6cm}
-\beta_1\mathcal D_1
\left(Q(r)Q'(r)\right)'
\bigr],
\label{eq:Q_connection1}
\end{align}
and
\begin{align}
0=&\;6\mu
\left(
\beta_1-4h_{10}-2h_{11}-3h_{16}+6h_{17}-2h_9
\right)-2\beta_1
\left(
h_{10}-h_{11}-12h_{12}+3h_{19}-h_9
\right)rQ^2(r)
\nonumber\\
&+r^2Q(r)
\big[
2\left(
3h_{11}-3h_{10}+6h_{16}+6h_{17}
-18h_{18}+15h_{19}-6h_9+2\beta_1
\right)
\nonumber\\
&\hspace{1.6cm}
-2\beta_1
\left(
10h_{10}-10h_{11}-48h_{12}
-3h_{16}+6h_{17}+8h_9-\beta_1
\right)Q'(r)
\big]
\nonumber\\
&+r^3
\big[\left(
36h_{17}-30h_{10}-6h_{11}-12h_{16}
-12h_{18}+18h_{19}-24h_9+13\beta_1
\right)Q'(r)-6m_t^2
\nonumber\\
&\hspace{1.6cm}
-2\beta_1\mathcal D_2
\left(Q(r)Q'(r)\right)'
\big],
\label{eq:Q_connection2}
\end{align}
with
\begin{align}
\mathcal D_1={}&
20h_{10}-2h_{11}-24h_{12}
-3h_{16}+6h_{17}+3h_{19}+16h_9-4\beta_1\,,\label{eq:Q_D1}\\
\mathcal D_2={}&
10h_{10}-h_{11}-12h_{12}
+3h_{16}-6h_{17}-3h_{19}+8h_9-2\beta_1\,.
\label{eq:Q_D2}
\end{align}
Accordingly, the linear combination
\begin{equation}
    2\mathcal D_2\,\mathcal C^{(1)}
    -\mathcal D_1\,\mathcal C^{(2)}
    =0\,,
    \label{eq:Q_first_order_combination}
\end{equation}
constitutes a first-order differential equation for $Q(r)$.

We now consider the branch for which the derivative dependence in Eq.~\eqref{eq:Q_first_order_combination} occurs only through $Q(r)Q'(r)$, 
which requires the coefficient of the term $Q'(r)$ to vanish
\begin{align}
0=&\;
2\left(6h_{18}-6h_{17}-3h_{19}-2\beta_1\right)
\left(
10h_{10}-h_{11}-12h_{12}+8h_9-2\beta_1
\right)
\nonumber\\
&+3\left(
h_{16}-2h_{17}-h_{19}
\right)
\left(
10h_{10}-10h_{11}-48h_{12}+8h_9+\beta_1
\right).
\label{eq:Q_first_order_condition}
\end{align}
In the general case
\begin{equation}
    6h_{16}-18h_{17}+6h_{18}-9h_{19}-2\beta_1\neq0\,,
    \label{eq:Q_generic_condition}
\end{equation}
Eq.~\eqref{eq:Q_first_order_condition} determines the Lagrangian coefficient $h_{12}$ as
\begin{align}
h_{12}
={}&
\frac{1}{
24\left(
6h_{16}-18h_{17}+6h_{18}-9h_{19}-2\beta_1
\right)}
\bigl\{
8\beta_1^2
-\beta_1
\bigr[
40h_{10}-4h_{11}+32h_9-3\left(
h_{16}+6h_{17}-8h_{18}+3h_{19}
\right)
\bigr]
\nonumber\\
&\hspace{1.2cm}
+6\bigl[
\left(5h_{10}+4h_9\right)
\left(
h_{16}-6h_{17}+4h_{18}-3h_{19}
\right)
+h_{11}
\left(
-5h_{16}+12h_{17}-2h_{18}+6h_{19}
\right)
\bigr]
\bigr\}.
\label{eq:Q_h12}
\end{align}
With the constraint~\eqref{eq:Q_first_order_condition} fixing the Lagrangian coefficient $h_{12}$, the independent system can be then taken as Eq.~\eqref{eq:Q_first_order_combination} together with either one of the original connection equations. We now analyse the branch for which this system can be reduced to a single first-order differential equation
for $Q(r)$. The complementary case, in which the differential reduction degenerates, will be treated separately in Sec.~\ref{subsec:Q-algebraic-case}.

\subsection{Differential branch: black holes with Lambert $W$ metric corrections}
\label{subsec:Q-differential-case}

\subsubsection{Seed solution with the even-parity components of the tensor mode}
\label{subsubsec:Lambert-Q-sector}

We first obtain an analytical solution for the even-parity components of the tensor mode within the differential branch, maintaining the constraint~\eqref{eq:Q_support}. This solution will serve then as a seed configuration to construct the complete geometry including a primary torsion hair, by switching on the axial mode and the odd-parity components of the tensor mode described by the functions $Y(r)$ and $U(r)$.

Upon using Eq.~\eqref{eq:Q_h12} and assuming
\begin{equation}
6h_{16}-18h_{17}+6h_{18}-9h_{19}-2\beta_1\neq0\,,
\quad
h_{16}-2h_{17}-h_{19}\neq0\,,
\label{eq:Q_differential_conditions}
\end{equation}
the first-order differential equation~\eqref{eq:Q_first_order_combination} can be written as
\begin{equation}
\beta_1\lambda_1 rQ^2(r)-3\lambda_0 r^2Q(r)
-\beta_1\lambda_2 r^2Q'(r)Q(r)+\mu\left(2\lambda_1+\lambda_2\right)-6m_{t}^2
\left(
6h_{16}-18h_{17}+6h_{18}-9h_{19}-2\beta_1
\right)r^3=0\,,
\label{eq:Q_first_order_lambda}
\end{equation}
with
\begin{align}
\lambda_0=&\;
\beta_1
\left(
16h_{10}+8h_{11}
-18h_{16}-18h_{17}+54h_{18}-45h_{19}+8h_9
\right)-6\beta_1^2
\nonumber\\
&
+\,12\left[
h_{10}\left(6h_{16}+2h_{17}-14h_{18}+11h_{19}\right)
+4h_{11}\left(h_{17}-h_{18}+h_{19}\right)
+h_9\left(6h_{16}-2h_{17}-10h_{18}+7h_{19}\right)
\right],\quad
\label{eq:Q_lambda0}
\\[1ex]
\lambda_1=&\;
8\beta_1^2
-3\beta_1
\left(
12h_{10}-h_{16}-6h_{17}+8h_{18}+6h_{19}+12h_9
\right)
\nonumber\\
&
+18\left[
h_{10}\left(h_{16}-8h_{17}+6h_{18}+h_{19}\right)
+h_{11}\left(-h_{16}+2h_{17}+2h_{19}\right)
+h_9\left(2h_{16}-10h_{17}+6h_{18}-h_{19}\right)
\right],
\label{eq:Q_lambda1}
\\[1ex]
\lambda_2=&\;
3\beta_1
\left(
40h_{10}+8h_{11}
-17h_{16}+6h_{17}+28h_{18}-6h_{19}+32h_9
\right)-28\beta_1^2
\nonumber\\
&
+18\left(5h_{10}+h_{11}+4h_9\right)
\left(h_{16}+2h_{17}-4h_{18}+2h_{19}\right).
\label{eq:Q_lambda2}
\end{align}

We now consider the generic case
\begin{equation}
    \lambda_0\neq0\,,
    \quad
    6h_{16}-18h_{17}+6h_{18}-9h_{19}-2\beta_1\neq0\,,
    \label{eq:Q_nondegenerate_conditions}
\end{equation}
and introduce the two dimensionless combinations
\begin{equation}
    n_1=\frac{\lambda_1}{3\lambda_0}\,,
    \quad
    n_2=-\,\frac{\lambda_2}{6\lambda_0}\,.
    \label{eq:Q_n1n2}
\end{equation}
which re-expresses Eq.~\eqref{eq:Q_first_order_lambda} as
\begin{equation}
 rQ(r)
 =
 \frac{2\mu\left(n_1-n_2\right)}{r}
 +\beta_1 n_1 Q^{2}(r)
 +2\beta_1 n_2 rQ'(r)Q(r)
 -\frac{2m_t^2}{\lambda_0}
 \left(
 6h_{16}-18h_{17}+6h_{18}-9h_{19}-2\beta_1
 \right)r^2.
 \label{eq:Q_first_order-n12}
\end{equation}

As for the remaining connection equation, we substitute Eq.~\eqref{eq:Q_first_order-n12} into $\mathcal C^{(1)}=0$ and restrict the parameter space to a subspace on which the former directly implies the latter:
\begin{align}
h_5={}&
2h_{12}-\frac{h_6}{2}+\frac{h_7}{2}+\frac{h_8}{4}
-\frac{
12n_1^2-12n_1n_2-n_1+18n_2^2+19n_2+1
}{
24n_1\left(n_1-n_2-1\right)
}\,h_{19}\,,\label{eq:Q_general_couplingsA}
\\
h_9={}&\frac{
3n_1^2+15n_1n_2+14n_1+4n_2+1
}{
12n_1\left(n_1-n_2-1\right)
}\,h_{19}-4h_{12}\,,\\
h_{10}={}&
4h_{12}
+\frac{
6n_1^2-42n_1n_2-23n_1+18n_2^2+20n_2+3
}{
12n_1\left(n_1-n_2-1\right)
}\,h_{19}\,,
\\
h_{11}={}&\frac{
6n_1^2-6n_1n_2-5n_1+9n_2^2+8n_2+1
}{
6n_1\left(n_1-n_2-1\right)
}\,h_{19}-4h_{12}\,,
\\
h_{16}={}&
2h_{18}-2h_{19}
+\frac{2n_2+1}{2n_1}\,h_{19}\,,
\\
h_{17}={}&
h_{18}-\frac32h_{19}
+\frac{4n_2+1}{4n_1}\,h_{19}\,,
\\
\beta_1={}&
\frac{
3\left(2n_1-4n_2-1\right)
}{
2n_1
}\,h_{19}\,.
\label{eq:Q_general_couplings}
\end{align}
In particular,
\begin{equation}
    h_{16}-2h_{17}-h_{19}
    =
    -\,\frac{n_2}{n_1}h_{19}\,,
    \label{eq:Q_n2_relation}
\end{equation}
so the branch considered here corresponds to $n_2\neq0$. Using Expressions~\eqref{eq:Q_general_couplingsA}-\eqref{eq:Q_general_couplings}, the first-order equation reduces to
\begin{equation}
    rQ(r)
    =
    \frac{2\mu\left(n_1-n_2\right)}{r}
    +\beta_1 n_1 Q^2(r)
    +2\beta_1 n_2 rQ'(r)Q(r)
    -\frac{4n_1\left(n_1-n_2-1\right)}
    {h_{19}\left(9n_1+3n_2-1\right)}\,m_t^2r^2\,,
    \label{eq:Q_affine_closure}
\end{equation}
while $\mathcal C^{(2)}=0$ and $\mathcal E=0$ are also satisfied if this equation holds. The present branch of solutions requires then
\begin{equation}
    h_{19}\neq0\,, \quad n_1\neq0\,,
    \quad
    n_2\neq0\,,
    \quad
    n_1-n_2-1\neq0\,,
    \quad
    9n_1+3n_2-1\neq0\,, \quad 2n_1-4n_2-1\neq0\,.
    \label{eq:Q_general_exclusions}
\end{equation}
For convenience, we further use the parametrisation
\begin{equation}
    n_1=\rho\nu\,,
    \quad
    n_2=\nu\,,
    \quad
    \nu\neq0\,,
    \label{eq:Q_rhonu}
\end{equation}
which re-expresses the coefficient $\beta_1$ as
\begin{equation}
    \beta_1
    =
    \frac{3h_{19}(2\rho\nu-4\nu-1)}
    {2\rho\nu}\,,
    \label{eq:Q_differential_beta}
\end{equation}
while Eq.~\eqref{eq:Q_affine_closure} takes the form
\begin{equation}
    rQ(r)
    =
    \frac{2\mu\nu\left(\rho-1\right)}{r}
    +\beta_1\rho\nu Q^2(r)
    +2\beta_1\nu rQ'(r)Q(r)
    -\frac{4\rho\nu\left(\rho\nu-\nu-1\right)}
    {h_{19}\left(9\rho\nu+3\nu-1\right)}\,m_t^2r^2\,.
    \label{eq:Q_differential_closure}
\end{equation}

Defining
\begin{equation}
    u(r)=\frac{Q(r)}{r}\,,
    \label{eq:Q_u_definition}
\end{equation}
Eq.~\eqref{eq:Q_differential_closure} becomes
\begin{equation}
    2\beta_1\nu\,r u'(r) u(r)
    +\beta_1\nu\left(\rho+2\right)u^2(r)-u(r)
    +\frac{2\nu \mu\left(\rho-1\right)}{r^3}
    -\frac{4\rho\nu\left(\rho\nu-\nu-1\right)}
    {h_{19}\left(9\rho\nu+3\nu-1\right)}\,m_t^2
    =0\,.
    \label{eq:Q_u_master}
\end{equation}
For generic values of $\rho$, this expression constitutes a nonlinear first-order equation. We then focus on the massless tensor sector,
\begin{equation}
    m_t^2=0,
    \label{eq:Q_Lambert_mt}
\end{equation}
for which an analytical solution can be obtained.

For $\rho\neq1$, the leading radial dependence in Eq.~\eqref{eq:Q_u_master} motivates the redefinition
\begin{equation}
    u(r)
    =
    \frac{2\nu \mu\left(\rho-1\right)}{r^3}
    \left[1+\mathcal W(r)\right],
    \label{eq:Q_general_w}
\end{equation}
which rewrites the expression as
\begin{equation}
    4\mu\beta_1\nu^2\left(\rho-1\right)\,
    r\left(1+\mathcal W(r)\right)\mathcal W\,'(r)
    +2\mu\beta_1\nu^2\left(\rho-1\right)\left(\rho-4\right)\left(1+\mathcal W(r)\right)^2
    -r^3\mathcal W(r)=0\,.
    \label{eq:Q_rho_w}
\end{equation}
The term proportional to $\left(1+\mathcal W(r)\right)^2$ vanishes for
\begin{equation}
    \rho=4\,,
    \label{eq:Q_rho4}
\end{equation}
which makes the equation directly integrable. For this branch, the respective Lagrangian coefficients can be obtained by substituting $n_1=4\nu$ and $n_2=\nu$ into Eqs.~\eqref{eq:Q_general_couplingsA}-\eqref{eq:Q_general_couplings}:
\begin{align}
    \beta_1&=\frac{3h_{19}\left(4\nu-1\right)}{8\nu}\,,\label{eq:Lambert-inherited-couplingsA}
\\
h_5&= 2h_{12} - \frac{h_6}{2} + \frac{h_7}{2} + \frac{h_8}{4}- \frac{162\nu^2 + 15\nu + 1}{96\nu(3\nu - 1)}\,h_{19}\,,
\\
 h_9&=\frac{108\nu^2+60\nu+1}
 {48\nu\left(3\nu-1\right)}\,h_{19}-4h_{12}\,,
\quad
 h_{10}=4h_{12}
 +\frac{1-24\nu-18\nu^2}
 {16\nu\left(3\nu-1\right)}\,h_{19}\,,
\\
 h_{11}&=\frac{81\nu^2-12\nu+1}
 {24\nu\left(3\nu-1\right)}\,h_{19}-4h_{12}\,,
\quad
 h_{16}=2h_{18}
 +\frac{1-14\nu}{8\nu}\,h_{19}\,,
\\
 h_{17}&=h_{18}
 +\frac{1-20\nu}{16\nu}\,h_{19}\,.
\label{eq:Lambert-inherited-couplingsB}
\end{align}
In addition, by defining the parameter
\begin{equation}
    \ell^2
    \equiv
    -\,9\beta_1\nu^2
    =
    \frac{27}{8}\nu h_{19}\left(1-4\nu\right)>0\,,
    \label{eq:Q_Lambert_ell}
\end{equation}
Eq.~\eqref{eq:Q_rho_w} acquires the compact expression
\begin{equation}
    \mathcal W\,'(r)
    =
    -\,\frac{3r^2}{4\mu\ell^2}
    \frac{\mathcal W(r)}{1+\mathcal W(r)}\,.
    \label{eq:Q_Lambert_w_eq}
\end{equation}
Integration gives
\begin{equation}
    \mathcal W(r)+\ln |\mathcal W(r)|
    =
    \ln|\kappa_{3}|
    -\frac{r^3}{4\mu\ell^2}\,,
\end{equation}
where $\kappa_{3}$ represents the torsion hair. At this stage, we can also identify the integration constant $\mu$ with the mass parameter $m$, as follows from the asymptotic behaviour of the metric function, which means
\begin{equation}
    \mathcal W(r)
    =
    W_0\Bigl(
        \kappa_{3}
        e^{
        -\,r^3/(4m\ell^2)}
    \Bigr)\,.
    \label{eq:Q_Lambert_W}
\end{equation}
Thus, for $\rho=4$, the solution for the torsion function describing the even-parity components of the tensor mode reads
\begin{equation}
    Q(r)
    =
    \frac{6\nu m}{r^2}
    \left[1+\mathcal W(r)\right],
    \label{eq:Q_Lambert_profile}
\end{equation}
which at large distances presents the asymptotic behaviour
\begin{equation}
    Q(r)\approx \frac{6\nu m}{r^2}\,.
\end{equation}

\subsubsection{Complete solution with the odd-parity components of the axial and tensor modes}
\label{subsubsec:Lambert-UY-sector}

We now complete the torsion configuration by switching on the remaining odd-parity components of this sector, for which we assume
\begin{equation}
    U(r)=-\,\frac{Y(r)}{2r^2}\,,
\end{equation}
namely
\begin{equation}
    U(r)=-\,\frac{1}{2r^2}\left(\kappa_{1}+\kappa_{2}r^3\right).
\end{equation}
To focus on the branch in which the axial mode decays asymptotically, we set $\kappa_{2}=0$, which reduces the torsion functions as
\begin{equation}
    U(r)=-\,\frac{\kappa_{1}}{2r^2}\,,
    \quad
    Y(r)=\kappa_{1}\,.
    \label{eq:Lambert-UY-charge}
\end{equation}
Thus, the integration constant $\kappa_{1}$ represents the additional torsion hair arising from this sector.

Substituting into the full set of field equations the seed solution given by Expression~\eqref{eq:Q_Lambert_W}, together with Expressions~\eqref{eq:Q_Lambert_profile} and \eqref{eq:Lambert-UY-charge}, these equations simply provide additional algebraic conditions on the remaining Lagrangian coefficients:
\begin{align}
  h_3 &= R_3(\nu)\,h_{19}\,,\quad 
 h_4 = R_4(\nu)\,h_{19}\,,\quad 
 h_5 = h_{12}-\frac{h_6}{2}+\frac{h_7}{2}
        +R_5(\nu)\,h_{19}\,, \quad h_8 = -4h_{12}+R_8(\nu)\,h_{19}\,,\label{eq:Lambert-charged-couplings1}
\\
 h_{15} &= -\frac{h_{18}}{3}+R_{15}(\nu)\,h_{19}\,,\quad
 h_{20} = -\frac{h_{24}}{2}+R_{20}(\nu)\,h_{19}\,, \quad h_{21} = h_{22}-h_{24}+h_{26}
           +R_{21}(\nu)\,h_{19}\,,\label{eq:Lambert-charged-couplings2}\\
  h_{23} &= 2h_{24}+R_{23}(\nu)h_{19}\,,\quad 
 h_{25} = R_{25}(\nu)\,h_{19}\,, \quad m_S^2 = m_T^2=0\,,
\label{eq:Lambert-charged-couplings3}
\end{align}
where the parameters $R_i$ are defined as
\begin{align}
R_3(\nu)={}&
-\frac{
3240\nu^4+5514\nu^3+1081\nu^2+337\nu-10
}{
864\nu(3\nu-1)\Delta_\nu
}\,,
\\
R_4(\nu)={}&
\frac{
6480\nu^4+9156\nu^3+8762\nu^2-79\nu-5
}{
1728\nu(3\nu-1)\Delta_\nu
}\,,
\\
R_5(\nu)={}&
-\frac{
24336\nu^4+34272\nu^3-27433\nu^2+2472\nu-45
}{
96\nu(3\nu-1)\Delta_\nu
}\,,
\\
R_8(\nu)={}&
-\frac{
21744\nu^4+43104\nu^3-27419\nu^2+2453\nu-50
}{
24\nu(3\nu-1)\Delta_\nu
}\,,
\\
R_{15}(\nu)={}&
\frac{
18720\nu^4+2400\nu^3-5454\nu^2-207\nu+35
}{
288\nu(3\nu-1)\Delta_\nu
}\,,
\\
R_{20}(\nu)={}&
-\frac{
44352\nu^4+74640\nu^3-51156\nu^2+3431\nu-50
}{
288\nu(3\nu-1)\Delta_\nu
}\,,
\\
R_{21}(\nu)={}&
-\frac{
40896\nu^4+65112\nu^3-41794\nu^2+2866\nu-55
}{
576\nu(3\nu-1)\Delta_\nu
}\,,
\\
R_{23}(\nu)={}&
\frac{
396\nu^3+633\nu^2-426\nu+10
}{
4(3\nu-1)\Delta_\nu
}\,,
\\
R_{25}(\nu)={}&
\frac{
2376\nu^4+4422\nu^3-4756\nu^2+311\nu-5
}{
72\nu(3\nu-1)\Delta_\nu
}\,,
\label{eq:Lambert-R-functions}
\end{align}
with
\begin{equation}
    \Delta_\nu\equiv16\nu^2-56\nu+5\,.
    \label{eq:Lambert-Delta-nu}
\end{equation}
Hence, on top of the constraints provided by Expression~\eqref{eq:Q_general_exclusions} for the seed solution constructed from the even-parity components of the tensor mode, including the odd-parity components of the axial and tensor modes requires $\Delta_\nu\neq0$.

The corresponding metric function~\eqref{eq:general_metric_integrated}
takes then the compact form
\begin{equation}
 \Psi(r)
 =
 1-\frac{2m}{r}
 +\frac{4m^2\ell^2}{r^4}
   \left[1+\mathcal W(r)\right]^2
 +\frac{h_{19}\,\mathcal P_\nu}
 {96\nu\left(3\nu-1\right)\Delta_\nu}
 \frac{\kappa_{1}^2}{r^4}\,,
\label{eq:charged-Lambert-metric}
\end{equation}
where
\begin{equation}
 \mathcal P_\nu
 =
 6912\nu^4+10800\nu^3-7176\nu^2+471\nu-5\,.
\label{eq:charged-Lambert-Pnu}
\end{equation}
The solution therefore carries two independent torsion hairs, $\kappa_{3}$ and $\kappa_{1}$. The former is encoded in the Lambert $W$ function~\eqref{eq:Q_Lambert_W}, whereas the latter gives rise to an inverse-quartic correction in the metric function. Indeed, at large distances, this function takes the asymptotic form \begin{equation}
 \Psi(r)
 \approx
 1-\frac{2m}{r}
 +\frac{1}{r^4}
 \left[
 4m^2\ell^2
 +\frac{h_{19}\mathcal P_\nu}
 {96\nu\left(3\nu-1\right)\Delta_\nu}\kappa_{1}^2
 \right]+\frac{8\kappa_{3}m^2\ell^2}{r^4} e^{-\,r^3/(4m\ell^2)}\,,
\end{equation}
which shows that the effects of the torsion hair $\kappa_{3}$ are exponentially suppressed in the asymptotic region.

The horizon structure of the solution is further discussed in Appendix~\ref{app:horizon-structure}. In the presence of both torsion hairs, the numerical analysis reveals horizonless, one-horizon and two-horizon configurations separated by an extremal branch. No configurations with more than two horizons were found within the parameter ranges explored.

\subsection{Algebraic branch: Boulware-Deser-like geometry with a cosmological constant}
\label{subsec:Q-algebraic-case}

\subsubsection{Seed solution with the even-parity components of the tensor mode}

We now return to the second possibility identified in the first-order reduction of the field equations for the even-parity tensor sector. The Lambert $W$ solution
studied above belongs to the branch $h_{16}-2h_{17}-h_{19}\neq0$, for which the connection equations retain a nontrivial differential dependence on the function $Q(r)$. The complementary case leads to a qualitatively different behaviour: the differential relation degenerates and the radial dependence of $Q(r)$ is instead determined algebraically. As in the previous branch, we first solve this configuration for the even-parity components of the tensor mode and then switch on the additional odd-parity components of the axial and tensor modes. Therefore, we impose
\begin{equation}
    h_{16}-2h_{17}-h_{19}=0\,.
    \label{eq:Q_algebraic_condition}
\end{equation}
Note that this possibility was excluded in deriving Eq.~\eqref{eq:Q_first_order_lambda}, where the corresponding combination was assumed to be nonzero. The algebraic case must therefore be obtained directly from the original connection equations \eqref{eq:Q_connection1} and \eqref{eq:Q_connection2}.

After imposing Eq.~\eqref{eq:Q_algebraic_condition} together with Eq.~\eqref{eq:Q_h12}, the remaining connection equations admit a branch in which the radial derivative of $Q(r)$ drops out. The field equations then impose a set of algebraic relations among the Lagrangian coefficients, which can be written as
\begin{align}\label{eq:Q_algebraic_couplings_Nfree_1st}
h_5={}&
2h_{12}-\frac{h_6}{2}+\frac{h_7}{2}+\frac{h_8}{4}
-\frac{
2\beta_1^2-9\beta_1h_{19}+63h_{19}^2
}{
36\left(2\beta_1-3h_{19}\right)
}\,,
\\
h_9={}&
-4h_{12}
+\frac{
4\beta_1^2-108\beta_1h_{19}+315h_{19}^2
}{
36\left(2\beta_1-3h_{19}\right)
}\,,
\\
h_{10}={}&
4h_{12}
+\frac{
2\beta_1^2+11\beta_1h_{19}-42h_{19}^2
}{
6\left(2\beta_1-3h_{19}\right)
}\,,
\\
h_{11}={}&
-4h_{12}
+\frac{
\beta_1\left(2\beta_1+3h_{19}\right)
}{
9\left(2\beta_1-3h_{19}\right)
}\,,
\\
h_{16}={}&
2h_{18}-h_{19}-\frac{\beta_1}{3}\,,
\\
h_{17}={}&
h_{18}-h_{19}-\frac{\beta_1}{6}\,.
\label{eq:Q_algebraic_couplings_Nfree}
\end{align}
For convenience, let us reparametrise the relations above by introducing the
dimensionless parameter\footnote{Note that although the parameter $N$ enters explicitly in the metric and torsion functions, it does not constitute an independent integration constant of the solutions.}:
\begin{equation}
    N\equiv
    \frac{3h_{19}}{2\left(3h_{19}-\beta_1\right)}\,,
    \label{eq:Q_algebraic_Ndef}
\end{equation}
which provides
\begin{align}
h_5={}&
2h_{12}-\frac{h_6}{2}+\frac{h_7}{2}+\frac{h_8}{4}
-\frac{12N^2-N+1}{24N\left(N-1\right)}\,h_{19}\,,\label{eq:Q_algebraic_couplingsA}
\\
h_9={}&
-4h_{12}
+\frac{3N^2+14N+1}{12N\left(N-1\right)}\,h_{19}\,,
\\
h_{10}={}&
4h_{12}
+\frac{6N^2-23N+3}{12N\left(N-1\right)}\,h_{19}\,,
\\
h_{11}={}&
-4h_{12}
+\frac{6N^2-5N+1}{6N\left(N-1\right)}\,h_{19}\,,
\\
h_{16}={}&
2h_{18}
+\frac{1-4N}{2N}\,h_{19}\,,
\\
h_{17}={}&
h_{18}
+\frac{1-6N}{4N}\,h_{19}\,,
\\
\beta_1={}&
\frac{3h_{19}\left(2N-1\right)}{2N}\,.
\label{eq:Q_algebraic_couplingsB}
\end{align}
This parametrisation requires $N\neq0,1$. In addition, the value $N=1/2$, for which the algebraic equation derived below becomes linear, defines a separate degenerate branch that will not be considered here. On the other hand, the values $N=1/9$ and $N=2/3$ correspond to exceptional reductions and will be treated separately below.

After imposing the above relations, the derivatives of $Q(r)$ can be eliminated between the two remaining connection equations. For $N\neq2/3$, we find
\begin{equation}
    \mathcal C^{(1)}
    -\frac{1-3N}{4-6N}\mathcal C^{(2)}=0
    =
    4\mu h_{19}N\left(9N-1\right)
    -8m_t^2N\left(N-1\right)r^3
    +h_{19}\left(9N-1\right)rQ(r)
    \left[
    3h_{19}\left(2N-1\right)Q(r)-2r
    \right].
    \label{eq:Q_algebraic_equation}
\end{equation}

As can be seen, Eq.~\eqref{eq:Q_algebraic_equation} is a quadratic equation for the torsion function $Q(r)$. For $N\neq 1/9$, the two possible roots are
\begin{equation}
    Q_{\pm}(r)
    =
    \frac{r}{3h_{19}\left(2N-1\right)}
    \left[
    1\pm
    \sqrt{
    1-\frac{12h_{19}N\left(2N-1\right)\mu}{r^3}
    +\frac{24N\left(N-1\right)\left(2N-1\right)}
    {9N-1}\,m_t^2
    }
    \right].
    \label{eq:Q_algebraic_roots}
\end{equation}
Thus, in contrast with the differential branch, the radial dependence of $Q(r)$ is fixed algebraically and no additional integration is required. The important distinction between the two cases is therefore already visible at the level of the connection equations. In the differential branch, $Q(r)$ obeys a nonlinear first-order equation whose integrable member gives the Lambert $W$ profile, whereas the present branch is determined by the two algebraic roots~\eqref{eq:Q_algebraic_roots}. After the remaining torsion sector is included below, the same quadratic structure gives rise to the Boulware-Deser-type square-root geometry.

As previously mentioned, there are exceptional values of $N$ that must be treated independently. For $N=1/9$, Eq.~\eqref{eq:Q_algebraic_equation} simply reduces to $m_t^2=0$ and the radial dependence of $Q(r)$ must be obtained from one of the original connection equations. Substituting then $N=1/9$ and $m_t^2=0$ into $\mathcal C^{(1)}=0$ and integrating this equation gives
\begin{equation}
    \frac76h_{19}Q^2(r)+rQ(r)-\frac{2\mu}{9r}
    =\kappa_{4} r^2,
    \label{eq:Q_N19_first_integral}
\end{equation}
where $\kappa_{4}$ is an integration constant.  Solving this quadratic equation provides the roots
\begin{equation}
    Q_{\pm}(r)
    =
    \frac{3r}{7h_{19}}
    \left(
    -\,1\pm
    \sqrt{
    1+\frac{14}{3}h_{19}\kappa_{4}
    +\frac{28h_{19}\mu}{27r^3}
    }\,
    \right).
    \label{eq:Q_N19_roots}
\end{equation}

The second exceptional value is $N=2/3$. In this case, the coefficient multiplying $\mathcal C^{(2)}$ in Eq.~\eqref{eq:Q_algebraic_equation} is singular and the elimination leading to Eq.~\eqref{eq:Q_algebraic_equation} cannot be performed. Returning directly to the original connection equations and imposing them simultaneously gives instead
\begin{equation}
    Q_{\pm}(r)
    =
    \frac{r}{h_{19}}
    \left(
    1\pm
    \sqrt{
    1-\frac{16}{45}m_t^2
    -\frac{8h_{19}\mu}{3r^3}
    }\,
    \right),
    \label{eq:Q_N23_roots}
\end{equation}
with no additional integration constant associated with this branch.

\subsubsection{Complete solution with the odd-parity components of the axial and tensor modes}
\label{subsubsec:EGB-UY-sector}

We now switch on the remaining odd-parity components of the axial and tensor modes. As in the Lambert $W$ branch, we impose the simplifying relation $Y(r)+2r^2U(r)=0$. Using Expression~\eqref{eqYsol} and restricting to the branch $\kappa_{2}=0$, this gives again the torsion profile~\eqref{eq:Lambert-UY-charge} for the aforementioned odd-parity components of the axial and tensor modes.

In order to solve the corresponding field equations, we first consider the generic algebraic branch where $N\neq1/9$ and discuss the exceptional value $N=1/9$ separately below.

For $N\neq1/9$, it is convenient to introduce
\begin{equation}
    \chi
    \equiv
    -\,\frac{4\left(N-1\right)m_t^2}
    {h_{19}\left(9N-1\right)}\,.
    \label{eq:EGB-chi}
\end{equation}
The algebraic relation obtained for the torsion function $Q(r)$ can then be written as
\begin{equation}
    Q(r)
    =
    \frac{N}{r}
    \left(
        1-\Psi(r)+\chi r^2
    \right).
    \label{eq:EGB-Q-closure-charged}
\end{equation}
Substituting Eqs.~\eqref{eq:Lambert-UY-charge} and \eqref{eq:EGB-Q-closure-charged} into the complete field equations, together with Expressions~\eqref{eq:Q_algebraic_couplingsA}-\eqref{eq:Q_algebraic_couplingsB}, the remaining equations provide a further set of relations for the Lagrangian coefficients
\begin{align}
 h_3&=R_3(N)\,h_{19}\,,\label{eq:EGB-charged-extra-couplingsA}
\quad
 h_4=R_4(N)\,h_{19}\,,
\quad
 h_8=R_8(N)\,h_{19}-4h_{12}\,,
\\
 h_{15}&=R_{15}(N)\,h_{19}-\frac{h_{18}}{3}\,,
\quad
 h_{20}=R_{20}(N)\,h_{19}-\frac{h_{24}}{2}\,,
\\
 h_{21}&=h_{22}-h_{24}+h_{26}
          +R_{21}(N)\,h_{19}\,,
\quad
 h_{23}=2h_{24}+R_{23}(N)\,h_{19}\,,
\\
 h_{25}&=R_{25}(N)\,h_{19}\,,
\quad
 m_S^2=
 \frac{12N^2-424N+55}
 {6\left(9N-1\right)\left(24N-5\right)}\,m_t^2 \,,\label{eq:EGB-charged-extra-couplingsB}
\end{align}
with
\begin{align}
 R_3(N)&=
 -\,\frac{36N^3-408N^2-111N+20}
 {432N\left(N-1\right)\left(24N-5\right)}\,,
\\
 R_4(N)&=
 \frac{36N^3-840N^2+27N+10}
 {864N\left(N-1\right)\left(24N-5\right)}\,,
\\
 R_8(N)&=
 -\,\frac{336N^3+1571N^2-595N+50}
 {6N\left(N-1\right)\left(24N-5\right)}\,,
\\
 R_{15}(N)&=
 \frac{42N^2-13N+7}
 {72N\left(N-1\right)}\,,
\\
 R_{20}(N)&=
 -\,\frac{462N^3+1445N^2-405N+25}
 {36N\left(N-1\right)\left(24N-5\right)}\,,
\\
 R_{21}(N)&=
 -\,\frac{1248N^3+2818N^2-774N+55}
 {144N\left(N-1\right)\left(24N-5\right)}\,,
\\
 R_{23}(N)&=
 \frac{66N^2+161N-20}
 {6\left(N-1\right)\left(24N-5\right)}\,,
\\
 R_{25}(N)&=
 \frac{9N-1}{18N\left(N-1\right)}\,.
\label{eq:EGB-R-functions}
\end{align}
Thus, the present completion introduces the additional constraint $N\neq5/24$. The particular case $N=5/24$ would therefore require a separate treatment of the field equations and will not be considered here.

In summary, the complete torsion configuration is therefore
\begin{equation}
 P(r)=0\,,\quad
 Q(r)=\frac{N}{r}\left(1-\Psi(r)+\chi r^2\right),
 \quad
 U(r)=-\,\frac{\kappa_{1}}{2r^2}\,,
 \quad
 Y(r)=\kappa_{1}\,.
 \label{eq:EGB-generic-torsion-final}
\end{equation}
For convenience, we introduce the combinations
\begin{equation}
 \mathcal A
 =
 3N\left(2N-1\right)h_{19}\,,
 \quad
 \mathcal C
 =
 \frac{h_{19}\left(6N^2+13N-1\right)}
 {6N\left(N-1\right)}\,,
 \quad
 \Delta
 =
 \sqrt{1-2\mathcal A\chi}\,.
 \label{eq:EGB-generic-combinations}
\end{equation}
For the generic real branch considered below, we assume $\Delta>0$, or equivalently $1-2\mathcal A\chi>0$. Note that the limiting case $\Delta=0$ is degenerate and is not covered by the normalisation $\mu=m\Delta$ or by the asymptotic expansion used below. 

The remaining tetrad equation can then be integrated exactly. After normalising the Schwarzschild term by introducing the mass parameter
$m$ through $\mu=m\Delta$, the two algebraic branches are
\begin{equation}
 \Psi_{\pm}(r)
 =
 1+\frac{r^2}{\mathcal A}
 \left(
 \chi\mathcal A-1
 \pm
 \sqrt{
 \Delta^2
 -\frac{4\mathcal A m\Delta}{r^3}
 +\frac{\mathcal A\mathcal C\kappa_{1}^2}{2r^6}
 }\,
 \right).
 \label{eq:EGB-generic-metric-final}
\end{equation}
The physical domain of the solution is therefore restricted to regions where the argument of the square root is nonnegative, with a simple zero of this quantity corresponding
generically to a branch singularity rather than to a Killing horizon.
On the other hand, the branch continuously connected to the GR limit corresponds to the plus sign, which behaves at large distances as
\begin{equation}
 \Psi_{+}(r)
 \approx
 1 -\frac{2m}{r}
 +\frac{\chi\mathcal A-1+\Delta}{\mathcal A}\,r^2
 +\frac{1}{\Delta}
 \left(
    \frac{\mathcal C\kappa_{1}^2}{4}
    -2\mathcal A m^2
 \right)\frac{1}{r^4}\,.
 \label{eq:EGB-generic-asymptotic}
\end{equation}
Thereby, the parameter $\chi$ controls the asymptotic curvature and plays the role of an effective cosmological constant.  In contrast, $\kappa_{1}$ is an independent integration constant and does not modify the mass parameter $m$, while its leading backreaction on the geometry appears at order $r^{-4}$.


The square-root structure in Expression~\eqref{eq:EGB-generic-metric-final} is reminiscent of the two-branch black hole solutions originally found in the framework of Einstein-Gauss-Bonnet gravity~\cite{Boulware:1985wk}, as well as of their more recent four-dimensional counterparts and regularised realisations~\cite{Glavan:2019inb,Lu:2020iav,Fernandes:2020nbq,Aoki:2020lig}. The similarity, however, concerns only the form of the background metric. Here, the square-root geometry arises directly from the backreaction effects of a dynamical torsion field in a four-dimensional RC space-time, without invoking a singular dimensional limit or a regularisation of the Gauss-Bonnet term. Moreover, the additional $r^{-4}$ contribution is sourced by the independent torsion hair $\kappa_{1}$. Therefore, the present solution describes a Boulware-Deser-like geometry carrying short-range primary torsion hair.

We finally discuss the two exceptional values for the parameter $N$. For $N=1/9$, the generic reduction cannot be used because the factor $9N-1$ vanishes. Thus, we return directly to the original field equations, keeping \;\;\;\;\;\;\;\;\;\;\;\;\;$U(r)=-\,\kappa_{1}/(2r^2)$ and $Y(r)=\kappa_{1}$. In the complete exceptional branch, the additional constant that arises in the seed configuration is related by the remaining field equations to the mass parameter of the axial mode of torsion. Since we are interested here in integration constants that remain independent of the Lagrangian coefficients, we restrict to the subbranch $m_S^2=0$. For $h_{19}\neq0$, the complete compatibility conditions then imply $\kappa_{4}=0$. Thus, the field equations are consistently solved by the relations
\begin{align}
 h_3&=-\,\frac{31}{1152}h_{19}\,,\quad
 h_4=\frac{31}{2304}h_{19}\,,\quad
 h_{14}=-\,\frac{31}{576}h_{19}\,,\quad
 h_{15}=\frac{31}{288}h_{19}\,,
 \label{eq:EGB-N19-special-couplingsA}
 \\[1mm]
 h_5&=h_{12}-\frac12h_6+\frac12h_7-\frac{23}{96}h_{19}\,,
 \quad
 h_8=-\,4h_{12}-\frac{65}{24}h_{19}\,,
 \label{eq:EGB-N19-special-couplingsB}
 \\[1mm]
 h_9&=-\,4h_{12}-\frac{35}{16}h_{19}\,,\quad
 h_{10}=4h_{12}-\frac{7}{16}h_{19}\,,\quad
 h_{11}=-\,4h_{12}-\frac78h_{19}\,,
 \label{eq:EGB-N19-special-couplingsC}
 \\[1mm]
 h_{16}&=-\,\frac{157}{48}h_{19}\,,\quad
 h_{17}=-\,\frac{205}{96}h_{19}\,,\quad
 h_{18}=-\,\frac{277}{96}h_{19}\,,
 \label{eq:EGB-N19-special-couplingsD}
 \\[1mm]
 h_{20}&=-\,\frac12h_{24}+\frac{53}{288}h_{19},\quad
 h_{21}=h_{22}-h_{24}+h_{26}-\frac{191}{1152}h_{19}\,,
 \label{eq:EGB-N19-special-couplingsE}
 \\[1mm]
 h_{23}&=2h_{24}-\frac{5}{48}h_{19}\,,\quad
 h_{25}=0\,,\quad
 m_t^2=m_S^2=0\,,
 \label{eq:EGB-N19-special-couplings}
\end{align}
as well as by the algebraic expression
\begin{equation}
 rQ(r)+\frac{7}{6}h_{19}Q^2(r)-\frac{2\mu}{9r}
 -\frac{7h_{19}\kappa_{1}^2}{288r^4}=0\,.
 \label{eq:EGB-N19-first-integral}
\end{equation}
Combining this relation with the metric function~\eqref{eq:general_metric_integrated} gives the two Boulware-Deser-type branches
\begin{equation}
 \Psi_{\pm}(r)
 =
 1+\frac{27r^2}{7h_{19}}
 \left(
 1\pm
 \sqrt{
 1+\frac{28h_{19}m}{27r^3}
 +\frac{49h_{19}^2\kappa_{1}^2}{432r^6}
 }\,
 \right),
 \label{eq:EGB-N19-metric}
\end{equation}
where we have directly identified the integration constant $\mu=m$ with the mass parameter. Note that the branch continuously connected with the GR limit corresponds to the minus sign and is asymptotically flat. In addition, the constant $\kappa_{1}$ remains independent, representing a genuine primary torsion hair that produces a short-range deformation of the geometry.

The second exceptional value is $N=2/3$. As discussed previously, this case must be treated separately in the seed sector because the particular elimination leading to Eq.~\eqref{eq:Q_algebraic_equation} becomes singular. Nevertheless, the complete solution has a regular limit at $N=2/3$. Thus, this branch is obtained
by evaluating Expressions~\eqref{eq:EGB-chi}-\eqref{eq:EGB-generic-metric-final} at $N=2/3$. In particular, we have
$\chi=4m_t^2/(15h_{19})$,
$\mathcal A=2h_{19}/3$,
$\mathcal C=-\,31h_{19}/4$, and
$\Delta=\sqrt{1-16m_t^2/45}$, whereas the integration constant $\kappa_{1}$ remains as an independent primary torsion hair. In any case, for $\kappa_{1}=0$, the torsion function $Q(r)$ reduces to Expression~\eqref{eq:Q_N23_roots}, with the two signs corresponding to the two algebraic branches.

\section{Odd-parity axial-tensor sector: regular black holes with primary torsion hair}
\label{general-regular-axial-tensor}

Regarding the search for completely regular solutions, we first consider the odd-parity axial and tensor sector, described by the torsion functions
\begin{equation}
 P(r)=Q(r)=0\,,
 \quad
 Y(r)=\kappa_{1}+\kappa_{2} r^3\,, \quad U(r) \neq 0\,.
 \label{eq:general-regular-support}
\end{equation}
Specifically, the axial mode~\eqref{eq:Svector_PQ} is completely regular at the origin if $\kappa_{1}=0$, whereas the absence of divergences in the metric function~\eqref{eq:general_metric_integrated} demands $\mu=0$.

Under these conditions, we then perform a rank-one reduction in the field equations and find regular configurations with primary torsion hair.

\subsection{Rank-one reduction in the field equations}
\label{subsec:general-regular-reduction}
In the present case, the system of equations consists of four independent connection equations and one remaining tetrad equation. To solve this system, we first require that the connection equations $\mathcal C^{(3)}$, $\mathcal C^{(4)}$ and the reduced tetrad equation $\mathcal E$ are satisfied for any value of the torsion function $U(r)$. In the branch $h_{10}+h_{11}\neq0$, a particular coefficient matching solves these equations by the relations
\begin{align}
 h_3&=h_4=h_{25}=m_t^2=m_S^2=0\,,
 \label{eq:general-regular-identitiesB}\\
 h_{12}&=\frac{h_{10}}6-\frac{h_{11}}{12}\,,\quad
 h_{23}=2h_{24}\,,\quad
 h_9=\frac{h_{10}+4h_{11}}3\,,\quad
 h_8=\frac{2h_{10}+8h_{11}}3+4h_5+2h_6-2h_7\,.
 \label{eq:general-regular-identitiesA}
\end{align}
which means that the parameters entering the metric function~\eqref{eq:general_metric_integrated} become
\begin{equation}
 \alpha_3=54\left(h_{10}+h_{11}\right),
 \quad
 \alpha_4=18\left(h_{10}+h_{11}\right),
 \quad
 \beta_1=\frac{14}{3}\left(h_{10}+h_{11}\right),
 \quad
 \beta_2=-\,6\left(h_{10}+h_{11}\right).
 \label{eq:general-regular-alpha-relations}
\end{equation}

Thus, the other two connection equations $\mathcal C^{(1)}=0$ and $\mathcal C^{(2)}=0$ can be respectively written as
\begin{align}
    &\gamma_1U^2(r)+\gamma_2rU'(r)U(r)+\gamma_3\kappa_{2}rU(r)+\gamma_4\kappa_{2}r^2U'(r)
 +\gamma_5\kappa_{2}^2r^2=0\,,
 \label{eq:general-regular-C1}\\
 &\gamma_6U^2(r)+\gamma_7rU'(r)U(r)
 +\gamma_8\kappa_{2}rU(r)+\gamma_9\kappa_{2}r^2U'(r)
 +\gamma_{10}\kappa_{2}^2r^2=0\,,
 \label{eq:general-regular-C2}
\end{align}
with $\gamma_1,\ldots,\gamma_{10}$ denoting cumbersome combinations of the remaining Lagrangian coefficients. In the general case $h_{10}+h_{11}\neq0$, these equations are independent and impose different restrictions on the function $U(r)$.
Thereby, we solve the equations for a particular parameter space in which they become proportional and reduce to a single independent first-order equation
\begin{equation}
 \left(\gamma_6,\gamma_7,\gamma_8,\gamma_9,\gamma_{10}\right)
 =\sigma\left(\gamma_1,\gamma_2,\gamma_3,\gamma_4,\gamma_5\right),
 \label{eq:general-regular-rank-one}
\end{equation}
where $\sigma$ is a proportionality constant.

Furthermore, focusing on asymptotically flat geometries and on the classes of models where $h_{10}+h_{11} \neq 0$ and $\gamma_{4} \neq 0$, the torsion function $U(r)$ must satisfy the following boundary conditions:
\begin{equation}
 rU(r)\longrightarrow0\,,
 \quad
 r^2U'(r)\longrightarrow0\,,
 \quad
 r\longrightarrow\infty\,.
 \label{eq:general-regular-asymptotic-U}
\end{equation}
Accordingly, Eqs.~\eqref{eq:general-regular-C1}-\eqref{eq:general-regular-C2} constrain the parameters as
\begin{equation}
 \gamma_5=\gamma_{10}=0\,,
 \label{eq:general-regular-gamma5}
\end{equation}
whereas the asymptotic form of the torsion function is given by
\begin{equation}
 U(r)\propto r^{-\gamma_3/\gamma_4},
\end{equation}
with $\gamma_3/\gamma_4>1$. Then, we further select the conventional $1/r$ fall-off in the metric function at large distances, which sets
\begin{equation}
 \gamma_3=2\gamma_4\,.
 \label{eq:general-regular-gamma3}
\end{equation}
Additionally, the absence of backreaction effects associated with the axial mode of torsion, which could otherwise violate asymptotic flatness, requires
\begin{equation}
 h_{11}=-\,\frac{1}{4}\left(4h_{10}+9c_{1}\right).
 \label{eq:general-regular-AF}
\end{equation}

Combining the proportionality condition~\eqref{eq:general-regular-rank-one} with Eqs.~\eqref{eq:general-regular-gamma5} and
\eqref{eq:general-regular-gamma3}, it is then possible to obtain explicit expressions for the constrained Lagrangian coefficients. Specifically, in the generic case where $\sigma \neq 1$ and $\sigma \neq -\,2$, they read\footnote{Note that the value $\sigma=1$ corresponds to a separate algebraic parametrisation and does not lead to a new form of the solution. In contrast, for $\sigma=-\,2$, the field equations
imply $h_{10}+h_{11}=0$, which eliminates the backreaction contribution of torsion.}:
\begin{align}
m_t^2&=m_S^2=h_3=h_4=h_{25}=0\,,
\quad
h_{11}=-\,\frac14\left(4h_{10}+9c_1\right),
\label{eq:general-regular-complete-couplingsA}
\\
h_{12}&=\frac{h_{10}}4+\frac{3}{16}c_1\,,
\quad
h_{23}=2h_{24}\,,
\quad
h_9=-\,h_{10}-3c_1\,,
\\
h_8&=\frac{3(\sigma+2)}{\sigma-1}\left(2h_{18}-h_{19}\right)
+\frac{3(19\sigma+89)}{8(\sigma-1)}c_1-h_{10}\,,
\\
h_6&=4\left(h_{21}-h_{22}+h_{24}-h_{26}\right)
+\frac{2(\sigma+2)}{\sigma-1}\left(2h_{18}-h_{19}\right)
+\frac{9(5\sigma+7)}{4(\sigma-1)}c_1\,,
\\
h_{15}&=\frac{3(5\sigma-17)}{8(\sigma+2)}c_1
-\frac{5}{12}\left(2h_{18}-h_{19}\right),
\quad
h_{16}=h_{17}
+\frac12\left(2h_{18}-h_{19}\right)
+\frac{27(\sigma-1)}{8(\sigma+2)}c_1\,,
\\
h_5&=\frac{h_7}{2}+\frac{h_{10}}4
-2\left(h_{21}-h_{22}+h_{24}-h_{26}\right)
-\frac{\sigma+2}{4(\sigma-1)}\left(2h_{18}-h_{19}\right)
-\frac{3(25\sigma+11)}{32(\sigma-1)}c_1\,,
\\
h_{13}&=
\frac{\sigma-1}{18(\sigma+2)}
\left(h_{21}-h_{22}+h_{24}-h_{26}\right)
+\frac{7}{144}\left(2h_{18}-h_{19}\right)
+\frac{25-\sigma}{16(\sigma+2)}c_1\,,
\\
h_{14}&=
\frac{\sigma-1}{9(\sigma+2)}
\left(h_{21}-h_{22}+h_{24}-h_{26}\right)
-\frac{11}{72}\left(2h_{18}-h_{19}\right)
+\frac{7\sigma-13}{4(\sigma+2)}c_1\,,
\\
h_{20}&=-\frac{h_{24}}2
-\frac73\left(h_{21}-h_{22}+h_{24}-h_{26}\right)
-\frac{\sigma+2}{6(\sigma-1)}\left(2h_{18}-h_{19}\right)
-\frac{3(13\sigma-1)}{16(\sigma-1)}c_1\,,
\label{eq:general-regular-complete-couplings}
\end{align}
which provides the following nonvanishing parameters $\gamma_{i}$:
\begin{align}
    \gamma_1={}&
 \frac{243}{\left(\sigma-1\right)\left(\sigma+2\right)}
 \left[
 8\left(\sigma+2\right)\left(2h_{18}-h_{19}\right)
 +9\left(\sigma+11\right)c_1
 \right],\\
 \gamma_2={}&
 \frac{243}{\left(\sigma-1\right)\left(\sigma+2\right)}
 \left[
 8\left(\sigma+2\right)\left(2h_{18}-h_{19}\right)
 +9\left(7\sigma+5\right)c_1
 \right],\\
 \gamma_4={}&
 \frac{432}{\left(\sigma-1\right)\left(\sigma+2\right)}
 \left[
 \left(\sigma+2\right)\left(2h_{18}-h_{19}\right)
 -\left(\sigma-1\right)\left(h_{21}-h_{22}+h_{24}-h_{26}\right)
 +\frac92\left(\sigma+2\right)c_1
 \right],\\
 \gamma_3={}&2\gamma_4\,,
 \quad
 \gamma_6=\sigma\gamma_1\,,
 \quad
 \gamma_7=\sigma\gamma_2\,,
 \quad
 \gamma_8=2\sigma\gamma_4\,,
 \quad
 \gamma_9=\sigma\gamma_4\,.
 \label{eq:general-regular-gammas}
\end{align}
A convenient parametrisation of this family is then given by taking
\begin{equation}
 c_1,\ h_1,\ h_2,\ h_7,\ h_{10},\ h_{17},\ h_{18},\ h_{19},\
 h_{21},\ h_{22},\ h_{24},\ h_{26},\ m_T^2\,,
 \label{eq:general-regular-free-couplings}
\end{equation}
together with $\sigma$.  Once these quantities are specified, all the remaining Lagrangian coefficients are fixed by Eqs.~\eqref{eq:general-regular-complete-couplingsA}-\eqref{eq:general-regular-complete-couplings}.

Thereby, the only field equation describing the torsion function $U(r)$ within this branch is reduced to
\begin{equation}
 \gamma_1U^2(r)+\gamma_2rU'(r)U(r)
 +2\gamma_4\kappa_{2}rU(r)+\gamma_4\kappa_{2}r^2U'(r)=0\,.
 \label{eq:general-regular-master}
\end{equation}
In order to solve Eq.~\eqref{eq:general-regular-master}, we introduce the auxiliary variable
\begin{equation}
 x(r)=-\,\frac{\gamma_1+\gamma_2}{3\gamma_4}
 \frac{U(r)}{\kappa_{2}r}\,,
 \label{eq:general-regular-x}
\end{equation}
which transforms the expression into a separable first-order differential equation, whose integration yields the implicit solution
\begin{equation}
 x(r)[1-x(r)]^{\frac{2\gamma_2-\gamma_1}{\gamma_1+\gamma_2}}
 =\left(\frac{\kappa_{3}}{r}\right)^3\,,
 \label{eq:general-regular-ximplicit}
\end{equation}
where $\kappa_{3}$ is an integration constant. In particular, for $\kappa_{3} > 0$, the resulting solution takes the following asymptotic form at large distances:
\begin{equation}
 x(r)\approx\left(\frac{\kappa_{3}}{r}\right)^3\,,
 \quad
 U(r)\approx
 -\,\frac{3\gamma_4}{\gamma_1+\gamma_2}
 \kappa_{2}\kappa_{3}^3r^{-2}\,,
 \label{eq:general-regular-infinity-U}
\end{equation}
while exhibiting the following behaviour near the centre:
\begin{equation}
 x(r)\approx
 1-\left(\frac{\kappa_{3}}{r}\right)^{
 -\,\frac{3(\gamma_1+\gamma_2)}{\gamma_1-2\gamma_2}}\,,
 \quad
 U(r)\approx
 -\,\frac{3\gamma_4}{\gamma_1+\gamma_2}\kappa_{2} r\,.
 \label{eq:general-regular-centre-U}
\end{equation}
Thus, the function $x(r)$ decays at spatial infinity and acquires a regular value at the origin
\begin{equation}
    x(0) = 1\,,
\end{equation}
provided
\begin{equation}
    \frac{\gamma_1+\gamma_2}{\gamma_1-2\gamma_2}>0\,,
 \quad
 \gamma_1+\gamma_2\neq0\,.
 \label{eq:general-regular-condition}
\end{equation}

Indeed, this configuration gives rise to a regular black hole solution endowed with a primary torsion hair, as shown below.

\subsection{Metric, regularity and primary hair}
\label{subsec:general-regular-metric}

With the torsion function $U(r)$ determined by the implicit solution~\eqref{eq:general-regular-ximplicit}, the metric function reads
\begin{equation}
 \Psi(r)=1
 +\frac{27c_1\gamma_4\kappa_{2}^2}{\gamma_1+\gamma_2}r^2x(r)
 -\frac{243c_1\gamma_4^2\kappa_{2}^2}
 {2\left(\gamma_1+\gamma_2\right)^2}r^2x^{2}(r)\,.
 \label{eq:general-regular-metric-couplings}
\end{equation}
Taking into account the asymptotic behaviour at spatial infinity given by Expression~\eqref{eq:general-regular-infinity-U}, the ADM mass is
\begin{equation}
 m_{\rm ADM}=-\,\frac{27c_1\gamma_4}{2\left(\gamma_1+\gamma_2\right)}
 \kappa_{2}^2\kappa_{3}^3\,,
 \label{eq:general-regular-mass}
\end{equation}
which re-expresses the metric function as
\begin{equation}
 \Psi(r)=1-\frac{2m_{\rm ADM}}{\kappa_{3}^3}r^2x(r)
 +\frac{9\gamma_4}{\gamma_1+\gamma_2}
 \frac{m_{\rm ADM}}{\kappa_{3}^3}r^2x^{2}(r)\,,
 \label{eq:general-regular-metric}
\end{equation}
Hence, for $\kappa_{3}>0$ and $\kappa_{2}\neq0$, a positive ADM mass requires
\begin{equation}
 \frac{c_1\gamma_4}{\gamma_1+\gamma_2}<0\,.
 \label{eq:positive-mass-condition}
\end{equation}

Note that the asymptotic role of the primary torsion hair $\kappa_3$ can be made explicit by
expanding the implicit relation for the auxiliary function $x(r)$. At large radius,
\begin{equation}
x(r)\approx\left(\frac{\kappa_3}{r}\right)^3
+\frac{2\gamma_2-\gamma_1}{\gamma_1+\gamma_2}
\left(\frac{\kappa_3}{r}\right)^6\,,
\label{eq:regular-x-asymptotic-next}
\end{equation}
yielding
\begin{equation}
\Psi(r)\approx1-\frac{2m_{\rm ADM}}{r}
+
\frac{\left(2\gamma_1-4\gamma_2+9\gamma_4\right)m_{\rm ADM}\kappa_3^3}
{\left(\gamma_1+\gamma_2\right)r^4}\,.
\label{eq:regular-metric-asymptotic-next}
\end{equation}
Thus, $\kappa_3$ does not renormalise the ADM term and its first metric
correction generically appears at order $r^{-4}$. On the other hand, the axial mode of torsion does not decay at large distances, but instead grows linearly in the asymptotic region
\begin{equation}
S_\mu=2\kappa_2r
\left(1,-\,\frac{1}{\Psi(r)},0,0\right).
\label{eq:regular-axial-asymptotic}
\end{equation}
At the level of
the local exact solution, $m_{\rm ADM}$ and $\kappa_3$ can be varied
independently by adjusting $\kappa_2$ through
Eq.~\eqref{eq:general-regular-mass}. Whether $\kappa_3$ defines an
independent canonical hair within a fixed asymptotic phase space depends
on the admissible RC boundary conditions and on the associated conserved
charges.

Furthermore, Expressions~\eqref{eq:general-regular-centre-U} and~\eqref{eq:general-regular-condition} set the following regular form in the central region:
\begin{equation}
 \Psi(r) \approx 1+\frac{m_{\rm ADM}}{\kappa_{3}^3}
 \left(\frac{9\gamma_4}{\gamma_1+\gamma_2}-2\right)r^2+\frac{2m_{\rm ADM}}{\kappa_{3}^{3(2\gamma_1-\gamma_2)/(\gamma_1-2\gamma_2)}}\left(1-\frac{9\gamma_4}{\gamma_1+\gamma_2} \right)r^{(5\gamma_1-\gamma_2)/(\gamma_1-2\gamma_2)}\,.
 \label{eq:general-regular-centre-metric}
\end{equation}
Hence, for a positive mass, the geometry presents a de Sitter core if $9\gamma_4/(\gamma_1+\gamma_2)<2$, an anti-de Sitter core if $9\gamma_4/(\gamma_1+\gamma_2)>2$ and approaches the Minkowski space-time if $9\gamma_4/(\gamma_1+\gamma_2)=2$, with completely regular curvature and torsion tensors in all these cases. As $m_{\rm ADM}/\kappa_{3}$ is increased, the solution passes from a horizonless regular geometry through an extremal configuration to a black hole with an inner and an outer horizon. The corresponding horizon structure is analysed in Appendix~\ref{app:horizon-structure}, where no additional pairs of horizons were found over the range of parameters considered. Likewise, it is useful to distinguish finite curvature from smooth
extendibility through the centre. Defining the parameter
\begin{equation}
p\equiv\frac{3\left(\gamma_1+\gamma_2\right)}{\gamma_1-2\gamma_2}>0\,,
\label{eq:regular-central-exponent}
\end{equation}
the leading corrections to the metric function $\Psi(r)$ near the centre scale as
$r^2$ and $r^{2+p}$. Thus, the condition $p>0$
ensures that the curvature remains finite at the centre. However, finite
curvature alone does not guarantee that the metric is smooth when extended
through $r=0$; this also depends on the powers of $r$ appearing in the
near-centre expansion. In the algebraic sequence of
Sec.~\ref{subsec:general-regular-cases} one has $p=3n$, so the even-$n$
members, and in particular the explicit $n=2$ branch, possess an even-power
central expansion. Bounded curvature alone does not establish a unique
extension or causal geodesic completeness
~\cite{Carballo-Rubio:2019fnb,Zhou:2022yio}.

Given the regular behaviour of the solutions, it is then worthwhile to assess possible violations of the conditions of the singularity theorems of PG theory~\cite{Cembranos:2016xqx}. Specifically, considering null and timelike vectors
\begin{equation}
    n^\mu=\left(\frac{1}{\sqrt{2\Psi(r)}},0,
              \frac{1}{\sqrt{2}\,r},0\right), \quad t^\mu=\left(\frac{1}{\sqrt{\Psi(r)}},0,0,0\right),
\end{equation}
the corresponding null and timelike convergence conditions can be directly evaluated as
\begin{align}
    R_{\mu\nu}n^{\mu}n^{\nu}&=\frac{1}{4r^{2}}\left[r^{2}\Psi''(r)+2\left(1-\Psi(r)\right)\right] \equiv \mathcal{N}(r)\,,\label{null_cond}\\
    R_{\mu\nu}t^{\mu}t^{\nu}&=\frac{1}{2r}\left(r\Psi''(r)+2\Psi'(r)\right) \equiv \mathcal{T}(r)\,,\label{timelike_cond}
\end{align}
which, near the central region, provides the approximate expressions
\begin{align}
    \mathcal{N}(r) &\approx \frac{9\left(\gamma_1+\gamma_2\right)\left(2\gamma_1-\gamma_2\right)m_{\rm ADM}}{2\left(\gamma_1-2\gamma_2\right)^2\kappa_{3}^{3(2\gamma_1-\gamma_2)/(\gamma_1-2\gamma_2)}} \left(1-\frac{9\gamma_4}{\gamma_1+\gamma_2}\right) r^{3(\gamma_1+\gamma_2)/(\gamma_1-2\gamma_2)}\,,\\
    \mathcal{T}(r) &\approx \frac{3m_{\rm ADM}}{\kappa_{3}^3}\left(\frac{9\gamma_4}{\gamma_1+\gamma_2}-2\right) + \frac{3\left(5\gamma_1-\gamma_2\right)\left(2\gamma_1-\gamma_2\right)m_{\rm ADM}}{\left(\gamma_1-2\gamma_2\right)^2\kappa_{3}^{3(2\gamma_1-\gamma_2)/(\gamma_1-2\gamma_2)}} \left(1-\frac{9\gamma_4}{\gamma_1+\gamma_2}\right) r^{3(\gamma_1+\gamma_2)/(\gamma_1-2\gamma_2)}\,.
\end{align}
Consequently, it is clear that the present regular black hole solutions violate at least one of the causal convergence conditions of the singularity theorems in a neighbourhood of the centre, as outlined in Table~\ref{tab:NTcond}.

\begin{table}[H]
\centering
\renewcommand{\arraystretch}{1.2}
\setlength{\tabcolsep}{8pt}
\setlength{\arrayrulewidth}{0.5pt}
\begin{tabular}{|c|c|c|c|}
\hline
\textbf{Core}
&
\textbf{Parameter range}
&
\textbf{Central behaviour}
&
\textbf{Failed condition}
\\
\hline\hline
de Sitter
&
$9\gamma_4/(\gamma_1+\gamma_2)<2$
&
$\mathcal{T}(r)<0$
&
Timelike convergence
\\
\hline
Minkowski
&
$9\gamma_4/(\gamma_1+\gamma_2)=2$
&
$\mathcal{N}(r)<0$\,,
$\mathcal{T}(r)<0$
&
Null and timelike convergence
\\
\hline
anti-de Sitter
&
$9\gamma_4/(\gamma_1+\gamma_2)>2$
&
$\mathcal{N}(r)<0$
&
Null convergence
\\
\hline
\end{tabular}
\caption{Behaviour of the null and timelike convergence conditions in the central region of the regular black hole solutions.}
\label{tab:NTcond}
\end{table}

Overall, the present solutions represent regular black holes with a metric structure determined by the function~\eqref{eq:general-regular-metric}, therefore characterised by an ADM mass $m_{\rm ADM}$ and a primary torsion hair $\kappa_{3}$. It is worthwhile to stress that the simultaneous occurrence of central regularity and continuous primary hair remains unusual. In representative scalar-tensor formulations, black holes with primary scalar hair are generically singular, while removal of the central singularity requires a relation between the scalar parameter and the ADM mass~\cite{Bakopoulos:2023sdm,Karakasis:2023hni,Myung:2025afs}. Similarly, regularity can be achieved in nonlinear realisations of vector-tensor theories, but it commonly constrains the independence of the vector hair~\cite{Ayon-Beato:2000mjt,Bronnikov:2000vy}. A remarkable exception is given in Ref.~\cite{Eichhorn:2025pgy}, by a Hayward-type black hole retaining a primary vector hair, for any value of the mass and the coupling constants of the theory. Therefore, our solutions provide a further example of regular black hole configurations supporting an independent primary hair, in this case associated with the torsion field.

\subsection{Explicit algebraic subfamilies}
\label{subsec:general-regular-cases}

The implicit solution~\eqref{eq:general-regular-ximplicit} admits several distinguished exact branches, governed by specific constraints on the parameter space of the theory. For instance, a rational branch is realised by imposing the vanishing constraint $\gamma_2=0$, which reduces the metric and auxiliary functions to rational expressions
\begin{equation}
 x(r)=\frac{\kappa_{3}^3}{r^3+\kappa_{3}^3}\,,
 \quad
 \Psi(r)=1-\frac{2m_{\rm ADM}r^2}{r^3+\kappa_{3}^3}
 +\frac{9\gamma_4}{\gamma_1}
 \frac{m_{\rm ADM}\kappa_{3}^3r^2}{\left(r^3+\kappa_{3}^3\right)^2}\,.
 \label{eq:general-regular-rational}
\end{equation}
Within this branch, imposing the specific relation $\gamma_1=-\,3\gamma_4$, yields particularly simple expressions
\begin{equation}
 U(r)=\frac{\kappa_{2}\kappa_{3}^3r}{r^3+\kappa_{3}^3}\,,
 \quad
 \Psi(r)=1-
 \frac{m_{\rm ADM}r^2\left(2r^3+5\kappa_{3}^3\right)}
 {\left(r^3+\kappa_{3}^3\right)^2}\,.
 \label{eq:general-regular-known}
\end{equation}
Alternatively, a ratio $9\gamma_4/\gamma_1=2$ gives rise to a metric structure with a Minkowski-core limit
\begin{equation}
 \Psi(r)=1-\frac{2m_{\rm ADM}r^5}{\left(r^3+\kappa_{3}^3\right)^2}\,.
 \label{eq:general-regular-rational-Minkowski}
\end{equation}

A broader algebraic sequence follows by requiring
\begin{equation}
 \left(n-1\right)\gamma_1=\left(2n+1\right)\gamma_2\,,
 \quad n=1,2,3,\ldots
 \label{eq:general-regular-ncondition}
\end{equation}
which determines the auxiliary function algebraically by
\begin{equation}
 \left(\frac{r^3}{\kappa_{3}^3}\right)^n x_n^{n}(r)+x_n(r)-1=0\,.
 \label{eq:general-regular-npolynomial}
\end{equation}
Specifically, the linear case $n=1$ describes the previous rational solution~\eqref{eq:general-regular-rational}. For the quadratic relation given by $n=2$, we have $\gamma_1=5\gamma_2$ and the auxiliary function takes the form
\begin{equation}
 x_2(r)=\frac{2\kappa_{3}^3}{\kappa_{3}^3+\sqrt{\kappa_{3}^6+4r^6}}\,.
 \label{eq:general-regular-n2}
\end{equation}
Thus, it approaches the central regular region as
$x_2 (r) \approx 1-r^6/\kappa_{3}^6$, more rapidly than the previous case. Because $x_2(r)$ depends only on $r^6$, this member has an even-power
expansion at the centre and provides a particularly clean representative
for a future analysis of smooth extendibility and geodesic completeness. On the other hand, the cubic relation provided by $n=3$ can be expressed via Cardano's formula, whereas higher values of the parameter $n$ provide the corresponding algebraic extensions. A particularly simple subfamily is obtained when the de Sitter factor in the central region vanishes
\begin{equation}
 \frac{9\gamma_4}{\gamma_1+\gamma_2}=2\,,
 \label{eq:general-regular-Minkowski-condition}
\end{equation}
which yields a metric function that takes Minkowski values at the origin
\begin{equation}
 \Psi_n(r)=1-\frac{2m_{\rm ADM}}{\kappa_{3}^3}r^2x_n(r)\left[1-x_n(r)\right].
 \label{eq:general-regular-Minkowski-family}
\end{equation}
In particular, for $n=2$, the metric function near the centre takes the form
\begin{equation}
 \Psi_2(r)\approx 1-\frac{2m_{\rm ADM}}{\kappa_{3}^9}r^8\,.
 \label{eq:general-regular-n2-Minkowski}
\end{equation}

Overall, these configurations represent explicit examples of the general family of solutions defined by Expression~\eqref{eq:general-regular-ximplicit}.

\section{General vector-axial-tensor sector: regular Hayward-like black holes with secondary torsion hair}
\label{general-regularAll}

Finally, let us study the case in which all modes of torsion are present. Throughout this section, each dimensionful quantity is replaced by its numerical value in Planck units, and we use the same symbol for the resulting dimensionless variable. We set $\kappa_1=0$ and restrict the analysis to the normalised subbranch $\kappa_2=1$.
In general, the simultaneous presence of the vector, axial and tensor modes leads to a highly nonlinear system of equations, but a tractable system can be isolated for the torsion functions
\begin{equation}
    P(r)=Q(r)=r\,, \quad Y(r)=r^3\,, \label{eq:regularAll-support}
\end{equation}
while focusing on completely regular configurations sets again $\mu=0$ in the metric function~\eqref{eq:general_metric_integrated}.

Then, the four independent connection equations and the remaining tetrad equation provide an overdetermined system of differential equations for the function $U(r)$. In order to isolate a tractable branch, we require the five equations to share a single differential equation for arbitrary values of $U(r)$. Performing the corresponding coefficient matching directly in terms of the original Lagrangian coefficients, and using $\alpha_4$ as a convenient parameter, gives the following coupling subspace:
\begin{align}\label{eq:regularAll-h-couplings_1st}
h_{1} &=
-\,\frac{
3645+18009\alpha_{4}+198\alpha_{4}^{2}+28\alpha_{4}^{3}
}{
162\left(9+17\alpha_{4}\right)
}\,,
\\
h_{2} &=
\frac{
3645+18063\alpha_{4}+300\alpha_{4}^{2}+28\alpha_{4}^{3}
}{
324\left(9+17\alpha_{4}\right)
}\,,
\\
h_{3} &=
\frac{
1215+2025\alpha_{4}+36\alpha_{4}^{2}-32\alpha_{4}^{3}
}{
216\left(9+17\alpha_{4}\right)
}\,,
\\
h_{4} &=
-\,\frac{
1215+2043\alpha_{4}+70\alpha_{4}^{2}-32\alpha_{4}^{3}
}{
432\left(9+17\alpha_{4}\right)
}\,,
\\
h_{5} &= -\,\frac{\alpha_4}{27}-2c_1+h_{12}+\frac{h_7}{2}\,,
\\
h_{6} &= 4c_{1}+\frac{2\alpha_{4}}{27}\,,
\\
h_{8} &= -\,4h_{12}\,,
\quad
h_{9}=-\,4h_{12}\,,
\quad
h_{10}=4h_{12}\,,
\quad
h_{11}=-\,4h_{12}\,,
\\
h_{13}
&=
\frac{1}{5832\left(9+17\alpha_4\right)^2}
\bigl[
196830
-1944m_S^2\left(9+17\alpha_4\right)^2+81c_1\left(9+17\alpha_4\right)
\left(
1431+2469\alpha_4+104\alpha_4^2-32\alpha_4^3
\right)
\nonumber\\
&\hspace{1.3cm}
+3572829\alpha_4
+5498928\alpha_4^2
+319743\alpha_4^3
+24818\alpha_4^4
+4276\alpha_4^5
-1664\alpha_4^6
\bigr]\,,
\\
h_{14} &=
\frac{
\alpha_{4}
\left(
3645+8397\alpha_{4}
+90\alpha_{4}^{2}
+92\alpha_{4}^{3}
\right)
}{
2916\left(9+17\alpha_{4}\right)
}\,,
\\
h_{15} &=
\frac{
\alpha_{4}
\left(
81-1062\alpha_{4}
-39\alpha_{4}^{2}
+8\alpha_{4}^{3}
\right)
}{
243\left(9+17\alpha_{4}\right)
}\,,
\\
h_{16} &=
-\,\frac{
22\alpha_{4}\left(81+\alpha_{4}^{2}\right)
}{
81\left(9+17\alpha_{4}\right)
}\,,
\\
h_{17} &=
\frac{
\alpha_{4}
\left(
-1782+9\alpha_{4}-5\alpha_{4}^{2}
\right)
}{
81\left(9+17\alpha_{4}\right)
}\,,
\\
h_{18} &=
\frac{
\alpha_{4}
\left(
297-3\alpha_{4}-2\alpha_{4}^{2}
\right)
}{
27\left(9+17\alpha_{4}\right)
}\,,
\\
h_{19} &=
\frac{
2\alpha_{4}
\left(297-3\alpha_{4}-2\alpha_{4}^{2}
\right)
}{
27\left(9+17\alpha_{4}\right)
}\,,
\\
h_{20} &=
-\,\frac{h_{24}}{2}
+
\frac{
\alpha_{4}
\left(
486+621\alpha_{4}
+3\alpha_{4}^{2}
+2\alpha_{4}^{3}
\right)
}{
486\left(9+17\alpha_{4}\right)
}\,,
\\
h_{21} &=
h_{22}-h_{24}+h_{26}
+\frac{\alpha_{4}}{12}\,,
\\
h_{23} &=
2h_{24}-\frac{\alpha_{4}}{9}\,,
\\
h_{25} &= 0\,,
\\
m_{t}^{2} &=
\frac{
\alpha_{4}
\left(9\alpha_{4}-16\alpha_{4}^{2}-2673
\right)
}{
81\left(9+17\alpha_{4}\right)
}\,,
\\
m_T^2
&=
4m_S^2
-\frac{1}{729\left(9+17\alpha_4\right)^2}
\bigl[
590490
+9043245\alpha_4
+27677214\alpha_4^2
+1424223\alpha_4^3
\nonumber\\
&\hspace{1.3cm}
+478608\alpha_4^4
+18564\alpha_4^5
-1768\alpha_4^6
+81c_1\left(9+17\alpha_4\right)
\left(
3645+12150\alpha_4+357\alpha_4^2-34\alpha_4^3
\right)
\bigr]\,.\label{eq:regularAll-h-couplings}
\end{align}
 
Thus, after substituting the above relations into the definition of $\alpha_4$, the latter reduces identically to $\alpha_4=\alpha_4$ and therefore does not impose any additional constraint. A convenient set of independent free parameters is
\begin{equation}
 \alpha_4,\quad
 c_1,\quad
 h_7,\quad
 h_{12},\quad
 h_{22},\quad
 h_{24},\quad
 h_{26},\quad
 m_S^2 .
 \label{eq:regularAll-free-couplings}
\end{equation}
In particular, none of the black-hole integration constants enters the Lagrangian coefficients.

On this coupling subspace, the combinations entering the metric function and the reduced field equations simplify to
\begin{align}
 \alpha_1
 &=
 -\frac{
 12\alpha_4^3
 +70\alpha_4^2
 +5607\alpha_4
 +1215
 }{
 3\left(17\alpha_4+9\right)
 }\,,\quad
 \alpha_2
 =
 -\frac{
 2\alpha_4
 \left(
 2\alpha_4^2+3\alpha_4-297
 \right)
 }{
 3\left(17\alpha_4+9\right)
 }\,,
 \label{eq:regularAll-beta2A}
\\
 \beta_2
 &=
 -\frac{
 32\alpha_4^3
 +32\alpha_4^2
 -1989\alpha_4
 -1215
 }{
 6\left(17\alpha_4+9\right)
 }\,, \quad \alpha_3=\beta_1=0\,.
 \label{eq:regularAll-beta2B}
\end{align}
The nondegenerate branch considered here requires
\begin{equation}
 \alpha_4\neq0\,,
 \quad
 17\alpha_4+9\neq0\,.
 \label{eq:regularAll-nondegenerate}
\end{equation}

With these relations imposed, all algebraically independent radial structures cancel and the complete reduced system factorises as
\begin{equation}
 \bigl(
 \mathcal C^{(1)}(r),
 \mathcal C^{(2)}(r),
 \mathcal C^{(3)}(r),
 \mathcal C^{(4)}(r),
 \mathcal E(r)
 \bigr)
 =
 -\,\frac{4}{3}\alpha_4^2\,\mathcal G_{H}(r)
 \bigl(
 r^4,-\,2r^4,-\,2r^3,
 r^3,-\,72r^9U(r)
 \bigr)\,,
 \label{eq:regularAll-factorisation}
\end{equation}
with
\begin{equation}
 \mathcal G_{H}(r)
 =
 r\left(r^2U'(r)+2rU(r)-3U^{2}(r)\right)'-2\left(r^2U'(r)+2rU(r)-3U^{2}(r)\right).
 \label{eq:regularAll-GH}
\end{equation}
Therefore, since $\alpha_4\neq0$, the five reduced field equations collapse to the second-order differential equation $\mathcal G_H(r)=0$. Integrating once gives
\begin{equation}
 r^2U'(r)+2rU(r)-3U^{2}(r)
 =
 \kappa_{3} r^2,
 \label{eq:regularAll-U-first}
\end{equation}
where $\kappa_{3}$ is an integration constant. The remaining differential equation can be integrated exactly. For
$\kappa_{3}<3/4$, by defining
\begin{equation}
 \nu\equiv\sqrt{1-\frac{4\kappa_{3}}{3}}\,,
\end{equation}
the corresponding solution can be written as
\begin{equation}
 U(r)
 =
 r\left[
 \frac{1-\nu}{2}
 +
 \frac{\nu}{1+\left(r/\kappa_{4}\right)^{3\nu}}
 \right],
 \quad
 \kappa_{4}>0\,,
 \label{eq:regularAll-U}
\end{equation}
where $\kappa_{4}$ is an integration constant. Substitution into the metric function~\eqref{eq:general_metric_integrated} gives
\begin{equation}
 \Psi(r)
 =
 1-\frac{\Lambda_\infty}{3}r^2
 +
 \frac{2\alpha_4\nu}{9}
 \frac{\kappa_{4}^{3\nu}r^2}
 {r^{3\nu}+\kappa_{4}^{3\nu}}\,,
 \label{eq:regularAll-Psi}
\end{equation}
where
\begin{equation}
 \Lambda_\infty
 =
 -\,\frac{9}{2}c_1
 -
 \frac{
 \alpha_4\left(52\alpha_4^2+114\alpha_4+2430\right)
 }{
 18\left(17\alpha_4+9\right)
 }
 -
 \frac{\alpha_4}{3}\left(1-\nu\right)\,.
 \label{eq:regularAll-Lambda}
\end{equation}

The branch \eqref{eq:regularAll-Psi} is regular at the centre. Indeed,
for $r\rightarrow0$, one has
\begin{equation}
U(r)\approx\frac{1+\nu}{2}r\,,
\quad
\Psi(r)\approx1-\frac{\Lambda_0}{3}r^2\,,
\end{equation}
with
\begin{equation}
 \Lambda_0
 =
 \Lambda_\infty-\frac{2\alpha_4\nu}{3}\,.
\end{equation}
Therefore, the torsion and metric functions remain finite at $r=0$, while the usual Schwarzschild $1/r$ singularity is absent. The centre is consequently a regular constant-curvature core, which is de Sitter, Minkowski or anti-de Sitter, according to the sign of $\Lambda_0$. Likewise, near the centre, the torsion functions $P(r)$, $Q(r)$ and $U(r)$ vanish linearly, whereas $Y(r)$ vanishes cubically, so that the orthonormal torsion and contortion components remain finite and vanish at the origin. Together with $\Psi(r)\approx1-\Lambda_0 r^2/3$, this ensures that the derivative and quadratic contortion contributions to the RC curvature remain finite. Hence, the complete RC geometry is regular at the centre. 

In the particular case, where the constant-curvature contribution is eliminated by imposing
\begin{equation}
 \Lambda_\infty=0\,,
\end{equation}
the remaining torsion correction provides the asymptotic form
\begin{equation}
 \Psi(r)-1
\approx
\frac{2\alpha_4\nu\kappa_4^{3\nu}}{9}r^{2-3\nu}\,.
\end{equation}
Thereby, the metric tensor approaches the Minkowski space-time at large distances, provided that $\nu>2/3$. The standard Schwarzschild $1/r$ fall-off is recovered for $\nu=1$, or equivalently $\kappa_{3}=0$. In this case,
\begin{equation}
 U(r)=\frac{\kappa_{4}^3r}{r^3+\kappa_{4}^3}\,,
 \quad
 \Psi(r)
 =
 1+
 \frac{2\alpha_4\kappa_{4}^3r^2}
 {9(r^3+\kappa_{4}^3)}\,.
\end{equation}
Introducing the mass parameter as
\begin{equation}
 m=-\,\frac{\alpha_4\kappa_{4}^3}{9}\,,\label{HaywardConst}
\end{equation}
the metric function takes the Hayward-like form
\begin{equation}
 \Psi(r)
 =
 1-\frac{2mr^2}{r^3+\kappa_{4}^3}\,.\label{HaywardSol}
\end{equation}

Since the parameter $\alpha_4$ acquires a particular value once the gravitational theory is specified, Expression~\eqref{HaywardConst} fixes the integration constant $\kappa_{4}$ in terms of the mass. Therefore, the present Hayward-like solution describes a regular black hole characterised by a secondary torsion hair. In fact, the rigidity of this result allows a simpler analysis of the causal convergence conditions~\eqref{null_cond}-\eqref{timelike_cond}, which near the centre take the form
\begin{equation}
    \mathcal{N}(r) \approx \frac{9m}{\kappa_{4}^6} r^3\,, \quad \mathcal{T}(r) \approx -\,\frac{6m}{\kappa_{4}^3}\,.
\end{equation}
Hence, while the null convergence condition holds in this case, the timelike convergence condition turns out to be violated.

\section{Conclusions}\label{sec:conclusions}

In the present work, we have investigated extensions of the Holst quadratic model of PG theory by including into the gravitational action cubic order invariants defined from mixing terms of the curvature and torsion tensors. The presence of these cubic invariants significantly enriches the vacuum structure of the theory, which is no longer constrained by the Birkhoff theorem of GR, allowing for new exact static and spherically symmetric black hole solutions with dynamical torsion.

By imposing the set of conditions~\eqref{reg1}-\eqref{reg2} to avoid divergences in the torsion tensor at the roots of the metric functions, as well as the reciprocal case~\eqref{reciprocal_metric}, we are able to systematically solve the field equations and find exact black hole solutions across distinct torsion sectors and models within the present Holst-type class.

The exact solutions obtained in the different torsion sectors are
summarised in Tables~\ref{tab:BH-singular-summary} and~\ref{tab:BH-summary}. Depending on the specific irreducible modes and parity components of the torsion field involved in the analysis, we have identified several distinct classes of modified geometries. Specifically, in the odd-parity tensor sector, the backreaction effects of the tensor mode generate Kiselev-like black holes endowed with a primary tensor hair. By expanding the analysis to the even-parity vector-tensor sector, we have determined different models that admit exact black hole solutions with algebraic and Lambert $W$ metric corrections. Furthermore, the exploration of the combined even and odd-parity axial-tensor sector reveals a richer space-time phenomenology. By systematically solving the differential and algebraic branches of the field equations for different choices of the Lagrangian coefficients, we have obtained black hole configurations featuring additional Lambert $W$ corrections as well as Boulware-Deser-like geometries. A particularly distinctive feature of the vector-tensor Lambert $W$
branch is the existence of a finite range of values of the black hole parameters admitting three event
horizons, including a critical configuration in which the three horizons coalesce.

Crucially, our investigation of the final two sectors demonstrates that the presence of dynamical torsion can fundamentally resolve the black hole singularities within this class of models. By focusing on the purely odd-parity axial-tensor sector and on the broader general vector-axial-tensor sector, we have derived exact regular black hole solutions, which manifestly violate the convergence conditions of the singularity theorems of PG theory~\cite{Cembranos:2016xqx}. Upon fixing the Lagrangian coefficients to the values that admit the solutions, in the purely odd-parity axial-tensor sector the general form of the metric function is characterised by a mass parameter and an independent primary torsion hair, as shown in Expression~\eqref{eq:general-regular-metric}. By contrast, the solution found in the general vector-axial-tensor sector exhibits a Hayward-like geometry given by the  metric function~\eqref{HaywardSol}, with the torsion parameter constrained by Expression~\eqref{HaywardConst}. Hence, in this case no additional
independent parameter remains, but the torsion hair is secondary. 

The horizon structure presented in Appendix~\ref{app:horizon-structure} further confirms that the new geometries found in our work contain genuine black hole regions. Most notably, the Lambert $W$ branch of the even-parity vector-tensor sector is characterised by a three-horizon structure, with two inner horizons in addition to the outer event horizon and a critical configuration in which all three horizons coalesce. On the other hand, the Lambert $W$ branch arising in the even and odd-parity axial-tensor sector, as well as the regular black hole solutions with primary torsion hair, display at most two horizons in the parameter ranges explored. A complementary question concerns the maximal extension of the regular solutions and whether the resulting space-times are
causally geodesically complete, since a finite curvature in the central region does
not by itself establish these global properties~\cite{Carballo-Rubio:2019fnb,Zhou:2022yio}.

In summary, our results highlight that extending the Holst quadratic model with cubic order invariants not only diversifies the phenomenological landscape of exact vacuum geometries, but also provides a promising theoretical pathway to singularity resolution without the need for exotic matter sources. Several directions for future work include performing a thorough stability analysis of the vector, axial and tensor modes of torsion for this class of models, in order to identify the theoretically most viable subclasses and further constrain the allowed values of the Lagrangian coefficients. Once the perturbative stability of the relevant branches is established,
the computation of their quasinormal-mode spectrum would provide a natural
next step towards characterising their ringdown phenomenology and possible
observational signatures~\cite{Kokkotas:1999bd}. In addition, the thermodynamic properties of these modified geometries warrant careful investigation. Because the presence of dynamical torsion and cubic order invariants in the gravitational action generically alters the conserved charges and their conjugate potentials, establishing a consistent first law of black hole thermodynamics requires a dedicated analysis within the Hamiltonian formalism of PG theory~\cite{Blagojevic:2019gsd,Blagojevic:2019bqg,Blagojevic:2021mli,Blagojevic:2021pqp,Blagojevic:2022etm,Cvetkovic:2022qpt}, together with an examination of the near-horizon geometries and their asymptotic symmetry structures in the extremal limits~\cite{Cvetkovic:2022qpt,Cvetkovic:2025rpj}. Another natural extension of the present framework would be to incorporate parity violating interactions in the gravitational action, which may also lead to further modifications of the black hole solutions~\cite{Baekler:2011jt,Obukhov:2020hlp}. Finally, the search for rotating counterparts exhibiting a gravitational spin-orbit interaction beyond the Kerr geometry is especially relevant. This result would provide an explicit violation of the Kerr hypothesis and would allow the backreaction effects of torsion on black hole singularities to be assessed under the influence of this interaction~\cite{Herdeiro:2022yle}. Further studies along these lines are currently underway.

\begin{table}[H]
\centering

\renewcommand{\arraystretch}{1.03}
\setlength{\tabcolsep}{3pt}
\setlength{\arrayrulewidth}{0.5pt}
\scriptsize

\setlength{\abovedisplayskip}{3pt}
\setlength{\belowdisplayskip}{3pt}
\setlength{\abovedisplayshortskip}{3pt}
\setlength{\belowdisplayshortskip}{3pt}

\begin{tabular}{
|p{0.14\textwidth}|
 p{0.53\textwidth}|
 p{0.24\textwidth}|
}
\hline

\textbf{Branch}
&
\textbf{Torsion and metric functions}
&
\textbf{Physical content}
\\
\hline\hline

\branchcell{2.20cm}{Kiselev}
&
\begin{minipage}[t]{\linewidth}\vspace{0.5mm}\raggedright
\textbf{Torsion:}
\[
P(r)=Q(r)=Y(r)=0,
\qquad
U(r)=\kappa_{3}r^\gamma .
\]

\textbf{Metric:}
\[
\Psi(r)=1-\frac{2m}{r}
+\frac{9}{2}(h_9+h_{10})\kappa_{3}^{\,2}r^{2\gamma}.
\]
\vspace{0.5mm}
\end{minipage}
&
\begin{minipage}[t][2.20cm][c]{\linewidth}\raggedright
\textbf{Primary tensor hair.}

Independent parameters:
\[
m,\ \kappa_{3}.
\]

$\gamma$ is fixed by the theory.

Theory:
Eqs.~\eqref{eq:kiselev_h8}-\eqref{eq:kiselev_gamma}.
\end{minipage}
\\
\hline

\branchcell{3.55cm}{$P$--$Q$\\Lambert $W$}
&
\begin{minipage}[t]{\linewidth}\vspace{0.5mm}\raggedright
\textbf{Torsion:}
\[
U(r)=Y(r)=0,
\]
\[
P(r)=
-\,\frac{\kappa_{4}}{4}
\left[5W_0\!\left(\frac{\kappa_{3}}{r}\right)+9\right]
e^{\frac13W_0(\kappa_{3}/r)},
\]
\[
Q(r)=
\frac{3\kappa_{4}}{4}
\left[W_0\!\left(\frac{\kappa_{3}}{r}\right)+1\right]
e^{\frac13W_0(\kappa_{3}/r)} .
\]

\textbf{Metric:}
\[
\Psi(r)=1-\frac{2m}{r}
+\frac{9(h_{17}-h_{18})}{8}\kappa_{4}^{\,2}
W_0^2\!\left(\frac{\kappa_{3}}{r}\right)
e^{\frac23W_0(\kappa_{3}/r)}.
\]
\vspace{0.5mm}
\end{minipage}
&
\begin{minipage}[t][3.55cm][c]{\linewidth}\raggedright
\textbf{Lambert $W$ vector--tensor branch.}

Independent parameters:
\[
m,\ \kappa_{3},\ \kappa_{4}.
\]

The metric is asymptotically Reissner--Nordstr\"om-like. A finite
parameter region contains three Killing horizons.

Theory:
Eqs.~\eqref{hcase2a}--\eqref{eq:PQmixed_etazeta} and
\eqref{eq:PQmixed_critical_relation}.
\end{minipage}
\\
\hline

\branchcell{3.55cm}{$P$--$Q$\\algebraic}
&
\begin{minipage}[t]{\linewidth}\vspace{0.5mm}\raggedright
\textbf{Torsion ($n=2$):}
\[
U(r)=Y(r)=0,
\]
\[
P(r)=\kappa_{4}
\left(
1+2\sqrt{1-\frac{\kappa_{3}}{r}}
\right),
\qquad
Q(r)=-\,\kappa_{4}
\sqrt{1-\frac{\kappa_{3}}{r}} .
\]

\textbf{Metric:}
\[
\Psi(r)=1-\frac{2m}{r}
-2\kappa_{4}^{\,2}(h_{17}-h_{18})
\left(
\frac{\kappa_{3}}{2r}
+\sqrt{1-\frac{\kappa_{3}}{r}}-1
\right).
\]
\vspace{0.5mm}
\end{minipage}
&
\begin{minipage}[t][3.55cm][c]{\linewidth}\raggedright
\textbf{Algebraic vector--tensor branch.}

Independent parameters:
\[
m,\ \kappa_{3},\ \kappa_{4}.
\]

The explicit $n=2$ representative is real for $r>\kappa_{3}$ and
is asymptotically Reissner--Nordstr\"om-like.

Theory:
Eqs.~\eqref{hcase2a}--\eqref{eq:PQmixed_etazeta} and
\eqref{eq:PQmixed_nratio}, with $n=2$.

\end{minipage}
\\
\hline

\branchcell{4.75cm}{Lambert $W$}
&
\begin{minipage}[t]{\linewidth}\vspace{0.5mm}\raggedright
\textbf{Torsion:}
\[
P(r)=0,\qquad
Q(r)=\frac{6\nu m}{r^2}\bigl[1+\mathcal W(r)\bigr],
\]
\[
U(r)=-\,\frac{\kappa_{1}}{2r^2},
\qquad
Y(r)=\kappa_{1}.
\]

\textbf{Metric:}
\[
\Psi(r)=1-\frac{2m}{r}
+\frac{4m^2\ell^{\,2}}{r^4}\bigl[1+\mathcal W(r)\bigr]^2
+\frac{h_{19}\mathcal P_\nu}{96\nu\left(3\nu-1\right)\Delta_\nu}
\frac{\kappa_{1}^2}{r^4}.
\]

\textbf{Auxiliary definitions:}
\begin{equation*}
\mathcal W(r)=
W_0\!\Biggl[
\kappa_{3}
\exp\!\left(-\,\frac{r^3}{4m\ell^{\,2}}\right)
\Biggr],
\end{equation*}
\[
\ell^{\,2}
=-\,\frac{27}{8}\nu h_{19}\bigl(4\nu-1\bigr),
\qquad
\Delta_\nu=16\nu^2-56\nu+5,
\]
\[
\mathcal P_\nu=
6912\nu^4+10800\nu^3-7176\nu^2+471\nu-5.
\]
\vspace{0.5mm}
\end{minipage}
&
\begin{minipage}[t][4.75cm][c]{\linewidth}\raggedright
\textbf{Two independent torsion hairs.}

Independent parameters:
\begin{equation*}
m,\ \kappa_{3},\ \kappa_{1}.
\end{equation*}

$\kappa_{3}$ yields the Lambert $W$ deformation; $\kappa_{1}$ is
the additional axial torsion hair.

Theory:
Eqs.~\eqref{eq:Lambert-inherited-couplingsA}-\eqref{eq:Lambert-inherited-couplingsB},
\eqref{eq:Lambert-charged-couplings1}-\eqref{eq:Lambert-R-functions}.
\end{minipage}
\\
\hline

\branchcell{4.35cm}{Boulware-Deser}
&
\begin{minipage}[t]{\linewidth}\vspace{0.5mm}\raggedright
\textbf{Torsion:}
\[
P(r)=0,\qquad
Q(r)=\frac{N}{r}\bigl(1-\Psi(r)+\chi r^2\bigr),
\]
\[
U(r)=-\,\frac{\kappa_{1}}{2r^2},
\qquad
Y(r)=\kappa_{1}.
\]

\textbf{Metric:}
\[
\Psi_{\pm}(r)=
1+\frac{r^2}{\mathcal A}
\Biggl(
\chi\mathcal A-1
\pm
\sqrt{
\Delta^2
-\frac{4\mathcal A m\Delta}{r^3}
+\frac{\mathcal A\mathcal C\kappa_{1}^2}{2r^6}}\,
\Biggr).
\]

\textbf{Auxiliary definitions:}
\[
\chi=-\,\frac{4\bigl(N-1\bigr)m_t^2}{h_{19}\left(9N-1\right)},
\qquad
\mathcal A=3N\bigl(2N-1\bigr)h_{19},
\]
\[
\mathcal C=
\frac{h_{19}\bigl(6N^2+13N-1\bigr)}{6N\bigl(N-1\bigr)},
\qquad
\Delta=\sqrt{1-2\mathcal A\chi}.
\]
\vspace{0.5mm}
\end{minipage}
&
\begin{minipage}[t][4.35cm][c]{\linewidth}\raggedright
\textbf{Boulware-Deser-type geometry with primary short-range torsion hair.}

Independent parameters:
\[
m,\ \kappa_{1}.
\]

Not asymptotically flat in general; $\chi$ controls the effective
asymptotic curvature.

Theory:
Eqs.~\eqref{eq:Q_algebraic_couplingsA}-\eqref{eq:Q_algebraic_couplingsB},~\eqref{eq:EGB-charged-extra-couplingsA}-\eqref{eq:EGB-charged-extra-couplingsB}.
\end{minipage}
\\
\hline

\end{tabular}

\caption{\label{tab:BH-singular-summary}
Representative singular black hole solutions obtained in the extended class of Holst-type PG models.
For each branch, we display the torsion and metric functions, as well as its main physical content.
The equations quoted in the last column specify the corresponding restrictions on the Lagrangian coefficients.
Except for the universal constants $\kappa_{1}$ and $\kappa_{2}$ defined in
Eq.~\eqref{eqYsol}, labels $\kappa_{i\geq3}$ are branch-specific.}
\end{table}
\newpage

\newpage
\begin{table}[H]
\centering

\renewcommand{\arraystretch}{1.03}
\setlength{\tabcolsep}{3pt}
\setlength{\arrayrulewidth}{0.5pt}
\scriptsize

\setlength{\abovedisplayskip}{3pt}
\setlength{\belowdisplayskip}{3pt}
\setlength{\abovedisplayshortskip}{3pt}
\setlength{\belowdisplayshortskip}{3pt}

\begin{tabular}{
|p{0.14\textwidth}|
 p{0.53\textwidth}|
 p{0.24\textwidth}|
}
\hline

\textbf{Branch}
&
\textbf{Torsion and metric functions}
&
\textbf{Physical content}
\\
\hline\hline

\branchcell{3.55cm}{Regular\\axial-tensor}
&
\begin{minipage}[t]{\linewidth}\vspace{0.5mm}\raggedright
\textbf{Torsion:}
\[
P(r)=Q(r)=0,
\qquad
Y(r)=\kappa_{2} r^3,
\]
\[
U(r)=
\frac{\kappa_{2}\kappa_{3}^3 r}{r^3+\kappa_{3}^3}.
\]

\textbf{Metric:}
\[
\Psi(r)=
1-
\frac{m_{\rm ADM}r^2
\bigl(2r^3+5\kappa_{3}^3\bigr)}
{\bigl(r^3+\kappa_{3}^3\bigr)^2}.
\]
\vspace{0.5mm}
\end{minipage}
&
\begin{minipage}[t][3.55cm][c]{\linewidth}\raggedright
\textbf{Regular black hole with primary torsion hair.}

Independent parameters:
\[
m_{\rm ADM},\ \kappa_{3}.
\]

The scale $\kappa_{3}$ remains independent after fixing the ADM mass and
controls the regular short-distance geometry.

Representative shown:
$\gamma_2=0$, $\gamma_1=-\,3\gamma_4$.

Theory:
Eqs.~\eqref{eq:general-regular-AF}-\eqref{eq:general-regular-gammas};
explicit solution~\eqref{eq:general-regular-known}.
\end{minipage}
\\
\hline

\branchcell{3.25cm}{Hayward}
&
\begin{minipage}[t]{\linewidth}\vspace{0.5mm}\raggedright
\textbf{Torsion:}
\[
P(r)=Q(r)=r,
\qquad
Y(r)=r^3,
\]
\[
U(r)=\frac{\kappa_{4}^3r}{r^3+\kappa_{4}^3}.
\]

\textbf{Metric:}
\[
\Psi(r)=1-\frac{2mr^2}{r^3+\kappa_{4}^3}.
\]

\textbf{Parameter constraint:}
\[
m=-\,\frac{\alpha_4\kappa_{4}^3}{9}.
\]
\vspace{0.5mm}
\end{minipage}
&
\begin{minipage}[t][3.25cm][c]{\linewidth}\raggedright
\textbf{Regular black hole with all torsion modes switched on.}

Vector, axial and tensor torsion are simultaneously nonvanishing.

Secondary hair: for a fixed theory, $\kappa_{4}$ is fixed once $m$ is
specified.

Independent parameter:
\[
m.
\]

Theory:
Eqs.~\eqref{eq:regularAll-h-couplings_1st}-\eqref{eq:regularAll-nondegenerate}.
\end{minipage}
\\
\hline

\end{tabular}

\caption{\label{tab:BH-summary}
Representative regular black hole solutions obtained in the extended class of Holst-type PG models.
For each branch, we display the torsion and metric functions, as well as its main physical content.
The equations quoted in the last column specify the corresponding list of Lagrangian coefficients.
Except for the universal constants $\kappa_{1}$ and $\kappa_{2}$ defined in
Eq.~\eqref{eqYsol}, labels $\kappa_{i\geq3}$ are branch-specific.}
\end{table}

\noindent
\section*{Acknowledgements}

We would like to thank Jose Beltrán Jiménez for helpful discussions. The work of S.B. and M.M. is supported by the Institute for Basic
Science (IBS-R018-D3). The work of J.G.V. is supported by the
Institute for Basic Science (IBS-R003-D1).

\newpage

\appendix
\titleformat{\section}
{\centering\normalfont\bfseries}
{Appendix~\thesection.}{1em}{}
\section{Horizon structure of the new geometries}
\label{app:horizon-structure}

Several of the solutions obtained in the main text have metric functions that do not belong to the usual explicit black hole families. It is therefore useful to check explicitly that these branches contain genuine black hole regions and to determine whether the torsion corrections generate additional horizons. We restrict the discussion to the two different Lambert $W$ branches obtained in the even-parity vector sector and in the even and odd-parity axial-tensor sector, as well as to the regular black hole solutions with primary torsion hair arising in the odd-parity axial-tensor sector.

In general, the different horizons are determined by the positive roots of the metric function $\Psi(r)$. Degenerate configurations are given by imposing simultaneously $\Psi(r)=0$ and $\Psi'(r)=0$. 
For the numerical analysis, we first located sign changes of $\Psi(r)$ on a logarithmic radial grid and then refined each root with a one dimensional solver. We enlarged the radial interval and increased the grid resolution until
the number and positions of the roots were stable. The boundaries between regions with different numbers of horizons were also verified using the degenerate-horizon condition.

\subsection{Lambert $W$ branch in the even-parity vector-tensor sector}

We first consider the Lambert $W$ solution of Sec.~\ref{subsec:PQmixed}. For the globally real principal branch with $\kappa_{3}>0$, it is convenient to introduce
\begin{equation}
 x\equiv\frac{r}{m},\quad
 s\equiv\frac{\kappa_{3}}{m},\quad
 \mathcal A\equiv\frac{9}{8}(h_{17}-h_{18})\kappa_{4}^2\,,
\end{equation}
so that Eq.~\eqref{eq:PQmixed_Lambert_metric} becomes
\begin{equation}
 \Psi(x)=1-\frac{2}{x}
 +\mathcal A\,W_0^2\!\left(\frac{s}{x}\right)
 \exp\!\left[\frac{2}{3}W_0\!\left(\frac{s}{x}\right)\right].
 \label{eq:app-PQ-Lambert-dimless}
\end{equation}
The mass term dominates as $x\to0$, whereas the Lambert $W$ contribution is important at intermediate radii. This competition produces a horizon structure that is considerably richer than the standard solutions of GR. For $\mathcal A\leq0$, the numerical search gives a single horizon. However, for $\mathcal A>0$, a finite region of the parameter space contains three zeros of the metric function: an outer event horizon and two additional inner horizons.

\begin{figure}[H]
\centering
\begin{minipage}[t]{0.49\textwidth}
\centering
\includegraphics[width=\linewidth]{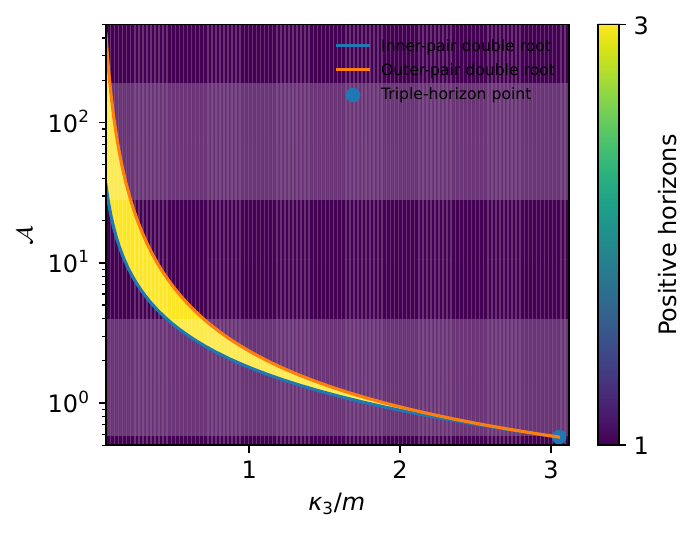}
\vspace{-2mm}
\centerline{\small (a)}
\end{minipage}
\hfill
\begin{minipage}[t]{0.49\textwidth}
\centering
\includegraphics[width=\linewidth]{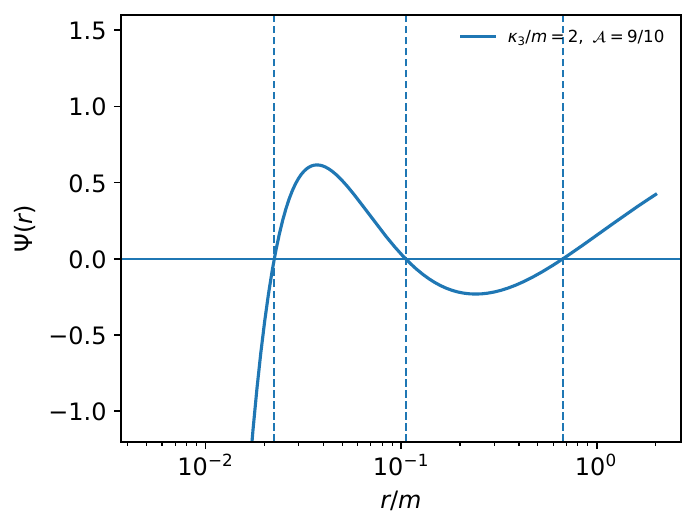}
\vspace{-2mm}
\centerline{\small (b)}
\end{minipage}
\caption{Horizon structure of the real principal Lambert $W$ branch arising in the even-parity vector-tensor sector. Panel (a) shows the number of roots of $\Psi(r)$ in the $(s,\mathcal A)$ plane. The three-horizon region lies between the two double-root boundaries, which meet at a critical configuration where all three horizons coalesce. Panel (b) shows a representative three-horizon geometry for $s=2$ and $\mathcal A=9/10$; the dashed vertical lines mark the three zeros of $\Psi(r)$.}
\label{fig:app-PQ-Lambert-horizons}
\end{figure}

The corresponding region is displayed in Fig.~\ref{fig:app-PQ-Lambert-horizons}. It is bounded by two curves on which a pair of horizons merges. Between these curves the three horizons are distinct; outside them only one horizon remains. The two boundaries meet at a critical point where the three roots coalesce. This three-horizon configuration is a genuine effect provided by the torsion field and occurs while the geometry remains asymptotically flat.

The real Lambert $W$ branches with negative argument are not included in Fig.~\ref{fig:app-PQ-Lambert-horizons}, since they cease to be real below a finite radius and therefore require a separate treatment.

\subsection{Lambert $W$ branch in the even and odd-parity axial-tensor sector}

We next turn to the Lambert $W$ branch in the even and odd-parity axial-tensor sector considered in Sec.~\ref{subsubsec:Lambert-UY-sector}. Defining
\begin{equation}
 x\equiv\frac{r}{m}\,,\quad
 \lambda\equiv\frac{\ell}{m}\,,\quad
 \mathfrak q\equiv
 \frac{h_{19}\mathcal P_\nu}{96\nu(3\nu-1)\Delta_\nu}
 \frac{\kappa_{1}^2}{m^4}\,,
\end{equation}
we can write Eq.~\eqref{eq:charged-Lambert-metric} as
\begin{equation}
 \Psi(x)=1-\frac{2}{x}
 +\frac{4\lambda^2[1+\mathcal W(x)]^2+\mathfrak q}{x^4}\,,
 \quad
 \mathcal W(x)=W_0\!\left[\kappa_{3}
 e^{-x^3/(4\lambda^2)}\right].
 \label{eq:app-charged-Lambert-dimless}
\end{equation}

In this case, the horizon pattern is simpler compared to the previous Lambert $W$ branch. In particular, no configuration with more than two roots was found in the numerical analysis, while there are instead three physically distinct regions. If the effective $r^{-4}$ contribution is sufficiently negative, the metric has one horizon. Increasing the parameter $\mathfrak q$ produces a two-horizon region, bounded above by an extremal curve where the inner and outer horizons merge. Finally, beyond this curve, the geometry is horizonless. The lower boundary of the two-horizon region corresponds to the inner zero being pushed to the centre rather than to a finite-radius extremal horizon.

\begin{figure}[H]
\centering
\includegraphics[width=0.6\textwidth]{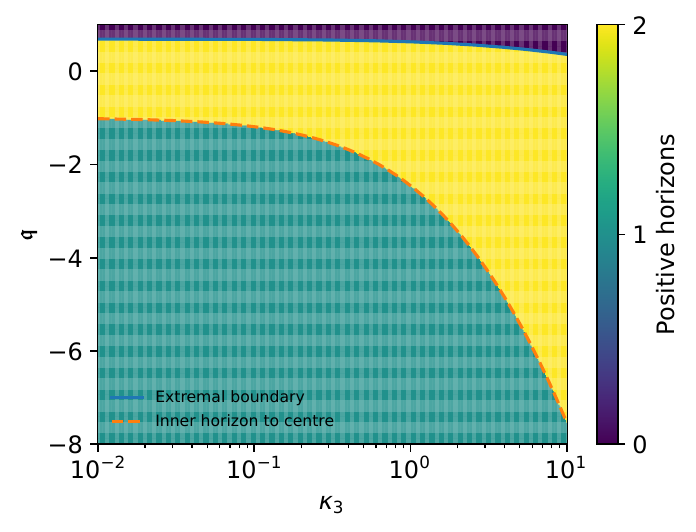}
\caption{Horizon structure of the Lambert $W$ branch in the even and odd-parity axial-tensor sector for the representative slice $\lambda=1/2$. The shaded regions contain zero, one or two horizons. The solid curve denotes the finite-radius extremal boundary, where the inner and outer horizons merge, while the dashed curve marks the transition at which the inner horizon is pushed to the centre.}
\label{fig:app-charged-Lambert-horizons}
\end{figure}

For reference, the finite-radius extremal condition simplifies to
\begin{equation}
 x_{\rm e}=\frac{3}{2}\left[1+\mathcal W(x_{\rm e})\right],
 \label{eq:app-charged-Lambert-extremal}
\end{equation}
which provides a convenient check of the numerical boundary without setting either torsion hair to zero. Figure~\ref{fig:app-charged-Lambert-horizons} shows a representative slice at $\lambda=1/2$. Varying $\lambda$ shifts the two boundaries but does not change the qualitative $0/1/2$-horizon structure found in the analysis.

\subsection{Regular black hole with primary torsion hair}

Finally, we consider the regular axial-tensor solutions of
Sec.~\ref{general-regular-axial-tensor}. In contrast with the
Lambert $W$ branches, the regularity conditions make the horizon structure
particularly simple. On the real branch selected in the main text,
$x(r)$ interpolates monotonically from $x=1$ at the centre to $x=0$ at
infinity. In the parameter ranges explored, after scaling the radius by
$\kappa_{3}$, the mass-dependent deformation of
Eq.~\eqref{eq:general-regular-metric} was found to have a single positive
maximum. Accordingly, increasing $m_{\rm ADM}/\kappa_{3}$ produces the
standard sequence: a horizonless regular geometry, one extremal double
horizon, and a black hole with inner and outer horizons. No
additional pairs of horizons were found when the exponent of the
implicit solution and the remaining coupling ratio were varied over
the regular branch.

\begin{figure}[H]
\centering
\includegraphics[width=0.68\textwidth]{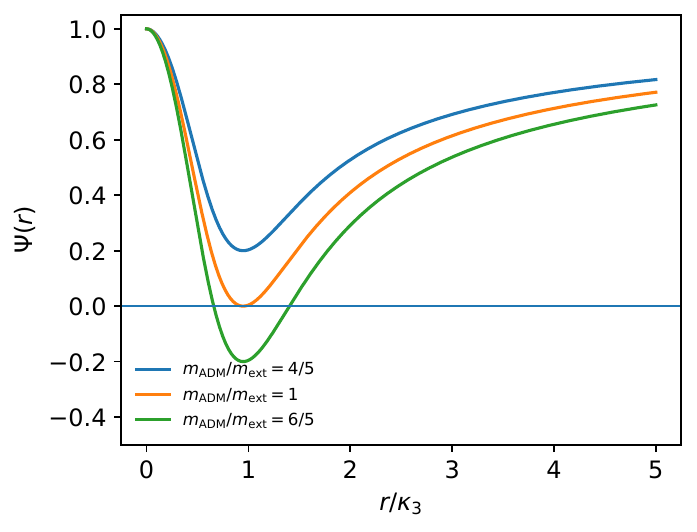}
\caption{Horizon structure of the regular axial-tensor representative in Expression~\eqref{eq:general-regular-known}. For $m_{\rm ADM}<m_{\rm ext}$ the geometry is horizonless, $m_{\rm ADM}=m_{\rm ext}$ gives a degenerate horizon, and $m_{\rm ADM}>m_{\rm ext}$ produces inner and outer horizons. The radial coordinate is normalised by the independent torsion hair $\kappa_{3}$, while the mass is shown relative to the extremal value.}
\label{fig:app-regular-axial-horizons}
\end{figure}

Figure~\ref{fig:app-regular-axial-horizons} illustrates this behaviour for the explicit algebraic subfamily given by Expression~\eqref{eq:general-regular-known}, which is also the regular axial-tensor solution displayed in Table~\ref{tab:BH-summary}. Rather than quoting a decimal critical mass, the curves are normalised directly by the extremal value $m_{\rm ext}$ determined by the conditions $\Psi(r)=\Psi'(r)=0$:
\begin{equation}
    m_{\rm ext}=\frac{3}{5}\kappa_{3}\left(\frac{3\sqrt{5}-5}{2}\right)^{1/3}\,.
\end{equation}
The result makes clear that the primary torsion hair $\kappa_{3}$ controls not only the regular core but also the separation between the two horizons once the black hole regime is reached.

The numerical analysis therefore confirms that the three new geometries considered here contain genuine black hole configurations. The main qualitative novelty is the Lambert $W$ branch arising in the even-parity vector-tensor sector, where the metric corrections provided by torsion can generate two additional inner horizons and a critical three-horizon merger. By contrast, the Lambert $W$ branch in the even and odd parity axial tensor sector has at most two horizons in the parameter ranges explored. The regular odd parity axial tensor solution follows the usual sequence from a horizonless geometry to an extremal black hole and then to a black hole with two horizons, while retaining an independent primary torsion hair and a regular core that can be de Sitter, Minkowski or anti-de Sitter depending on the couplings.

\newpage

\bibliographystyle{utphys}
\bibliography{references}

@article{Zhou:2022yio,
    author = "Zhou, Tian and Modesto, Leonardo",
    title = "{Geodesic incompleteness of some popular regular black holes}",
    eprint = "2208.02557",
    archivePrefix = "arXiv",
    primaryClass = "gr-qc",
    doi = "10.1103/PhysRevD.107.044016",
    journal = "Phys. Rev. D",
    volume = "107",
    number = "4",
    pages = "044016",
    year = "2023"
}

@article{Carballo-Rubio:2019fnb,
    author = "Carballo-Rubio, Ra{\'u}l and Di Filippo, Francesco and Liberati, Stefano and Visser, Matt",
    title = "{Geodesically complete black holes}",
    eprint = "1911.11200",
    archivePrefix = "arXiv",
    primaryClass = "gr-qc",
    doi = "10.1103/PhysRevD.101.084047",
    journal = "Phys. Rev. D",
    volume = "101",
    pages = "084047",
    year = "2020"
}

@article{Eichhorn:2025pgy,
    author = "Eichhorn, Astrid and Fernandes, Pedro G. S.",
    title = "{Regular black holes without mass-inflation instability and gravastars from modified gravity}",
    eprint = "2508.00686",
    archivePrefix = "arXiv",
    primaryClass = "gr-qc",
    doi = "10.1103/nqz2-88zf",
    journal = "Phys. Rev. D",
    volume = "113",
    number = "8",
    pages = "L081501",
    year = "2026"
}

@article{Bakopoulos:2023sdm,
    author = "Bakopoulos, Athanasios and Charmousis, Christos and Kanti, Panagiota and Lecoeur, Nicolas and Nakas, Theodoros",
    title = "{Black holes with primary scalar hair}",
    eprint = "2310.11919",
    archivePrefix = "arXiv",
    primaryClass = "gr-qc",
    doi = "10.1103/PhysRevD.109.024032",
    journal = "Phys. Rev. D",
    volume = "109",
    number = "2",
    pages = "024032",
    year = "2024"
}

@article{Ayon-Beato:2000mjt,
    author = "Ayón-Beato, Eloy and García, Alberto",
    title = "{The Bardeen model as a nonlinear magnetic monopole}",
    eprint = "gr-qc/0009077",
    archivePrefix = "arXiv",
    doi = "10.1016/S0370-2693(00)01125-4",
    journal = "Phys. Lett. B",
    volume = "493",
    pages = "149--152",
    year = "2000"
}

@article{Boulware:1985wk,
    author = "Boulware, David G. and Deser, Stanley",
    title = "{String Generated Gravity Models}",
    reportNumber = "DOE-ER-40048-27 P5",
    doi = "10.1103/PhysRevLett.55.2656",
    journal = "Phys. Rev. Lett.",
    volume = "55",
    pages = "2656",
    year = "1985"
}

@article{Glavan:2019inb,
    author = "Glavan, Dra{\v{z}}en and Lin, Chunshan",
    title = "{Einstein-Gauss-Bonnet Gravity in Four-Dimensional Spacetime}",
    eprint = "1905.03601",
    archivePrefix = "arXiv",
    primaryClass = "gr-qc",
    reportNumber = "CP3-19-24",
    doi = "10.1103/PhysRevLett.124.081301",
    journal = "Phys. Rev. Lett.",
    volume = "124",
    number = "8",
    pages = "081301",
    year = "2020"
}

@article{Lu:2020iav,
    author = "Lu, H. and Pang, Yi",
    title = "{Horndeski gravity as $D \rightarrow 4$ limit of Gauss-Bonnet}",
    eprint = "2003.11552",
    archivePrefix = "arXiv",
    primaryClass = "gr-qc",
    doi = "10.1016/j.physletb.2020.135717",
    journal = "Phys. Lett. B",
    volume = "809",
    pages = "135717",
    year = "2020"
}

@article{Fernandes:2020nbq,
    author = "Fernandes, Pedro G. S. and Carrilho, Pedro and Clifton, Timothy and Mulryne, David J.",
    title = "{Derivation of Regularized Field Equations for the Einstein-Gauss-Bonnet Theory in Four Dimensions}",
    eprint = "2004.08362",
    archivePrefix = "arXiv",
    primaryClass = "gr-qc",
    doi = "10.1103/PhysRevD.102.024025",
    journal = "Phys. Rev. D",
    volume = "102",
    number = "2",
    pages = "024025",
    year = "2020"
}

@article{Aoki:2020lig,
    author = "Aoki, Katsuki and Gorji, Mohammad Ali and Mukohyama, Shinji",
    title = "{A consistent theory of $D \to 4$ Einstein-Gauss-Bonnet gravity}",
    eprint = "2005.03859",
    archivePrefix = "arXiv",
    primaryClass = "gr-qc",
    reportNumber = "YITP-20-66, IPMU20-0050",
    doi = "10.1016/j.physletb.2020.135843",
    journal = "Phys. Lett. B",
    volume = "810",
    pages = "135843",
    year = "2020"
}

@article{Myung:2025afs,
    author = "Myung, Yun Soo and Nakas, Theodoros",
    title = "{Smarr formula for black holes with primary and secondary scalar hair}",
    eprint = "2505.02368",
    archivePrefix = "arXiv",
    primaryClass = "gr-qc",
    doi = "10.1103/qv3n-r8q9",
    journal = "Phys. Rev. D",
    volume = "112",
    number = "6",
    pages = "064002",
    year = "2025"
}

@article{Bronnikov:2000vy,
    author = "Bronnikov, K. A.",
    title = "{Regular magnetic black holes and monopoles from nonlinear electrodynamics}",
    eprint = "gr-qc/0006014",
    archivePrefix = "arXiv",
    doi = "10.1103/PhysRevD.63.044005",
    journal = "Phys. Rev. D",
    volume = "63",
    pages = "044005",
    year = "2001"
}

@article{Corless:1996zz,
    author = "Corless, R. M. and Gonnet, G. H. and Hare, D. E. G. and Jeffrey, D. J. and Knuth, D. E.",
    title = "{On the LambertW function}",
    doi = "10.1007/BF02124750",
    journal = "Adv. Comput. Math.",
    volume = "5",
    pages = "329--359",
    year = "1996"
}

@article{Karakasis:2023hni,
    author = "Karakasis, Thanasis and Mavromatos, Nick E. and Papantonopoulos, Eleftherios",
    title = "{Regular compact objects with scalar hair}",
    eprint = "2305.00058",
    archivePrefix = "arXiv",
    primaryClass = "gr-qc",
    doi = "10.1103/PhysRevD.108.024001",
    journal = "Phys. Rev. D",
    volume = "108",
    number = "2",
    pages = "024001",
    year = "2023"
}

@article{Will:2014kxa,
    author = "Will, Clifford M.",
    title = "{The Confrontation between General Relativity and Experiment}",
    eprint = "1403.7377",
    archivePrefix = "arXiv",
    primaryClass = "gr-qc",
    doi = "10.12942/lrr-2014-4",
    journal = "Living Rev. Rel.",
    volume = "17",
    pages = "4",
    year = "2014"
}

@book{Blagojevic:2013xpa,
    editor = "Blagojevi\'c, Milutin and Hehl, Friedrich W.",
    title = "{Gauge Theories of Gravitation}: {A Reader with Commentaries}",
    isbn = "978-1-84816-726-1",
    publisher = "World Scientific",
    address = "Singapore",
    year = "2013"
}

@book{ponomarev2017gauge,
	year = 2017,
	publisher = {Nauka, Moscow},
	author = "{V. N. Ponomarev, A. O. Barvinsky, and Yu. N. Obukhov}",
	title = {Gauge Approach and Quantization Methods in Gravity Theory}}

@article{Obukhov:2022khx,
    author = "Obukhov, Yuri N.",
    title = "{Poincar\'e gauge gravity primer}",
    eprint = "2206.05205",
    archivePrefix = "arXiv",
    primaryClass = "gr-qc",
    month = "6",
    year = "2022"
}

@article{Holst:1995pc,
    author = "Holst, Soren",
    title = "{Barbero's Hamiltonian derived from a generalized Hilbert-Palatini action}",
    eprint = "gr-qc/9511026",
    archivePrefix = "arXiv",
    reportNumber = "USITP-95-10",
    doi = "10.1103/PhysRevD.53.5966",
    journal = "Phys. Rev. D",
    volume = "53",
    pages = "5966--5969",
    year = "1996"
}

@article{Perez:2005pm,
    author = "Perez, Alejandro and Rovelli, Carlo",
    title = "{Physical effects of the Immirzi parameter}",
    eprint = "gr-qc/0505081",
    archivePrefix = "arXiv",
    doi = "10.1103/PhysRevD.73.044013",
    journal = "Phys. Rev. D",
    volume = "73",
    pages = "044013",
    year = "2006"
}

@article{Hohmann:2019fvf,
    author = "Hohmann, Manuel",
    archivePrefix = "arXiv",
    doi = "10.3390/sym12030453",
    eprint = "1912.12906",
    journal = "Symmetry",
    number = "3",
    pages = "453",
    primaryClass = "math-ph",
    title = "{Metric-Affine Geometries with Spherical Symmetry}",
    volume = "12",
    year = "2020"
}

@article{Kokkotas:1999bd,
    author = "Kokkotas, Kostas D. and Schmidt, Bernd G.",
    title = "{Quasinormal modes of stars and black holes}",
    eprint = "gr-qc/9909058",
    archivePrefix = "arXiv",
    doi = "10.12942/lrr-1999-2",
    journal = "Living Rev. Rel.",
    volume = "2",
    pages = "2",
    year = "1999"
}

@article{Ashtekar:2004eh,
      author         = "Ashtekar, Abhay and Lewandowski, Jerzy",
      title          = "{Background independent quantum gravity: A Status
                        report}",
      journal        = "Class. Quant. Grav.",
      volume         = "21",
      year           = "2004",
      pages          = "R53",
      doi            = "10.1088/0264-9381/21/15/R01",
      eprint         = "gr-qc/0404018",
      archivePrefix  = "arXiv",
      primaryClass   = "gr-qc",
      SLACcitation   = "%%CITATION = GR-QC/0404018;%%"
}

@article{McCrea:1992wa,
    author = "McCrea, J. D.",
    doi = "10.1088/0264-9381/9/2/018",
    journal = "Class. Quant. Grav.",
    pages = "553--568",
    title = "{Irreducible decompositions of non-metricity, torsion, curvature and Bianchi identities in metric-affine spacetimes}",
    volume = "9",
    year = "1992"
}

@article{Hehl:1976kj,
    author = "Hehl, F.W. and von der Heyde, P. and Kerlick, G.D. and Nester, J.M.",
    doi = "10.1103/RevModPhys.48.393",
    journal = "Rev. Mod. Phys.",
    pages = "393--416",
    title = "{General Relativity with Spin and Torsion: Foundations and Prospects}",
    volume = "48",
    year = "1976"
}

@article{Herdeiro:2022yle,
    author = "Herdeiro, Carlos A. R.",
    title = "{Black Holes: On the Universality of the Kerr Hypothesis}",
    eprint = "2204.05640",
    archivePrefix = "arXiv",
    primaryClass = "gr-qc",
    doi = "10.1007/978-3-031-31520-6_8",
    journal = "Lect. Notes Phys.",
    volume = "1017",
    pages = "315--331",
    year = "2023"
}

@article{Mielke:1981xe,
    author = "Mielke, Eckehard W.",
    title = "{Reduction of the Poincaré Gauge Field Equations by Means of a Duality Rotation}",
    reportNumber = "IC/81/210",
    doi = "10.1063/1.526172",
    journal = "J. Math. Phys.",
    volume = "25",
    pages = "663",
    year = "1984"
}

@article{Baekler:1981lkh,
    author = "Baekler, Peter",
    doi = "10.1016/0370-2693(81)90111-8",
    journal = "Phys. Lett. B",
    pages = "329--332",
    title = "{A spherically symmetric vacuum solution of the quadratic Poincar\'e gauge field theory of gravitation with newtonian and confinement potentials}",
    volume = "99",
    year = "1981"
}

@article{Obukhov:1987tz,
    author = "Obukhov, Yu.N. and Ponomarev, V.N. and Zhytnikov, V.V.",
    doi = "10.1007/BF00763457",
    journal = "Gen. Rel. Grav.",
    pages = "1107--1142",
    title = "{Quadratic Poincar\'e Gauge Theory of Gravity: A Comparison With the General Relativity Theory}",
    volume = "21",
    year = "1989"
}

@article{Cembranos:2017pcs,
    author = "Cembranos, Jose Alberto Ruiz and Gigante Valcarcel, Jorge",
    archivePrefix = "arXiv",
    doi = "10.1016/j.physletb.2018.01.081",
    eprint = "1708.00374",
    journal = "Phys. Lett. B",
    pages = "143--150",
    primaryClass = "gr-qc",
    title = "{Extended Reissner--Nordström solutions sourced by dynamical torsion}",
    volume = "779",
    year = "2018"
}

@article{Cembranos:2016gdt,
    author = "Cembranos, Jose Alberto Ruiz and Gigante Valcarcel, Jorge",
    archivePrefix = "arXiv",
    doi = "10.1088/1475-7516/2017/01/014",
    eprint = "1608.00062",
    journal = "JCAP",
    pages = "014",
    primaryClass = "gr-qc",
    title = "{New torsion black hole solutions in Poincar\'e gauge theory}",
    volume = "01",
    year = "2017"
}

@article{delaCruzDombriz:2021nrg,
    author = "{A. de la Cruz-Dombriz, F. J. Maldonado Torralba, and D. F. Mota}",
    title = "{Dark matter candidate from torsion}",
    eprint = "2112.03957",
    archivePrefix = "arXiv",
    primaryClass = "gr-qc",
    doi = "10.1016/j.physletb.2022.137488",
    journal = "Phys. Lett. B",
    volume = "834",
    pages = "137488",
    year = "2022"
}

@article{Casado-Turrion:2023omz,
    author = "Casado-Turri\'on, Adri\'an and de la Cruz-Dombriz, \'Alvaro and Jim\'enez-Cano, Alejandro and Maldonado Torralba, Francisco Jos\'e",
    title = "{Junction conditions in bi-scalar Poincar\'e gauge gravity}",
    eprint = "2303.01206",
    archivePrefix = "arXiv",
    primaryClass = "gr-qc",
    doi = "10.1088/1475-7516/2023/07/023",
    journal = "JCAP",
    volume = "07",
    pages = "023",
    year = "2023"
}

@article{Bakler:1988nq,
    author = "Baekler, P. and Gurses, M. and Hehl, F. W. and McCrea, J. D.",
    title = "{The exterior gravitational field of a charged spinning source in the Poincar\'{e} Gauge theory: A Kerr-Newman metric with dynamic torsion}",
    doi = "10.1016/0375-9601(88)90366-0",
    journal = "Phys. Lett. A",
    volume = "128",
    pages = "245--250",
    year = "1988"
}

@article{Neville:1978bk,
    author = "Neville, Donald E.",
    title = "{Gravity Lagrangian with ghost-free curvature-squared terms}",
    reportNumber = "Print-78-1155 (TEMPLE)",
    doi = "10.1103/PhysRevD.18.3535",
    journal = "Phys. Rev. D",
    volume = "18",
    pages = "3535",
    year = "1978"
}

@article{Sezgin:1979zf,
    author = "Sezgin, E. and van Nieuwenhuizen, P.",
    title = "{New ghost-free gravity Lagrangians with propagating torsion}",
    reportNumber = "ITP-SB-79-97",
    doi = "10.1103/PhysRevD.21.3269",
    journal = "Phys. Rev. D",
    volume = "21",
    pages = "3269",
    year = "1980"
}

@article{Sezgin:1981xs,
    author = "Sezgin, E.",
    doi = "10.1103/PhysRevD.24.1677",
    journal = "Phys. Rev. D",
    pages = "1677--1680",
    reportNumber = "ITP-SB-80-59",
    title = "{Class of ghost-free gravity Lagrangians with massive or massless propagating torsion}",
    volume = "24",
    year = "1981"
}

@article{Miyamoto:1983bf,
    author = "Miyamoto, S. and Nakano, T. and Ohtani, T. and Tamura, Y.",
    title = "{Linear approximation for the massless Lorentz gauge field}",
    doi = "10.1143/PTP.69.1236",
    journal = "Prog. Theor. Phys.",
    volume = "69",
    pages = "1236--1240",
    year = "1983"
}

@article{Fukui:1984gn,
    author = "Fukui, Masayasu and Masukawa, Junnichi",
    title = "{Massless torsion fields. II. The case $\alpha + 2a/3 = 0$}",
    reportNumber = "OUAM 84-6-1",
    doi = "10.1143/PTP.73.75",
    journal = "Prog. Theor. Phys.",
    volume = "73",
    pages = "75",
    year = "1985"
}

@article{Fukuma:1984cz,
    author = "Fukuma, Kazumi and Miyamoto, Shikao and Nakano, Tadao and Ohtani, Teruya and Tamura, Yoshinobu",
    doi = "10.1143/PTP.73.874",
    journal = "Prog. Theor. Phys.",
    pages = "874",
    reportNumber = "Print-84-0837 (OSAKA CITY)",
    title = "{Massless Lorentz gauge field consistent with Einstein’s gravitation theory. The case $\alpha + 3a/2 = \beta - 2a/3 = \gamma + 3a/2 = 0$}",
    volume = "73",
    year = "1985"
}

@article{Battiti:1985mu,
    author = "Battiti, R. and Toller, M.",
    doi = "10.1007/BF02746948",
    journal = "Lett. Nuovo Cim.",
    pages = "35",
    reportNumber = "Print-85-0691 (TRENTO)",
    title = "{Zero-mass normal modes in linearized Poincaré gauge theories}",
    volume = "44",
    year = "1985"
}

@article{Kuhfuss:1986rb,
    author = "Kuhfuss, R. and Nitsch, J.",
    doi = "10.1007/BF00763447",
    journal = "Gen. Rel. Grav.",
    pages = "1207",
    reportNumber = "MPA-223",
    title = "{Propagating Modes in Gauge Field Theories of Gravity}",
    volume = "18",
    year = "1986"
}

@article{Blagojevic:1986dm,
    author = "Blagojevi\'c, M. and Vasili\'c, M.",
    doi = "10.1103/PhysRevD.35.3748",
    journal = "Phys. Rev. D",
    pages = "3748",
    reportNumber = "Print-86-0954 (BELGRADE)",
    title = "{Extra gauge symmetries in a weak-field approximation of an $R + T^2 + R^2$ theory of gravity}",
    volume = "35",
    year = "1987"
}

@article{Baikov:1992uh,
    author = "Baikov, P. and Hayashi, M. and Nelipa, N. and Ostapchenko, S.",
    title = "{Ghost and tachyon free gauge invariant, Poincaré, affine and projective Lagrangians}",
    reportNumber = "SMC-12-91",
    doi = "10.1007/BF00759092",
    journal = "Gen. Rel. Grav.",
    volume = "24",
    pages = "867--880",
    year = "1992"
}

@article{Yo:1999ex,
    author = "Yo, Hwei-Jang and Nester, James M.",
    title = "{Hamiltonian analysis of Poincaré gauge theory scalar modes}",
    eprint = "gr-qc/9902032",
    archivePrefix = "arXiv",
    doi = "10.1142/S021827189900033X",
    journal = "Int. J. Mod. Phys. D",
    volume = "8",
    pages = "459--479",
    year = "1999"
}

@article{Yo:2001sy,
    author = "Yo, Hwei-Jang and Nester, James M.",
    archivePrefix = "arXiv",
    doi = "10.1142/S0218271802001998",
    eprint = "gr-qc/0112030",
    journal = "Int. J. Mod. Phys. D",
    pages = "747--780",
    title = "{Hamiltonian analysis of Poincar\'e gauge theory: higher spin modes}",
    volume = "11",
    year = "2002"
}

@article{Lin:2018awc,
    author = "Lin, Yun-Cherng and Hobson, Michael P. and Lasenby, Anthony N.",
    title = "{Ghost and tachyon free Poincar\'e gauge theories: A systematic approach}",
    eprint = "1812.02675",
    archivePrefix = "arXiv",
    primaryClass = "gr-qc",
    doi = "10.1103/PhysRevD.99.064001",
    journal = "Phys. Rev. D",
    volume = "99",
    number = "6",
    pages = "064001",
    year = "2019"
}

@article{Jimenez:2019qjc,
    author = "Beltrán Jiménez, Jose and Maldonado Torralba, Francisco José",
    title = "{Revisiting the stability of quadratic Poincaré gauge gravity}",
    eprint = "1910.07506",
    archivePrefix = "arXiv",
    primaryClass = "gr-qc",
    doi = "10.1140/epjc/s10052-020-8163-8",
    journal = "Eur. Phys. J. C",
    volume = "80",
    number = "7",
    pages = "611",
    year = "2020"
}

@article{Bahamonde:2024sqo,
    author = "Bahamonde, Sebastian and Gigante Valcarcel, Jorge",
    title = "{Stability of Poincar\'e gauge theory with cubic order invariants}",
    eprint = "2402.08937",
    archivePrefix = "arXiv",
    primaryClass = "gr-qc",
    doi = "10.1103/PhysRevD.109.104075",
    journal = "Phys. Rev. D",
    volume = "109",
    number = "10",
    pages = "104075",
    year = "2024"
}

@article{Audretsch:1981xn,
    author = "Audretsch, J.",
    title = "{Dirac Electron in Space-times With Torsion: Spinor Propagation, Spin Precession, and Nongeodesic Orbits}",
    doi = "10.1103/PhysRevD.24.1470",
    journal = "Phys. Rev. D",
    volume = "24",
    pages = "1470--1477",
    year = "1981"
}

@article{Bahamonde:2025fei,
    author = "Bahamonde, Sebastian and Gigante Valcarcel, Jorge",
    title = "{A gravitational spin-orbit interaction in Poincar{\'e} gauge theory}",
    eprint = "2508.20035",
    archivePrefix = "arXiv",
    primaryClass = "gr-qc",
    doi = "10.1016/j.physletb.2025.140126",
    journal = "Phys. Lett. B",
    volume = "873",
    pages = "140126",
    year = "2026"
}

@article{Yasskin:1980bu,
    author = "Yasskin, Philip B. and Stoeger, William R., S.J.",
    doi = "10.1103/PhysRevD.21.2081",
    journal = "Phys. Rev. D",
    pages = "2081",
    reportNumber = "HUTMP-79/B77, MdDP-PP-80-040",
    title = "{Propagation Equations for Test Bodies With Spin and Rotation in Theories of Gravity With Torsion}",
    volume = "21",
    year = "1980"
}

@article{Obukhov:2014fta,
    author = "Obukhov, Yuri N. and Silenko, Alexander J. and Teryaev, Oleg V.",
    title = "{Spin-torsion coupling and gravitational moments of Dirac fermions: theory and experimental bounds}",
    eprint = "1410.6197",
    archivePrefix = "arXiv",
    primaryClass = "hep-th",
    doi = "10.1103/PhysRevD.90.124068",
    journal = "Phys. Rev. D",
    volume = "90",
    number = "12",
    pages = "124068",
    year = "2014"
}

@article{Fadeev:2020gjk,
    author = "Fadeev, Pavel and Wang, Tao and Band, Y. B. and Budker, Dmitry and Graham, Peter W. and Sushkov, Alexander O. and Kimball, Derek F. Jackson",
    title = "{Gravity Probe Spin: Prospects for measuring general-relativistic precession of intrinsic spin using a ferromagnetic gyroscope}",
    eprint = "2006.09334",
    archivePrefix = "arXiv",
    primaryClass = "gr-qc",
    doi = "10.1103/PhysRevD.103.044056",
    journal = "Phys. Rev. D",
    volume = "103",
    number = "4",
    pages = "044056",
    year = "2021"
}

@article{Trukhanova:2022utt,
    author = "Trukhanova, Iv., Mariya and Andreev, Pavel and Obukhov, Yuri N.",
    title = "{Search for Manifestations of Spin{\textendash}Torsion Coupling}",
    eprint = "2212.13871",
    archivePrefix = "arXiv",
    primaryClass = "gr-qc",
    doi = "10.3390/universe9010038",
    journal = "Universe",
    volume = "9",
    number = "1",
    pages = "38",
    year = "2023"
}

@article{Baekler:2011jt,
    author = "Baekler, Peter and Hehl, Friedrich W.",
    title = "{Beyond Einstein-Cartan gravity: Quadratic torsion and curvature invariants with even and odd parity including all boundary terms}",
    eprint = "1105.3504",
    archivePrefix = "arXiv",
    primaryClass = "gr-qc",
    doi = "10.1088/0264-9381/28/21/215017",
    journal = "Class. Quant. Grav.",
    volume = "28",
    pages = "215017",
    year = "2011"
}

@article{Obukhov:2020hlp,
    author = "Obukhov, Yuri N.",
    title = "{Generalized Birkhoff theorem in the Poincar\'e gauge gravity theory}",
    eprint = "2009.00284",
    archivePrefix = "arXiv",
    primaryClass = "gr-qc",
    doi = "10.1103/PhysRevD.102.104059",
    journal = "Phys. Rev. D",
    volume = "102",
    number = "10",
    pages = "104059",
    year = "2020"
}

@article{Cembranos:2016xqx,
    author = "{J. A. R. Cembranos, J. Gigante Valcarcel, and F. J. Maldonado Torralba}",
    archivePrefix = "arXiv",
    doi = "10.1088/1475-7516/2017/04/021",
    eprint = "1609.07814",
    journal = "JCAP",
    pages = "021",
    primaryClass = "gr-qc",
    title = "{Singularities and n-dimensional black holes in torsion theories}",
    volume = "04",
    year = "2017"
}

@article{Hehl:1971qi,
    author = "Hehl, F. W. and Datta, B. K.",
    title = "{Nonlinear spinor equation and asymmetric connection in General Relativity}",
    doi = "10.1063/1.1665738",
    journal = "J. Math. Phys.",
    volume = "12",
    pages = "1334--1339",
    year = "1971"
}

@article{Cembranos:2018ipn,
    author = "{J. A. R. Cembranos, J. Gigante Valcarcel, and F. J. Maldonado Torralba}",
    title = "{Fermion dynamics in torsion theories}",
    eprint = "1805.09577",
    archivePrefix = "arXiv",
    primaryClass = "gr-qc",
    doi = "10.1088/1475-7516/2019/04/039",
    journal = "JCAP",
    volume = "04",
    pages = "039",
    year = "2019"
}

@article{tsamparlis1979cosmological,
  title={Cosmological principle and torsion},
  author={Tsamparlis, Michael},
  journal={Physics Letters A},
  volume={75},
  number={1-2},
  pages={27--28},
  year={1979},
  publisher={Elsevier}
}

@article{Chen:2009at,
    author = "Chen, Hsin and Ho, Fei-Hung and Nester, James M. and Wang, Chih-Hung and Yo, Hwei-Jang",
    title = "{Cosmological dynamics with propagating Lorentz connection modes of spin zero}",
    eprint = "0908.3323",
    archivePrefix = "arXiv",
    primaryClass = "gr-qc",
    doi = "10.1088/1475-7516/2009/10/027",
    journal = "JCAP",
    volume = "10",
    pages = "027",
    year = "2009"
}

@article{Blagojevic:2019gsd,
    author = "Blagojevi\'c, Milutin and Cvetkovi\'c, Branislav",
    title = "{Entropy in Poincar\'e gauge theory: Hamiltonian approach}",
    eprint = "1903.02263",
    archivePrefix = "arXiv",
    primaryClass = "gr-qc",
    doi = "10.1103/PhysRevD.99.104058",
    journal = "Phys. Rev. D",
    volume = "99",
    number = "10",
    pages = "104058",
    year = "2019"
}

@article{Blagojevic:2021pqp,
    author = "Blagojevi\'c, Milutin and Cvetkovi\'c, Branislav",
    title = "{Entropy of Reissner-Nordstr\"om-like black holes}",
    eprint = "2112.02099",
    archivePrefix = "arXiv",
    primaryClass = "gr-qc",
    doi = "10.1016/j.physletb.2021.136815",
    journal = "Phys. Lett. B",
    volume = "824",
    pages = "136815",
    year = "2022"
}

@article{Blagojevic:2019bqg,
    author = "Blagojevi\'c, M. and Cvetkovi\'c, B.",
    title = "{Hamiltonian approach to black hole entropy: Kerr-like spacetimes}",
    eprint = "1905.04928",
    archivePrefix = "arXiv",
    primaryClass = "gr-qc",
    doi = "10.1103/PhysRevD.100.044029",
    journal = "Phys. Rev. D",
    volume = "100",
    number = "4",
    pages = "044029",
    year = "2019"
}

@article{Blagojevic:2021mli,
    author = "Blagojevi\'c, M. and Cvetkovi\'c, B.",
    title = "{Thermodynamics of Riemannian Kerr-AdS black holes in Poincar\'e gauge theory}",
    eprint = "2103.00330",
    archivePrefix = "arXiv",
    primaryClass = "gr-qc",
    doi = "10.1016/j.physletb.2021.136242",
    journal = "Phys. Lett. B",
    volume = "816",
    pages = "136242",
    year = "2021"
}

@article{Blagojevic:2022etm,
    author = "Blagojevi\'c, Milutin and Cvetkovi\'c, Branislav",
    title = "{Entropy of Kerr-Newman-AdS black holes with torsion}",
    eprint = "2203.14696",
    archivePrefix = "arXiv",
    primaryClass = "gr-qc",
    doi = "10.1103/PhysRevD.105.104014",
    journal = "Phys. Rev. D",
    volume = "105",
    number = "10",
    pages = "104014",
    year = "2022"
}

@article{Cvetkovic:2022qpt,
    author = "Cvetkovi{\'c}, Branislav and Rakonjac, Danilo",
    title = "{Extremal Kerr black hole entropy in Poincar{\'e} gauge theory}",
    eprint = "2208.04383",
    archivePrefix = "arXiv",
    primaryClass = "gr-qc",
    doi = "10.1103/PhysRevD.107.044054",
    journal = "Phys. Rev. D",
    volume = "107",
    number = "4",
    pages = "044054",
    year = "2023"
}

@article{Cvetkovic:2025rpj,
    author = "Cvetkovi{\'c}, B. and Rakonjac, D.",
    title = "{Near-horizon geometry with torsion: Kerr-AdS black hole}",
    eprint = "2511.15106",
    archivePrefix = "arXiv",
    primaryClass = "gr-qc",
    doi = "10.1103/f5hg-kxrw",
    journal = "Phys. Rev. D",
    volume = "113",
    number = "12",
    pages = "124059",
    year = "2026"
}

@article{LIGOScientific:2016aoc,
    author = "Abbott, B. P. and others",
    collaboration = "LIGO Scientific, Virgo",
    title = "{Observation of Gravitational Waves from a Binary Black Hole Merger}",
    eprint = "1602.03837",
    archivePrefix = "arXiv",
    primaryClass = "gr-qc",
    reportNumber = "LIGO-P150914",
    doi = "10.1103/PhysRevLett.116.061102",
    journal = "Phys. Rev. Lett.",
    volume = "116",
    number = "6",
    pages = "061102",
    year = "2016"
}

@article{LIGOScientific:2017vwq,
    author = "Abbott, B. P. and others",
    collaboration = "LIGO Scientific, Virgo",
    title = "{GW170817: Observation of Gravitational Waves from a Binary Neutron Star Inspiral}",
    eprint = "1710.05832",
    archivePrefix = "arXiv",
    primaryClass = "gr-qc",
    reportNumber = "LIGO-P170817",
    doi = "10.1103/PhysRevLett.119.161101",
    journal = "Phys. Rev. Lett.",
    volume = "119",
    number = "16",
    pages = "161101",
    year = "2017"
}

@article{Bahamonde:2026vqm,
    author = "Bahamonde, Sebastian and Gigante Valcarcel, Jorge",
    title = "{Black hole superradiance in Poincar{\'e} gauge theory}",
    eprint = "2603.19140",
    archivePrefix = "arXiv",
    primaryClass = "gr-qc",
    month = "3",
    year = "2026"
}

@article{Kiselev:2002dx,
    author = "Kiselev, V. V.",
    title = "{Quintessence and black holes}",
    eprint = "gr-qc/0210040",
    archivePrefix = "arXiv",
    doi = "10.1088/0264-9381/20/6/310",
    journal = "Class. Quant. Grav.",
    volume = "20",
    pages = "1187--1198",
    year = "2003"
}

@article{Visser:2019brz,
    author = "Visser, Matt",
    title = "{The Kiselev black hole is neither perfect fluid, nor is it quintessence}",
    eprint = "1908.11058",
    archivePrefix = "arXiv",
    primaryClass = "gr-qc",
    doi = "10.1088/1361-6382/ab60b8",
    journal = "Class. Quant. Grav.",
    volume = "37",
    number = "4",
    pages = "045001",
    year = "2020"
}

\end{document}